\documentclass[longauth]{aa}
\usepackage{lipsum}

\usepackage{amsmath,amssymb,amsfonts}
\usepackage{graphicx}
\usepackage{txfonts}
\usepackage{enumerate} 
\usepackage{colordvi} 
\usepackage{bm}
\usepackage{epsfig}
\usepackage{hyperref}
\usepackage[dvipsnames]{xcolor}
\usepackage{color, colortbl}
\usepackage[T1]{fontenc} 
\usepackage{braket}
\usepackage{euclid}
\usepackage{placeins}
\usepackage{tabularx}
\usepackage{comment}
\graphicspath{{./Figures/}}
\usepackage[switch, mathlines]{lineno}
\usepackage[normalem]{ulem}
\usepackage{multirow}
\usepackage{csquotes}

\newcommand{\bfs}{\mbox{\boldmath$s$}}
\newcommand{\bfz}{\mbox{\boldmath$z$}}

\newcommand{\bfx}{\mbox{\boldmath$x$}}
\newcommand{\bfk}{\mbox{\boldmath$k$}}
\newcommand{\bfp}{\mbox{\boldmath$p$}}
\newcommand{\bfq}{\mbox{\boldmath$q$}}
\newcommand{\bfr}{\mbox{\boldmath$r$}}

\definecolor{amaranth}{rgb}{0.9, 0.17, 0.31}
\definecolor{forestgreen(web)}{rgb}{0.13, 0.55, 0.13}
\definecolor{lavender(web)}{rgb}{0.9, 0.9, 0.98}
\definecolor{cosmiclatte}{rgb}{1.0, 0.97, 0.91}
\definecolor{jonquil}{rgb}{0.98, 0.85, 0.37}
\definecolor{khaki(x11)(lightkhaki)}{rgb}{0.94, 0.9, 0.55}
\definecolor{thistle}{rgb}{0.85, 0.75, 0.85}

\newcommand{\xv}{\mathbf{x}}

\newcommand{\kv}{\mathbf{k}}

\newcommand{\qv}{\mathbf{q}}

\newcommand{\Refr}[1]{{\color{black}{{}~#1}}}

\defcitealias{Blanchard:2019oqi}{EC19}

\usepackage[compress,nameinlink,capitalise]{cleveref}
\crefname{equation}{Eq.}{Eqs.}
\Crefname{equation}{Equation}{Equations}
\crefname{chapter}{Chap.}{Chaps.}
\crefname{section}{Sect.}{Sects.}
\crefname{figure}{Fig.}{Figs.}
\Crefname{chapter}{Chapter}{Chapters}
\Crefname{section}{Section}{Sections}
\Crefname{figure}{Figure}{Figures}

\crefrangeformat{equation}{Eqs.~(#3#1#4--#5#2#6)}
\crefmultiformat{equation}{Eqs.~(#2#1#3}{, #2#1#3)}{, #2#1#3}{, #2#1#3)}

\begin{document}

\title{\Euclid preparation. XLIV. Modelling spectroscopic clustering on mildly nonlinear scales in beyond-\texorpdfstring{$\Lambda$}{L}CDM models}

\newcommand{\orcid}[1]{} 

\author{Euclid Collaboration: B.~Bose\orcid{0000-0003-1965-8614}\inst{\ref{aff1}}\thanks{\email{ben.bose@ed.ac.uk}}
\and P.~Carrilho\orcid{0000-0003-1339-0194}\inst{\ref{aff1}}
\and M.~Marinucci\orcid{0000-0003-1159-3756}\inst{\ref{aff2},\ref{aff3}}
\and C.~Moretti\orcid{0000-0003-3314-8936}\inst{\ref{aff4},\ref{aff1},\ref{aff5},\ref{aff6}}
\and M.~Pietroni\orcid{0000-0001-5480-5996}\inst{\ref{aff2},\ref{aff7}}
\and E.~Carella\inst{\ref{aff8},\ref{aff9}}
\and L.~Piga\orcid{0000-0003-2221-7406}\inst{\ref{aff2},\ref{aff7}}
\and B.~S.~Wright\orcid{0000-0001-6364-1639}\inst{\ref{aff10}}
\and F.~Vernizzi\orcid{0000-0003-3426-2802}\inst{\ref{aff11}}
\and C.~Carbone\orcid{0000-0003-0125-3563}\inst{\ref{aff9}}
\and S.~Casas\orcid{0000-0002-4751-5138}\inst{\ref{aff12}}
\and G.~D'Amico\inst{\ref{aff2},\ref{aff7}}
\and N.~Frusciante\inst{\ref{aff13}}
\and K.~Koyama\orcid{0000-0001-6727-6915}\inst{\ref{aff14}}
\and F.~Pace\orcid{0000-0001-8039-0480}\inst{\ref{aff15},\ref{aff16},\ref{aff17}}
\and A.~Pourtsidou\orcid{0000-0001-9110-5550}\inst{\ref{aff1},\ref{aff18}}
\and M.~Baldi\orcid{0000-0003-4145-1943}\inst{\ref{aff19},\ref{aff20},\ref{aff21}}
\and L.~F.~de~la~Bella\orcid{0000-0002-1064-3400}\inst{\ref{aff14}}
\and B.~Fiorini\orcid{0000-0002-0092-4321}\inst{\ref{aff14},\ref{aff10}}
\and C.~Giocoli\orcid{0000-0002-9590-7961}\inst{\ref{aff20},\ref{aff22}}
\and L.~Lombriser\inst{\ref{aff23}}
\and N.~Aghanim\inst{\ref{aff24}}
\and A.~Amara\inst{\ref{aff14},\ref{aff25}}
\and S.~Andreon\orcid{0000-0002-2041-8784}\inst{\ref{aff26}}
\and N.~Auricchio\orcid{0000-0003-4444-8651}\inst{\ref{aff20}}
\and S.~Bardelli\orcid{0000-0002-8900-0298}\inst{\ref{aff20}}
\and C.~Bodendorf\inst{\ref{aff27}}
\and D.~Bonino\inst{\ref{aff17}}
\and E.~Branchini\orcid{0000-0002-0808-6908}\inst{\ref{aff28},\ref{aff29},\ref{aff26}}
\and M.~Brescia\orcid{0000-0001-9506-5680}\inst{\ref{aff13},\ref{aff30},\ref{aff31}}
\and J.~Brinchmann\orcid{0000-0003-4359-8797}\inst{\ref{aff32}}
\and S.~Camera\orcid{0000-0003-3399-3574}\inst{\ref{aff15},\ref{aff16},\ref{aff17}}
\and V.~Capobianco\orcid{0000-0002-3309-7692}\inst{\ref{aff17}}
\and V.~F.~Cardone\inst{\ref{aff33},\ref{aff34}}
\and J.~Carretero\orcid{0000-0002-3130-0204}\inst{\ref{aff35},\ref{aff36}}
\and M.~Castellano\orcid{0000-0001-9875-8263}\inst{\ref{aff33}}
\and S.~Cavuoti\orcid{0000-0002-3787-4196}\inst{\ref{aff30},\ref{aff31}}
\and A.~Cimatti\inst{\ref{aff37}}
\and G.~Congedo\orcid{0000-0003-2508-0046}\inst{\ref{aff1}}
\and C.~J.~Conselice\inst{\ref{aff38}}
\and L.~Conversi\orcid{0000-0002-6710-8476}\inst{\ref{aff39},\ref{aff40}}
\and Y.~Copin\orcid{0000-0002-5317-7518}\inst{\ref{aff41}}
\and A.~Costille\inst{\ref{aff42}}
\and F.~Courbin\orcid{0000-0003-0758-6510}\inst{\ref{aff43}}
\and H.~M.~Courtois\orcid{0000-0003-0509-1776}\inst{\ref{aff44}}
\and A.~Da~Silva\orcid{0000-0002-6385-1609}\inst{\ref{aff45},\ref{aff46}}
\and H.~Degaudenzi\orcid{0000-0002-5887-6799}\inst{\ref{aff47}}
\and A.~M.~Di~Giorgio\orcid{0000-0002-4767-2360}\inst{\ref{aff48}}
\and F.~Dubath\orcid{0000-0002-6533-2810}\inst{\ref{aff47}}
\and C.~A.~J.~Duncan\inst{\ref{aff38},\ref{aff49}}
\and X.~Dupac\inst{\ref{aff40}}
\and S.~Dusini\orcid{0000-0002-1128-0664}\inst{\ref{aff50}}
\and M.~Farina\orcid{0000-0002-3089-7846}\inst{\ref{aff48}}
\and S.~Farrens\orcid{0000-0002-9594-9387}\inst{\ref{aff51}}
\and S.~Ferriol\inst{\ref{aff41}}
\and P.~Fosalba\orcid{0000-0002-1510-5214}\inst{\ref{aff52},\ref{aff53}}
\and M.~Frailis\orcid{0000-0002-7400-2135}\inst{\ref{aff54}}
\and E.~Franceschi\orcid{0000-0002-0585-6591}\inst{\ref{aff20}}
\and S.~Galeotta\orcid{0000-0002-3748-5115}\inst{\ref{aff54}}
\and B.~Garilli\orcid{0000-0001-7455-8750}\inst{\ref{aff9}}
\and B.~Gillis\orcid{0000-0002-4478-1270}\inst{\ref{aff1}}
\and A.~Grazian\orcid{0000-0002-5688-0663}\inst{\ref{aff55}}
\and F.~Grupp\inst{\ref{aff27},\ref{aff56}}
\and L.~Guzzo\orcid{0000-0001-8264-5192}\inst{\ref{aff8},\ref{aff26},\ref{aff57}}
\and S.~V.~H.~Haugan\orcid{0000-0001-9648-7260}\inst{\ref{aff58}}
\and F.~Hormuth\inst{\ref{aff59}}
\and A.~Hornstrup\orcid{0000-0002-3363-0936}\inst{\ref{aff60},\ref{aff61}}
\and K.~Jahnke\orcid{0000-0003-3804-2137}\inst{\ref{aff62}}
\and B.~Joachimi\orcid{0000-0001-7494-1303}\inst{\ref{aff63}}
\and E.~Keih\"anen\orcid{0000-0003-1804-7715}\inst{\ref{aff64}}
\and S.~Kermiche\orcid{0000-0002-0302-5735}\inst{\ref{aff65}}
\and A.~Kiessling\orcid{0000-0002-2590-1273}\inst{\ref{aff66}}
\and M.~Kilbinger\orcid{0000-0001-9513-7138}\inst{\ref{aff67}}
\and T.~Kitching\orcid{0000-0002-4061-4598}\inst{\ref{aff68}}
\and M.~Kunz\orcid{0000-0002-3052-7394}\inst{\ref{aff23}}
\and H.~Kurki-Suonio\orcid{0000-0002-4618-3063}\inst{\ref{aff69},\ref{aff70}}
\and S.~Ligori\orcid{0000-0003-4172-4606}\inst{\ref{aff17}}
\and P.~B.~Lilje\orcid{0000-0003-4324-7794}\inst{\ref{aff58}}
\and V.~Lindholm\orcid{0000-0003-2317-5471}\inst{\ref{aff69},\ref{aff70}}
\and I.~Lloro\inst{\ref{aff71}}
\and D.~Maino\inst{\ref{aff8},\ref{aff9},\ref{aff57}}
\and E.~Maiorano\orcid{0000-0003-2593-4355}\inst{\ref{aff20}}
\and O.~Mansutti\orcid{0000-0001-5758-4658}\inst{\ref{aff54}}
\and O.~Marggraf\orcid{0000-0001-7242-3852}\inst{\ref{aff72}}
\and K.~Markovic\orcid{0000-0001-6764-073X}\inst{\ref{aff66}}
\and N.~Martinet\orcid{0000-0003-2786-7790}\inst{\ref{aff42}}
\and F.~Marulli\orcid{0000-0002-8850-0303}\inst{\ref{aff73},\ref{aff20},\ref{aff21}}
\and R.~Massey\orcid{0000-0002-6085-3780}\inst{\ref{aff74}}
\and E.~Medinaceli\orcid{0000-0002-4040-7783}\inst{\ref{aff20}}
\and M.~Meneghetti\orcid{0000-0003-1225-7084}\inst{\ref{aff20},\ref{aff21}}
\and G.~Meylan\inst{\ref{aff43}}
\and M.~Moresco\orcid{0000-0002-7616-7136}\inst{\ref{aff73},\ref{aff20}}
\and L.~Moscardini\orcid{0000-0002-3473-6716}\inst{\ref{aff73},\ref{aff20},\ref{aff21}}
\and D.~F.~Mota\orcid{0000-0003-3141-142X}\inst{\ref{aff58}}
\and E.~Munari\orcid{0000-0002-1751-5946}\inst{\ref{aff54}}
\and S.-M.~Niemi\inst{\ref{aff75}}
\and C.~Padilla\orcid{0000-0001-7951-0166}\inst{\ref{aff35}}
\and S.~Paltani\inst{\ref{aff47}}
\and F.~Pasian\inst{\ref{aff54}}
\and K.~Pedersen\inst{\ref{aff76}}
\and W.~J.~Percival\orcid{0000-0002-0644-5727}\inst{\ref{aff77},\ref{aff78},\ref{aff79}}
\and V.~Pettorino\inst{\ref{aff80}}
\and S.~Pires\orcid{0000-0002-0249-2104}\inst{\ref{aff51}}
\and G.~Polenta\orcid{0000-0003-4067-9196}\inst{\ref{aff81}}
\and M.~Poncet\inst{\ref{aff82}}
\and L.~A.~Popa\inst{\ref{aff83}}
\and L.~Pozzetti\orcid{0000-0001-7085-0412}\inst{\ref{aff20}}
\and F.~Raison\orcid{0000-0002-7819-6918}\inst{\ref{aff27}}
\and A.~Renzi\orcid{0000-0001-9856-1970}\inst{\ref{aff84},\ref{aff50}}
\and J.~Rhodes\inst{\ref{aff66}}
\and G.~Riccio\inst{\ref{aff30}}
\and E.~Romelli\orcid{0000-0003-3069-9222}\inst{\ref{aff54}}
\and M.~Roncarelli\orcid{0000-0001-9587-7822}\inst{\ref{aff20}}
\and R.~Saglia\orcid{0000-0003-0378-7032}\inst{\ref{aff85},\ref{aff27}}
\and D.~Sapone\orcid{0000-0001-7089-4503}\inst{\ref{aff86}}
\and B.~Sartoris\inst{\ref{aff85},\ref{aff54}}
\and P.~Schneider\orcid{0000-0001-8561-2679}\inst{\ref{aff72}}
\and A.~Secroun\orcid{0000-0003-0505-3710}\inst{\ref{aff65}}
\and G.~Seidel\orcid{0000-0003-2907-353X}\inst{\ref{aff62}}
\and M.~Seiffert\orcid{0000-0002-7536-9393}\inst{\ref{aff66}}
\and S.~Serrano\orcid{0000-0002-0211-2861}\inst{\ref{aff52},\ref{aff87},\ref{aff88}}
\and C.~Sirignano\orcid{0000-0002-0995-7146}\inst{\ref{aff84},\ref{aff50}}
\and G.~Sirri\orcid{0000-0003-2626-2853}\inst{\ref{aff21}}
\and L.~Stanco\orcid{0000-0002-9706-5104}\inst{\ref{aff50}}
\and J.-L.~Starck\orcid{0000-0003-2177-7794}\inst{\ref{aff67}}
\and P.~Tallada-Cresp\'{i}\orcid{0000-0002-1336-8328}\inst{\ref{aff89},\ref{aff36}}
\and A.~N.~Taylor\inst{\ref{aff1}}
\and I.~Tereno\inst{\ref{aff45},\ref{aff90}}
\and R.~Toledo-Moreo\orcid{0000-0002-2997-4859}\inst{\ref{aff91}}
\and F.~Torradeflot\orcid{0000-0003-1160-1517}\inst{\ref{aff36},\ref{aff89}}
\and I.~Tutusaus\orcid{0000-0002-3199-0399}\inst{\ref{aff92}}
\and E.~A.~Valentijn\inst{\ref{aff93}}
\and L.~Valenziano\orcid{0000-0002-1170-0104}\inst{\ref{aff20},\ref{aff94}}
\and T.~Vassallo\orcid{0000-0001-6512-6358}\inst{\ref{aff85},\ref{aff54}}
\and A.~Veropalumbo\orcid{0000-0003-2387-1194}\inst{\ref{aff26},\ref{aff29}}
\and Y.~Wang\orcid{0000-0002-4749-2984}\inst{\ref{aff95}}
\and J.~Weller\orcid{0000-0002-8282-2010}\inst{\ref{aff85},\ref{aff27}}
\and G.~Zamorani\orcid{0000-0002-2318-301X}\inst{\ref{aff20}}
\and J.~Zoubian\inst{\ref{aff65}}
\and E.~Zucca\orcid{0000-0002-5845-8132}\inst{\ref{aff20}}
\and A.~Biviano\orcid{0000-0002-0857-0732}\inst{\ref{aff54},\ref{aff6}}
\and E.~Bozzo\orcid{0000-0002-8201-1525}\inst{\ref{aff47}}
\and C.~Burigana\orcid{0000-0002-3005-5796}\inst{\ref{aff96},\ref{aff94}}
\and C.~Colodro-Conde\inst{\ref{aff97}}
\and D.~Di~Ferdinando\inst{\ref{aff21}}
\and J.~Graci\'{a}-Carpio\inst{\ref{aff27}}
\and N.~Mauri\orcid{0000-0001-8196-1548}\inst{\ref{aff37},\ref{aff21}}
\and C.~Neissner\inst{\ref{aff35},\ref{aff36}}
\and Z.~Sakr\orcid{0000-0002-4823-3757}\inst{\ref{aff98},\ref{aff92},\ref{aff99}}
\and V.~Scottez\inst{\ref{aff100},\ref{aff101}}
\and M.~Tenti\orcid{0000-0002-4254-5901}\inst{\ref{aff21}}
\and M.~Viel\inst{\ref{aff6},\ref{aff54},\ref{aff4},\ref{aff102},\ref{aff5}}
\and M.~Wiesmann\inst{\ref{aff58}}
\and Y.~Akrami\orcid{0000-0002-2407-7956}\inst{\ref{aff103},\ref{aff104}}
\and V.~Allevato\orcid{0000-0001-7232-5152}\inst{\ref{aff30}}
\and S.~Anselmi\orcid{0000-0002-3579-9583}\inst{\ref{aff84},\ref{aff50},\ref{aff105}}
\and M.~Ballardini\inst{\ref{aff106},\ref{aff107},\ref{aff20}}
\and F.~Bernardeau\inst{\ref{aff11},\ref{aff108}}
\and S.~Borgani\orcid{0000-0001-6151-6439}\inst{\ref{aff109},\ref{aff6},\ref{aff54},\ref{aff102}}
\and S.~Bruton\orcid{0000-0002-6503-5218}\inst{\ref{aff110}}
\and R.~Cabanac\orcid{0000-0001-6679-2600}\inst{\ref{aff92}}
\and A.~Cappi\inst{\ref{aff20},\ref{aff111}}
\and C.~S.~Carvalho\inst{\ref{aff90}}
\and G.~Castignani\orcid{0000-0001-6831-0687}\inst{\ref{aff73},\ref{aff20}}
\and T.~Castro\orcid{0000-0002-6292-3228}\inst{\ref{aff54},\ref{aff102},\ref{aff6},\ref{aff5}}
\and G.~Ca\~{n}as-Herrera\orcid{0000-0003-2796-2149}\inst{\ref{aff75},\ref{aff112}}
\and K.~C.~Chambers\orcid{0000-0001-6965-7789}\inst{\ref{aff113}}
\and A.~R.~Cooray\orcid{0000-0002-3892-0190}\inst{\ref{aff114}}
\and J.~Coupon\inst{\ref{aff47}}
\and S.~Davini\inst{\ref{aff29}}
\and S.~de~la~Torre\inst{\ref{aff42}}
\and G.~De~Lucia\orcid{0000-0002-6220-9104}\inst{\ref{aff54}}
\and G.~Desprez\inst{\ref{aff115}}
\and S.~Di~Domizio\orcid{0000-0003-2863-5895}\inst{\ref{aff28},\ref{aff29}}
\and H.~Dole\inst{\ref{aff24}}
\and A.~D\'{i}az-S\'{a}nchez\orcid{0000-0003-0748-4768}\inst{\ref{aff116}}
\and J.~A.~Escartin~Vigo\inst{\ref{aff27}}
\and S.~Escoffier\orcid{0000-0002-2847-7498}\inst{\ref{aff65}}
\and P.~G.~Ferreira\inst{\ref{aff49}}
\and I.~Ferrero\orcid{0000-0002-1295-1132}\inst{\ref{aff58}}
\and F.~Finelli\orcid{0000-0002-6694-3269}\inst{\ref{aff20},\ref{aff94}}
\and L.~Gabarra\inst{\ref{aff84},\ref{aff50}}
\and K.~Ganga\orcid{0000-0001-8159-8208}\inst{\ref{aff117}}
\and J.~Garc\'ia-Bellido\orcid{0000-0002-9370-8360}\inst{\ref{aff103}}
\and F.~Giacomini\orcid{0000-0002-3129-2814}\inst{\ref{aff21}}
\and G.~Gozaliasl\orcid{0000-0002-0236-919X}\inst{\ref{aff118},\ref{aff69}}
\and D.~Guinet\orcid{0000-0002-8132-6509}\inst{\ref{aff41}}
\and A.~Hall\orcid{0000-0002-3139-8651}\inst{\ref{aff1}}
\and S.~Joudaki\orcid{0000-0001-8820-673X}\inst{\ref{aff14},\ref{aff77},\ref{aff78}}
\and J.~J.~E.~Kajava\orcid{0000-0002-3010-8333}\inst{\ref{aff119},\ref{aff120}}
\and V.~Kansal\inst{\ref{aff121},\ref{aff122},\ref{aff123}}
\and D.~Karagiannis\orcid{0000-0002-4927-0816}\inst{\ref{aff124}}
\and C.~C.~Kirkpatrick\inst{\ref{aff64}}
\and L.~Legrand\orcid{0000-0003-0610-5252}\inst{\ref{aff23}}
\and A.~Loureiro\orcid{0000-0002-4371-0876}\inst{\ref{aff125},\ref{aff126}}
\and J.~Macias-Perez\orcid{0000-0002-5385-2763}\inst{\ref{aff127}}
\and M.~Magliocchetti\orcid{0000-0001-9158-4838}\inst{\ref{aff48}}
\and R.~Maoli\orcid{0000-0002-6065-3025}\inst{\ref{aff128},\ref{aff33}}
\and M.~Martinelli\orcid{0000-0002-6943-7732}\inst{\ref{aff33},\ref{aff34}}
\and C.~J.~A.~P.~Martins\orcid{0000-0002-4886-9261}\inst{\ref{aff129},\ref{aff32}}
\and S.~Matthew\inst{\ref{aff1}}
\and M.~Maturi\orcid{0000-0002-3517-2422}\inst{\ref{aff98},\ref{aff130}}
\and L.~Maurin\orcid{0000-0002-8406-0857}\inst{\ref{aff24}}
\and R.~B.~Metcalf\orcid{0000-0003-3167-2574}\inst{\ref{aff73},\ref{aff20}}
\and M.~Migliaccio\inst{\ref{aff131},\ref{aff132}}
\and P.~Monaco\inst{\ref{aff109},\ref{aff54},\ref{aff102},\ref{aff6}}
\and G.~Morgante\inst{\ref{aff20}}
\and S.~Nadathur\orcid{0000-0001-9070-3102}\inst{\ref{aff14}}
\and Nicholas~A.~Walton\orcid{0000-0003-3983-8778}\inst{\ref{aff133}}
\and L.~Patrizii\inst{\ref{aff21}}
\and A.~Pezzotta\inst{\ref{aff27}}
\and V.~Popa\inst{\ref{aff83}}
\and C.~Porciani\orcid{0000-0002-7797-2508}\inst{\ref{aff72}}
\and D.~Potter\orcid{0000-0002-0757-5195}\inst{\ref{aff134}}
\and M.~P\"{o}ntinen\orcid{0000-0001-5442-2530}\inst{\ref{aff69}}
\and P.~Reimberg\orcid{0000-0003-3410-0280}\inst{\ref{aff100}}
\and P.-F.~Rocci\inst{\ref{aff24}}
\and A.~G.~S\'anchez\orcid{0000-0003-1198-831X}\inst{\ref{aff27}}
\and A.~Schneider\orcid{0000-0001-7055-8104}\inst{\ref{aff134}}
\and E.~Sefusatti\orcid{0000-0003-0473-1567}\inst{\ref{aff54},\ref{aff102},\ref{aff6}}
\and M.~Sereno\orcid{0000-0003-0302-0325}\inst{\ref{aff20},\ref{aff21}}
\and A.~Silvestri\orcid{0000-0001-6904-5061}\inst{\ref{aff112}}
\and A.~Spurio~Mancini\orcid{0000-0001-5698-0990}\inst{\ref{aff68}}
\and J.~Steinwagner\inst{\ref{aff27}}
\and G.~Testera\inst{\ref{aff29}}
\and R.~Teyssier\orcid{0000-0001-7689-0933}\inst{\ref{aff135}}
\and S.~Toft\orcid{0000-0003-3631-7176}\inst{\ref{aff61},\ref{aff136},\ref{aff137}}
\and S.~Tosi\orcid{0000-0002-7275-9193}\inst{\ref{aff28},\ref{aff29},\ref{aff26}}
\and A.~Troja\orcid{0000-0003-0239-4595}\inst{\ref{aff84},\ref{aff50}}
\and M.~Tucci\inst{\ref{aff47}}
\and J.~Valiviita\orcid{0000-0001-6225-3693}\inst{\ref{aff69},\ref{aff70}}
\and D.~Vergani\orcid{0000-0003-0898-2216}\inst{\ref{aff20}}}

\institute{Institute for Astronomy, University of Edinburgh, Royal Observatory, Blackford Hill, Edinburgh EH9 3HJ, UK\label{aff1}
\and
Dipartimento di Scienze Matematiche, Fisiche e Informatiche, Universit\`a di Parma, Viale delle Scienze 7/A 43124 Parma, Italy\label{aff2}
\and
Technion Israel Institute of Technology, Israel\label{aff3}
\and
SISSA, International School for Advanced Studies, Via Bonomea 265, 34136 Trieste TS, Italy\label{aff4}
\and
ICSC - Centro Nazionale di Ricerca in High Performance Computing, Big Data e Quantum Computing, Via Magnanelli 2, Bologna, Italy\label{aff5}
\and
IFPU, Institute for Fundamental Physics of the Universe, via Beirut 2, 34151 Trieste, Italy\label{aff6}
\and
INFN Gruppo Collegato di Parma, Viale delle Scienze 7/A 43124 Parma, Italy\label{aff7}
\and
Dipartimento di Fisica "Aldo Pontremoli", Universit\`a degli Studi di Milano, Via Celoria 16, 20133 Milano, Italy\label{aff8}
\and
INAF-IASF Milano, Via Alfonso Corti 12, 20133 Milano, Italy\label{aff9}
\and
School of Physics and Astronomy, Queen Mary University of London, Mile End Road, London E1 4NS, UK\label{aff10}
\and
Institut de Physique Th\'eorique, CEA, CNRS, Universit\'e Paris-Saclay 91191 Gif-sur-Yvette Cedex, France\label{aff11}
\and
Institute for Theoretical Particle Physics and Cosmology (TTK), RWTH Aachen University, 52056 Aachen, Germany\label{aff12}
\and
Department of Physics "E. Pancini", University Federico II, Via Cinthia 6, 80126, Napoli, Italy\label{aff13}
\and
Institute of Cosmology and Gravitation, University of Portsmouth, Portsmouth PO1 3FX, UK\label{aff14}
\and
Dipartimento di Fisica, Universit\`a degli Studi di Torino, Via P. Giuria 1, 10125 Torino, Italy\label{aff15}
\and
INFN-Sezione di Torino, Via P. Giuria 1, 10125 Torino, Italy\label{aff16}
\and
INAF-Osservatorio Astrofisico di Torino, Via Osservatorio 20, 10025 Pino Torinese (TO), Italy\label{aff17}
\and
Higgs Centre for Theoretical Physics, School of Physics and Astronomy, The University of Edinburgh, Edinburgh EH9 3FD, UK\label{aff18}
\and
Dipartimento di Fisica e Astronomia, Universit\`a di Bologna, Via Gobetti 93/2, 40129 Bologna, Italy\label{aff19}
\and
INAF-Osservatorio di Astrofisica e Scienza dello Spazio di Bologna, Via Piero Gobetti 93/3, 40129 Bologna, Italy\label{aff20}
\and
INFN-Sezione di Bologna, Viale Berti Pichat 6/2, 40127 Bologna, Italy\label{aff21}
\and
Istituto Nazionale di Fisica Nucleare, Sezione di Bologna, Via Irnerio 46, 40126 Bologna, Italy\label{aff22}
\and
Universit\'e de Gen\`eve, D\'epartement de Physique Th\'eorique and Centre for Astroparticle Physics, 24 quai Ernest-Ansermet, CH-1211 Gen\`eve 4, Switzerland\label{aff23}
\and
Universit\'e Paris-Saclay, CNRS, Institut d'astrophysique spatiale, 91405, Orsay, France\label{aff24}
\and
School of Mathematics and Physics, University of Surrey, Guildford, Surrey, GU2 7XH, UK\label{aff25}
\and
INAF-Osservatorio Astronomico di Brera, Via Brera 28, 20122 Milano, Italy\label{aff26}
\and
Max Planck Institute for Extraterrestrial Physics, Giessenbachstr. 1, 85748 Garching, Germany\label{aff27}
\and
Dipartimento di Fisica, Universit\`a di Genova, Via Dodecaneso 33, 16146, Genova, Italy\label{aff28}
\and
INFN-Sezione di Genova, Via Dodecaneso 33, 16146, Genova, Italy\label{aff29}
\and
INAF-Osservatorio Astronomico di Capodimonte, Via Moiariello 16, 80131 Napoli, Italy\label{aff30}
\and
INFN section of Naples, Via Cinthia 6, 80126, Napoli, Italy\label{aff31}
\and
Instituto de Astrof\'isica e Ci\^encias do Espa\c{c}o, Universidade do Porto, CAUP, Rua das Estrelas, PT4150-762 Porto, Portugal\label{aff32}
\and
INAF-Osservatorio Astronomico di Roma, Via Frascati 33, 00078 Monteporzio Catone, Italy\label{aff33}
\and
INFN-Sezione di Roma, Piazzale Aldo Moro, 2 - c/o Dipartimento di Fisica, Edificio G. Marconi, 00185 Roma, Italy\label{aff34}
\and
Institut de F\'{i}sica d'Altes Energies (IFAE), The Barcelona Institute of Science and Technology, Campus UAB, 08193 Bellaterra (Barcelona), Spain\label{aff35}
\and
Port d'Informaci\'{o} Cient\'{i}fica, Campus UAB, C. Albareda s/n, 08193 Bellaterra (Barcelona), Spain\label{aff36}
\and
Dipartimento di Fisica e Astronomia "Augusto Righi" - Alma Mater Studiorum Universit\`a di Bologna, Viale Berti Pichat 6/2, 40127 Bologna, Italy\label{aff37}
\and
Jodrell Bank Centre for Astrophysics, Department of Physics and Astronomy, University of Manchester, Oxford Road, Manchester M13 9PL, UK\label{aff38}
\and
European Space Agency/ESRIN, Largo Galileo Galilei 1, 00044 Frascati, Roma, Italy\label{aff39}
\and
ESAC/ESA, Camino Bajo del Castillo, s/n., Urb. Villafranca del Castillo, 28692 Villanueva de la Ca\~nada, Madrid, Spain\label{aff40}
\and
University of Lyon, Univ Claude Bernard Lyon 1, CNRS/IN2P3, IP2I Lyon, UMR 5822, 69622 Villeurbanne, France\label{aff41}
\and
Aix-Marseille Universit\'e, CNRS, CNES, LAM, Marseille, France\label{aff42}
\and
Institute of Physics, Laboratory of Astrophysics, Ecole Polytechnique F\'ed\'erale de Lausanne (EPFL), Observatoire de Sauverny, 1290 Versoix, Switzerland\label{aff43}
\and
UCB Lyon 1, CNRS/IN2P3, IUF, IP2I Lyon, 4 rue Enrico Fermi, 69622 Villeurbanne, France\label{aff44}
\and
Departamento de F\'isica, Faculdade de Ci\^encias, Universidade de Lisboa, Edif\'icio C8, Campo Grande, PT1749-016 Lisboa, Portugal\label{aff45}
\and
Instituto de Astrof\'isica e Ci\^encias do Espa\c{c}o, Faculdade de Ci\^encias, Universidade de Lisboa, Campo Grande, 1749-016 Lisboa, Portugal\label{aff46}
\and
Department of Astronomy, University of Geneva, ch. d'Ecogia 16, 1290 Versoix, Switzerland\label{aff47}
\and
INAF-Istituto di Astrofisica e Planetologia Spaziali, via del Fosso del Cavaliere, 100, 00100 Roma, Italy\label{aff48}
\and
Department of Physics, Oxford University, Keble Road, Oxford OX1 3RH, UK\label{aff49}
\and
INFN-Padova, Via Marzolo 8, 35131 Padova, Italy\label{aff50}
\and
Universit\'e Paris-Saclay, Universit\'e Paris Cit\'e, CEA, CNRS, AIM, 91191, Gif-sur-Yvette, France\label{aff51}
\and
Institut d'Estudis Espacials de Catalunya (IEEC), Carrer Gran Capit\'a 2-4, 08034 Barcelona, Spain\label{aff52}
\and
Institut de Ciencies de l'Espai (IEEC-CSIC), Campus UAB, Carrer de Can Magrans, s/n Cerdanyola del Vall\'es, 08193 Barcelona, Spain\label{aff53}
\and
INAF-Osservatorio Astronomico di Trieste, Via G. B. Tiepolo 11, 34143 Trieste, Italy\label{aff54}
\and
INAF-Osservatorio Astronomico di Padova, Via dell'Osservatorio 5, 35122 Padova, Italy\label{aff55}
\and
University Observatory, Faculty of Physics, Ludwig-Maximilians-Universit{\"a}t, Scheinerstr. 1, 81679 Munich, Germany\label{aff56}
\and
INFN-Sezione di Milano, Via Celoria 16, 20133 Milano, Italy\label{aff57}
\and
Institute of Theoretical Astrophysics, University of Oslo, P.O. Box 1029 Blindern, 0315 Oslo, Norway\label{aff58}
\and
von Hoerner \& Sulger GmbH, Schlo{\ss}Platz 8, 68723 Schwetzingen, Germany\label{aff59}
\and
Technical University of Denmark, Elektrovej 327, 2800 Kgs. Lyngby, Denmark\label{aff60}
\and
Cosmic Dawn Center (DAWN), Denmark\label{aff61}
\and
Max-Planck-Institut f\"ur Astronomie, K\"onigstuhl 17, 69117 Heidelberg, Germany\label{aff62}
\and
Department of Physics and Astronomy, University College London, Gower Street, London WC1E 6BT, UK\label{aff63}
\and
Department of Physics and Helsinki Institute of Physics, Gustaf H\"allstr\"omin katu 2, 00014 University of Helsinki, Finland\label{aff64}
\and
Aix-Marseille Universit\'e, CNRS/IN2P3, CPPM, Marseille, France\label{aff65}
\and
Jet Propulsion Laboratory, California Institute of Technology, 4800 Oak Grove Drive, Pasadena, CA, 91109, USA\label{aff66}
\and
AIM, CEA, CNRS, Universit\'{e} Paris-Saclay, Universit\'{e} de Paris, 91191 Gif-sur-Yvette, France\label{aff67}
\and
Mullard Space Science Laboratory, University College London, Holmbury St Mary, Dorking, Surrey RH5 6NT, UK\label{aff68}
\and
Department of Physics, P.O. Box 64, 00014 University of Helsinki, Finland\label{aff69}
\and
Helsinki Institute of Physics, Gustaf H{\"a}llstr{\"o}min katu 2, University of Helsinki, Helsinki, Finland\label{aff70}
\and
NOVA optical infrared instrumentation group at ASTRON, Oude Hoogeveensedijk 4, 7991PD, Dwingeloo, The Netherlands\label{aff71}
\and
Universit\"at Bonn, Argelander-Institut f\"ur Astronomie, Auf dem H\"ugel 71, 53121 Bonn, Germany\label{aff72}
\and
Dipartimento di Fisica e Astronomia "Augusto Righi" - Alma Mater Studiorum Universit\`a di Bologna, via Piero Gobetti 93/2, 40129 Bologna, Italy\label{aff73}
\and
Department of Physics, Institute for Computational Cosmology, Durham University, South Road, DH1 3LE, UK\label{aff74}
\and
European Space Agency/ESTEC, Keplerlaan 1, 2201 AZ Noordwijk, The Netherlands\label{aff75}
\and
Department of Physics and Astronomy, University of Aarhus, Ny Munkegade 120, DK-8000 Aarhus C, Denmark\label{aff76}
\and
Centre for Astrophysics, University of Waterloo, Waterloo, Ontario N2L 3G1, Canada\label{aff77}
\and
Department of Physics and Astronomy, University of Waterloo, Waterloo, Ontario N2L 3G1, Canada\label{aff78}
\and
Perimeter Institute for Theoretical Physics, Waterloo, Ontario N2L 2Y5, Canada\label{aff79}
\and
Universit\'e Paris-Saclay, Universit\'e Paris Cit\'e, CEA, CNRS, Astrophysique, Instrumentation et Mod\'elisation Paris-Saclay, 91191 Gif-sur-Yvette, France\label{aff80}
\and
Space Science Data Center, Italian Space Agency, via del Politecnico snc, 00133 Roma, Italy\label{aff81}
\and
Centre National d'Etudes Spatiales -- Centre spatial de Toulouse, 18 avenue Edouard Belin, 31401 Toulouse Cedex 9, France\label{aff82}
\and
Institute of Space Science, Str. Atomistilor, nr. 409 M\u{a}gurele, Ilfov, 077125, Romania\label{aff83}
\and
Dipartimento di Fisica e Astronomia "G. Galilei", Universit\`a di Padova, Via Marzolo 8, 35131 Padova, Italy\label{aff84}
\and
Universit\"ats-Sternwarte M\"unchen, Fakult\"at f\"ur Physik, Ludwig-Maximilians-Universit\"at M\"unchen, Scheinerstrasse 1, 81679 M\"unchen, Germany\label{aff85}
\and
Departamento de F\'isica, FCFM, Universidad de Chile, Blanco Encalada 2008, Santiago, Chile\label{aff86}
\and
Institute of Space Sciences (ICE, CSIC), Campus UAB, Carrer de Can Magrans, s/n, 08193 Barcelona, Spain\label{aff87}
\and
Satlantis, University Science Park, Sede Bld 48940, Leioa-Bilbao, Spain\label{aff88}
\and
Centro de Investigaciones Energ\'eticas, Medioambientales y Tecnol\'ogicas (CIEMAT), Avenida Complutense 40, 28040 Madrid, Spain\label{aff89}
\and
Instituto de Astrof\'isica e Ci\^encias do Espa\c{c}o, Faculdade de Ci\^encias, Universidade de Lisboa, Tapada da Ajuda, 1349-018 Lisboa, Portugal\label{aff90}
\and
Universidad Polit\'ecnica de Cartagena, Departamento de Electr\'onica y Tecnolog\'ia de Computadoras,  Plaza del Hospital 1, 30202 Cartagena, Spain\label{aff91}
\and
Institut de Recherche en Astrophysique et Plan\'etologie (IRAP), Universit\'e de Toulouse, CNRS, UPS, CNES, 14 Av. Edouard Belin, 31400 Toulouse, France\label{aff92}
\and
Kapteyn Astronomical Institute, University of Groningen, PO Box 800, 9700 AV Groningen, The Netherlands\label{aff93}
\and
INFN-Bologna, Via Irnerio 46, 40126 Bologna, Italy\label{aff94}
\and
Infrared Processing and Analysis Center, California Institute of Technology, Pasadena, CA 91125, USA\label{aff95}
\and
INAF, Istituto di Radioastronomia, Via Piero Gobetti 101, 40129 Bologna, Italy\label{aff96}
\and
Instituto de Astrof\'isica de Canarias, Calle V\'ia L\'actea s/n, 38204, San Crist\'obal de La Laguna, Tenerife, Spain\label{aff97}
\and
Institut f\"ur Theoretische Physik, University of Heidelberg, Philosophenweg 16, 69120 Heidelberg, Germany\label{aff98}
\and
Universit\'e St Joseph; Faculty of Sciences, Beirut, Lebanon\label{aff99}
\and
Institut d'Astrophysique de Paris, 98bis Boulevard Arago, 75014, Paris, France\label{aff100}
\and
Junia, EPA department, 41 Bd Vauban, 59800 Lille, France\label{aff101}
\and
INFN, Sezione di Trieste, Via Valerio 2, 34127 Trieste TS, Italy\label{aff102}
\and
Instituto de F\'isica Te\'orica UAM-CSIC, Campus de Cantoblanco, 28049 Madrid, Spain\label{aff103}
\and
CERCA/ISO, Department of Physics, Case Western Reserve University, 10900 Euclid Avenue, Cleveland, OH 44106, USA\label{aff104}
\and
Laboratoire Univers et Th\'eorie, Observatoire de Paris, Universit\'e PSL, Universit\'e Paris Cit\'e, CNRS, 92190 Meudon, France\label{aff105}
\and
Dipartimento di Fisica e Scienze della Terra, Universit\`a degli Studi di Ferrara, Via Giuseppe Saragat 1, 44122 Ferrara, Italy\label{aff106}
\and
Istituto Nazionale di Fisica Nucleare, Sezione di Ferrara, Via Giuseppe Saragat 1, 44122 Ferrara, Italy\label{aff107}
\and
Institut d'Astrophysique de Paris, UMR 7095, CNRS, and Sorbonne Universit\'e, 98 bis boulevard Arago, 75014 Paris, France\label{aff108}
\and
Dipartimento di Fisica - Sezione di Astronomia, Universit\`a di Trieste, Via Tiepolo 11, 34131 Trieste, Italy\label{aff109}
\and
Minnesota Institute for Astrophysics, University of Minnesota, 116 Church St SE, Minneapolis, MN 55455, USA\label{aff110}
\and
Universit\'e C\^{o}te d'Azur, Observatoire de la C\^{o}te d'Azur, CNRS, Laboratoire Lagrange, Bd de l'Observatoire, CS 34229, 06304 Nice cedex 4, France\label{aff111}
\and
Institute Lorentz, Leiden University, PO Box 9506, Leiden 2300 RA, The Netherlands\label{aff112}
\and
Institute for Astronomy, University of Hawaii, 2680 Woodlawn Drive, Honolulu, HI 96822, USA\label{aff113}
\and
Department of Physics \& Astronomy, University of California Irvine, Irvine CA 92697, USA\label{aff114}
\and
Department of Astronomy \& Physics and Institute for Computational Astrophysics, Saint Mary's University, 923 Robie Street, Halifax, Nova Scotia, B3H 3C3, Canada\label{aff115}
\and
Departamento F\'isica Aplicada, Universidad Polit\'ecnica de Cartagena, Campus Muralla del Mar, 30202 Cartagena, Murcia, Spain\label{aff116}
\and
Universit\'e Paris Cit\'e, CNRS, Astroparticule et Cosmologie, 75013 Paris, France\label{aff117}
\and
Department of Computer Science, Aalto University, PO Box 15400, Espoo, FI-00 076, Finland\label{aff118}
\and
Department of Physics and Astronomy, Vesilinnantie 5, 20014 University of Turku, Finland\label{aff119}
\and
Serco for European Space Agency (ESA), Camino bajo del Castillo, s/n, Urbanizacion Villafranca del Castillo, Villanueva de la Ca\~nada, 28692 Madrid, Spain\label{aff120}
\and
ARC Centre of Excellence for Dark Matter Particle Physics, Melbourne, Australia\label{aff121}
\and
Centre for Astrophysics \& Supercomputing, Swinburne University of Technology, Victoria 3122, Australia\label{aff122}
\and
W.M. Keck Observatory, 65-1120 Mamalahoa Hwy, Kamuela, HI, USA\label{aff123}
\and
Department of Physics and Astronomy, University of the Western Cape, Bellville, Cape Town, 7535, South Africa\label{aff124}
\and
Oskar Klein Centre for Cosmoparticle Physics, Department of Physics, Stockholm University, Stockholm, SE-106 91, Sweden\label{aff125}
\and
Astrophysics Group, Blackett Laboratory, Imperial College London, London SW7 2AZ, UK\label{aff126}
\and
Univ. Grenoble Alpes, CNRS, Grenoble INP, LPSC-IN2P3, 53, Avenue des Martyrs, 38000, Grenoble, France\label{aff127}
\and
Dipartimento di Fisica, Sapienza Universit\`a di Roma, Piazzale Aldo Moro 2, 00185 Roma, Italy\label{aff128}
\and
Centro de Astrof\'{\i}sica da Universidade do Porto, Rua das Estrelas, 4150-762 Porto, Portugal\label{aff129}
\and
Zentrum f\"ur Astronomie, Universit\"at Heidelberg, Philosophenweg 12, 69120 Heidelberg, Germany\label{aff130}
\and
Dipartimento di Fisica, Universit\`a di Roma Tor Vergata, Via della Ricerca Scientifica 1, Roma, Italy\label{aff131}
\and
INFN, Sezione di Roma 2, Via della Ricerca Scientifica 1, Roma, Italy\label{aff132}
\and
Institute of Astronomy, University of Cambridge, Madingley Road, Cambridge CB3 0HA, UK\label{aff133}
\and
Institute for Computational Science, University of Zurich, Winterthurerstrasse 190, 8057 Zurich, Switzerland\label{aff134}
\and
Department of Astrophysical Sciences, Peyton Hall, Princeton University, Princeton, NJ 08544, USA\label{aff135}
\and
Niels Bohr Institute, University of Copenhagen, Jagtvej 128, 2200 Copenhagen, Denmark\label{aff136}
\and
Cosmic Dawn Center (DAWN)\label{aff137}}

\date{\today}

\authorrunning{B. Bose et al.}

\titlerunning{\emph{Euclid} preparation. XLIV. Modelling nonlinear clustering beyond-$\Lambda$CDM }

\newpage
\clearpage 

\abstract
{The \Euclid space satellite mission will measure the large-scale clustering of galaxies at an unprecedented precision, providing a unique probe of modifications to the $\Lambda$CDM model.}
{We investigated the approximations needed to efficiently predict the large-scale clustering of matter and dark matter halos in the context of modified gravity and exotic dark energy scenarios. We examined the normal branch of the Dvali--Gabadadze--Porrati model, the Hu--Sawicki $f(R)$ model, a slowly evolving dark energy model, an interacting dark energy model, and massive neutrinos. For each, we tested approximations for the perturbative kernel calculations, including the omission of screening terms and the use of perturbative kernels based on the Einstein--de Sitter universe; we explored different infrared-resummation schemes, tracer bias models and a linear treatment of massive neutrinos; we investigated various approaches for dealing with redshift-space distortions and modelling the mildly nonlinear scales, namely the Taruya--Nishimishi--Saito prescription and the effective field theory of large-scale structure. This work provides a first validation of the various codes being considered by \Euclid for the spectroscopic clustering probe in beyond-$\Lambda$CDM scenarios.}
{We calculated and compared the $\chi^2$ statistic to assess the different modelling choices. This was done by fitting the spectroscopic clustering predictions to measurements from numerical simulations and perturbation theory-based mock data. We compared the behaviour of this statistic in the beyond-$\Lambda$CDM cases, as a function of the maximum scale included in the fit, to the baseline $\Lambda$CDM case.} 
{We find that the Einstein--de Sitter approximation without screening is surprisingly accurate for the modified gravity cases when comparing to the  halo clustering monopole and quadrupole obtained from simulations and mock data. Further, we find the same goodness-of-fit for both cases -- the one including and the one omitting non-standard physics in the predictions. Our results suggest that the inclusion of multiple redshift bins, higher-order multipoles, higher-order clustering statistics (such as the bispectrum), and photometric probes such as weak lensing, will be essential to extract information on massive neutrinos, modified gravity and dark energy. Additionally, we show that the three codes used in our analysis, namely,  \texttt{PBJ}, \texttt{Pybird} and \texttt{MG-Copter}, exhibit sub-percent agreement for $k\leq 0.5\,h\,{\rm Mpc}^{-1}$ across all the models. This consistency underscores their value as reliable tools.}
{}
   \maketitle
  


\section{Introduction}

The recently launched \Euclid mission will measure the positions and redshifts of billions of galaxies, granting us an unprecedented picture of the large-scale structure of the Universe \citep{EUCLID:2011zbd,Amendola:2016saw,2024arXiv240513491E}. This will serve as a fantastic experimental test of the standard model of cosmology, $\Lambda$CDM,\footnote{This model assumes the dominating energy components of the Universe are a constant dark energy, $\Lambda$, and cold dark matter (CDM), with gravity described by Einstein's general relativity (GR).} and as a means to constrain or detect deviations from it. In particular, the \Euclid mission will offer a unique opportunity to probe and constrain modifications in both the matter and gravitational sectors \citep{Euclid:2019clj}.

\Euclid can thus offer insight into a more fundamental description of gravity beyond General Relativity (GR), and perhaps into one that could account for the ongoing accelerated expansion of the Universe \citep{Perlmutter:1998np,Riess:1998cb}, one of the most poorly understood phenomena in Nature. It could also provide clues as to whether or not dark energy evolves over time, or whether or not it interacts with CDM, the latter of which can help solve emerging data-set tensions within $\Lambda$CDM \citep[see][for example]{Pourtsidou:2016ico}.

A key observable that \Euclid will provide is the power spectrum -- the Fourier transform of the 2-point correlation function -- of the galaxy distribution in redshift space. Given the immense number of galaxies that \Euclid will detect, the statistical uncertainty on the measured power spectrum will be tiny. In particular, \Euclid is set to measure more than twenty million galaxy redshifts \citep{Amendola:2016saw}.\footnote{See also \url{https://www.euclid-ec.org/public/core-science/}.} Since there will be more galaxy pairs found at small physical separations, the statistical uncertainty will be smaller as we move to the mildly nonlinear, small-scale, regime. There is therefore, potentially, a wealth of cosmological and gravitational information to be extracted by going beyond the linear, large-scale regime \citep[see, for example,][]{Bernardeau:2001qr, Nishimichi:2007xt, Lacasa:2019flz, Bose:2019ywu}.

In order to seize this opportunity, the precision of the data must be matched with the accuracy of the theoretical predictions, which becomes a challenge to maintain when nonlinear effects become important. Further, if modelling uncertainty is not well estimated, then a biased picture of the Universe will be inferred from the data \citep[see, for example,][]{Markovic:2019sva,MartinelliNL}. 

There exist various approaches to modelling large-scale structure observables on scales where nonlinear effects become significant. $N$-body simulations are powerful tools for this purpose. However, they are not well-suited for data-theory inference analyses due to their high computational cost. Another alternative is to employ faster, semi-analytical methods based on standard perturbation theory \citep[SPT, see][for a review]{Bernardeau:2001qr}, such as the effective field theory of large-scale structure \citep[EFTofLSS;][]{Baumann:2010tm, Carrasco:2012cv} or other related approaches \citep[see for example][]{Pueblas:2008uv, Pietroni:2011iz}. A procedure one can follow is to model all relevant physics within the chosen perturbative framework and then determine the range of scales where this approximate method remains valid --  where the theoretical uncertainty is smaller than the statistical and known systematic uncertainties -- by comparison with an $N$-body benchmark measurement \citep[see, for example,][]{Heitmann:2015xma,Heitmann:2019ytn,Bose:2019ywu,Bose:2017myh,Rossi:2020wxx}.

In this paper, we aim to quantify the uncertainty of various approximations and approaches to the nonlinear redshift space galaxy clustering power spectrum in beyond-$\Lambda$CDM cosmologies. The goal is to identify which such choices are worthy of further assessment and eventual use in computationally demanding, statistical analyses. 

In particular, we have considered four representative theories of gravity and dark energy that have been identified by the \Euclid consortium as target theories to test \citep{Amendola:2016saw}: the normal branch of the Dvali--Gabadadze--Porrati (DGP) model of gravity \citep{Dvali:2000hr}, $f(R)$ gravity \citep{Carroll:2003wy,Hu:2007nk}, a time-varying dark energy component \citep{Chevallier:2000qy,Linder:2002et}, and an interacting dark matter - dark energy model with pure momentum exchange \citep{Simpson:2010vh, Pourtsidou:2013nha}. Two gravity models provide examples of phenomenologically distinct types of screening,\footnote{Screening is a theoretical mechanism used to suppress additional forces coming from modifications to gravity in the Solar System \citep{Brax:2021wcv}, where experiments show theory must accord with GR's predictions \citep{Will:2018bme}.} these being the Vainshtein mechanism \citep[][in the case of DGP]{Vainshtein:1972sx} and the chameleon mechanism \citep[][in the case of $f(R)$ gravity]{Khoury:2003rn}. These two gravity models also provide distinct effects on the linear growth of structure, with $f(R)$ inducing a modification to growth that varies with physical scale while DGP induces a modification that is scale independent.

The dark sector models provide two phenomenological examples of possible deviations from a cosmological constant. The time-varying dark energy model is the simplest way to approximate and parameterise dark energy evolution as predicted by, for example, quintessence scalar field theories. The specific interacting dark energy model we have chosen represents more complex dynamics and has been shown to be able to address the tension between CMB and structure growth measurements \citep{Pourtsidou:2016ico, Carrilho:2022mon}.

On top of a modified gravitational or dark sector, we also considered massive neutrinos, the total mass of which \Euclid is expected to measure with a high significance \citep[see for example][]{Archidiacono:2017tlz}. The effects of a massive neutrino species becomes significant at the scales of interest and have demonstrated a notable degeneracy with modifications to gravity at low redshift \citep{He:2013qha,Baldi:2013iza,Garcia-Farieta:2019hal,Wright:2019qhf,Hagstotz:2019gsv,Moretti:2023drg}. 

An additional choice we considered is the redshift space distortion effect \citep{Kaiser:1987qv}. We looked at two prominent models: the Taruya--Nishimishi--Saito (TNS) model \citep{Taruya:2010mx} and the EFTofLSS model \citep{Perko:2016puo,Chudaykin:2020,DAmico:2019fhj}. We also examined choices for how to resum infrared modes and model tracer bias. 

This paper is structured as follows. In \cref{sec:models} we introduce the four theories of gravity and dark energy that we consider. In \cref{sec:nonlinear} we cover all the various nonlinear redshift-space power spectrum models and approximations. In \cref{sec:sims} we present the $N$-body simulations which we then use to quantify the validity of the various theoretical choices in \cref{sec:results}. We summarised these findings and our conclusions in \cref{sec:conclusions}.

\section{Perturbation theory beyond-\texorpdfstring{$\Lambda$}{L}CDM} \label{sec:models}

We consider a perturbed flat Friedmann--Lema\^{i}tre--Robertson--Walker (FLRW) universe with a background metric given by $\mathrm{d} s^2 = - c^2 \mathrm{d}t^2 + a^2(t) \delta_{ij} \mathrm{d}x^i \mathrm{d}x^j$, where $a$ represents the scale factor. The Hubble rate, $H$, is defined as $H \coloneqq \dot a /a$, where a dot denotes the derivative with respect to cosmic time $t$. We focus on scalar perturbations, and we adopt the Newtonian gauge where the perturbed FLRW metric can be written as
\begin{equation}
\mathrm{d} s^2 = - c^2 \, [1+ 2 \Phi(\bfx,t)] \, \mathrm{d}t^2 + a^2(t) \,  [1-2\Psi(\bfx,t)] \, \delta_{ij} \mathrm{d}x^i \mathrm{d}x^j \, , 
\end{equation}
where the gravitational potential,  $\Phi$,  appears in the time-time component of the metric. To express the continuity and Euler equations, we use the rescaled CDM velocity divergence, $\theta \coloneqq \nabla \cdot {\bf v}/ (a \, H)$, and the CDM density contrast, $\delta \coloneqq \delta \rho  /  \rho$, where $\rho$ and $\delta \rho$ are respectively the energy density background and perturbations. 

SPT assumes that density and velocity perturbations are small and can be expanded as 
\begin{equation}
\delta(\bfk,a) = \sum^{\infty}_{n=1}\delta_n(\bfk,a)\,, \quad \quad 
\theta(\bfk,a) = \sum^{\infty}_{n=1}\theta_n(\bfk,a)\,, \label{eq:ptexpansion}
\end{equation}
 where $\delta_n,\theta_n \sim \delta_1^n$, with $\delta_1$ being the linear theory solution. Explicitly we can write the $n^{\rm th}$ order perturbations in terms of scale and time dependent kernels $F_n$ and $G_n$, defined implicitly as 
\begin{align}
  \delta_n(\boldsymbol{k},a) & =  \frac{1}{(2 \pi)^{3 (n-1)}} \int \mathrm{d}^3\boldsymbol{k}_1 \cdots \mathrm{d}^3 \boldsymbol{k}_n \, \delta_{\rm D}(\boldsymbol{k}-\boldsymbol{k}_{1 \cdots n}) \nonumber  \\ &    \quad  \times 
 F_n(\boldsymbol{k}_1, \ldots ,\boldsymbol{k}_n,a) \, \delta_{\rm 1,i}(\boldsymbol{k}_1) \cdots \delta_{\rm 1,i}(\boldsymbol{k}_n) \, , \nonumber \\ 
    \theta_n(\boldsymbol{k},a) & =  \frac{1}{(2 \pi)^{3 (n-1)}}\int \mathrm{d}^3\boldsymbol{k}_1 \cdots \mathrm{d}^3 \boldsymbol{k}_n \, \delta_{\rm D}(\boldsymbol{k}-\boldsymbol{k}_{1 \cdots n})  \nonumber  \\ &   
  \quad \times G_n(\boldsymbol{k}_1, \dots ,\boldsymbol{k}_n,a)  \, \delta_{\rm 1,i}(\boldsymbol{k}_1) \cdots \delta_{\rm 1,i}(\boldsymbol{k}_n) \, , \label{eq:kernels}
\end{align}
where $\bfk_{1 \cdots n} = \bfk_1  + \ldots + \bfk_n$, $\delta_{\rm D}$ denotes the Dirac delta and a subscript `i' denotes a quantity computed at some early initial time. We define the linear growth rate of structure as
\begin{equation}
    f \coloneqq \frac{{\rm d} \, \ln \delta_1}{{\rm d} \ln a} = -G_1/F_1 \,,
\end{equation}
where in the second equality we have used the linear version of \cref{eq:Perturb1}. This can now clearly be both time and scale-dependent, as in $f(R)$ gravity (see \cref{sec:nl-fofr}). 

To solve for the kernels, $F_{n}$ and $G_{n}$, we can write down generic energy and momentum conservation equations for the CDM density  and rescaled velocity divergence perturbations, as well as the Poisson equation which relates the Newtonian gravitational potential ($\Phi$) to the density perturbation \footnote{The exact kernels can be obtained by solving these equations numerically as described in \cite{Taruya:2016jdt,Bose:2016qun} or by using a Fast Fourier Transform  decomposition.}
 \begin{align}
& a \, \delta'(\bfk,a)+\theta(\bfk,a)  \nonumber \\ &  = -
\int\frac{\mathrm{d}^3\bfk_1 \mathrm{d}^3\bfk_2}{(2\pi)^3} \, \delta_{\rm D}(\bfk-\bfk_{12}) \,
\alpha(\bfk_1,\bfk_2)\,\theta(\bfk_1,a) \, \delta(\bfk_2,a)\,,
\label{eq:Perturb1}\\
& a \, \theta'(\bfk,a)+
\left(2+ \frac{a H'(a)}{H(a)} + \frac{\Xi(a)}{H(a
)} \right)\theta(\bfk,a)
-\left(\frac{c \, k}{a\,H(a)}\right)^2\,  \Phi(\bfk,a) \nonumber \\ &  
=-\frac{1}{2}\int\frac{\mathrm{d}^3\bfk_1\mathrm{d}^3\bfk_2}{(2\pi)^3} \,
\delta_{\rm D}(\bfk-\bfk_{12}) \,
\beta(\bfk_1,\bfk_2)\,\theta(\bfk_1,a) \, \theta(\bfk_2,a)\,,
\label{eq:Perturb2} \\ 
& -\left(\frac{ c\, k}{a \, H(a)}\right)^2\Phi (\bfk,a) = 
\frac{3 \Omega_{\rm m}(a)}{2} \mu(k,a) \,\delta(\bfk,a) + S(\bfk,a)\,,
\label{eq:poisson1}
\end{align}
where a prime denotes a derivative with respect to the scale factor, $\Omega_{\rm m}$ is the total matter density fraction and $\Xi$ encodes the effects of a possible additional drag force, coming from, say, an interaction within the dark sector (see \cref{sec:nl-DS}).  The kernels $\alpha(\bfk_1,\bfk_2)$ and $\beta(\bfk_1,\bfk_2)$ are given by
\begin{align}
\alpha(\bfk_1,\bfk_2) &= 1+\frac{\bfk_1\cdot\bfk_2}{|\bfk_1|^2}\,, \\
\beta(\bfk_1,\bfk_2) & =
\frac{(\bfk_1\cdot\bfk_2)\left|\bfk_1+\bfk_2\right|^2}{|\bfk_1|^2|\bfk_2|^2}\,.
\label{alphabeta}
\end{align}
The function $\mu(k,a)$ expresses any linear modification to GR in the Poisson equation \citep{Bean:2010zq,Silvestri:2013ne,Pogosian:2010tj} and $S(\bfk,a)$ captures higher-order modifications, which include screening effects. Up to third order, we have
\begin{align}
 S(\bfk,a) & =  
\int\frac{\mathrm{d}^3\bfk_1\mathrm{d}^3\bfk_2}{(2\pi)^3}\,
\delta_{\rm D}(\bfk-\bfk_{12}) \, \gamma_2(\bfk_1, \bfk_2,a) \, 
\delta(\bfk_1,a)\,\delta(\bfk_2,a)  \nonumber \\ & \quad + 
\int\frac{\mathrm{d}^3\bfk_1\mathrm{d}^3\bfk_2\mathrm{d}^3\bfk_3}{(2\pi)^6} \,
 \delta_{\rm D}(\bfk-\bfk_{123}) \,
\gamma_3(\bfk_1, \bfk_2, \bfk_3,a)  \nonumber \\ &   \quad    \times 
\delta(\bfk_1,a)\,\delta(\bfk_2,a)\,\delta(\bfk_3,a)\,, 
\label{eq:Perturb3}
\end{align}
where the functions $\gamma_2$ and $\gamma_3$ are additional kernels coming from higher-order scalar field interactions in the Klein--Gordon equation \citep[see][for example]{Koyama:2009me}. For $\Lambda$CDM, we have $\mu(k,a)=1$ and $\Xi=\gamma_2 = \gamma_3 = 0$. We now look at the forms of these equations for various theories beyond-$\Lambda$CDM.


\subsection{\texorpdfstring{$w_0 w_a$}{w0wa}CDM} \label{sec:nl-w0wacdm}

For a dark energy fluid the equation of state parameter is defined as  $w = p_{\rm DE}/(\rho_{\rm DE} c^2)$, where $p_{\rm DE}$ and $\rho_{\rm DE}$ are the pressure and the background energy density of the fluid. In  $\Lambda$CDM, $w=-1$. In general, $w$ can be a function of time. One such  (phenomenological) form was proposed by Chevalier--Polarski--Linder  \citep[CPL,][]{Chevallier:2000qy,Linder:2002et}:
\begin{equation}
 w(a)=w_0 +(1-a)\,w_a\,,
 \label{eq:cplpar}
\end{equation}
where $w_0$ and $w_a$ are constants. This form is essentially the first order Taylor expansion of $w(a)$ about the present time, where we have chosen the present day scale factor to be $a=1$. Taking only the first order Taylor expansion also implies we assume the equation of state is slowly evolving. \cref{eq:cplpar} effectively changes the background expansion $H$ but leaves all other functional modifications to $\Lambda$CDM at their $\Lambda$CDM values, that is $\mu(k,a)=1$ and $\Xi=\gamma_2 = \gamma_3 = 0$.

\subsection{Dark Scattering model} \label{sec:nl-DS}

Dark Scattering is an interacting dark energy model in which dark energy exchanges momentum with dark matter without energy exchange~\citep{Simpson:2010vh, Pourtsidou:2013nha, Pourtsidou:2016ico, Baldi:2014ica, Baldi:2016zom, Bose:2017jjx}. The modifications to \cref{eq:Perturb3} are given by $\mu(k,a)=1$ and $\gamma_2 = \gamma_3 = 0$. On the other hand, we have a coupling between dark matter and dark energy perturbations which appears in the Euler equations for each species \citep{Simpson:2010vh,Baldi:2014ica}. 

In \cref{eq:Perturb1,eq:Perturb2} we made the assumption that the fluctuations of dark energy propagate at the speed of light, ensuring that they remain significantly smaller than the fluctuations of other species. This allows the rescaled velocity divergence of the dark energy fluid, $\theta_{\rm DE}$, to be effectively neglected in the Euler equation.  Then, the interaction introduces only an additional drag force  on the left-hand side of \cref{eq:Perturb2}, given by
\begin{equation}
\Xi(a)= [1+w(a)] \, \xi \, c \,  \rho_{\rm DE}(a) \, , 
\end{equation}
where $\xi\coloneqq\sigma_{\rm D}/(m_{\rm c} c^2)$. Here $m_{\rm c}$ is the dark matter particle mass and $\sigma_{\rm D}$ is the cross-section of the interaction. $\xi$ is treated as a free parameter of the theory. It is positive (or zero) and proportional to the strength of the modification to $\Lambda$CDM, with $\Lambda$CDM being recovered for $\xi \to 0$ or $w \to -1$.


\subsection{Dvali--Gabadadze--Porrati gravity} \label{sec:nl-DGP}

The Dvali--Gabadadze--Porrati (DGP) model has been introduced by \citet{Dvali:2000hr} and describes a model where matter lives on a four-dimensional brane which is embedded in a five-dimensional Minkowski background. This model is representative of the class of models exhibiting the Vainshtein mechanism, which screens additional scalar field-sourced forces at small physical scales \citep{Vainshtein:1972sx}. 

We consider the normal branch of DGP, which is free of ghost instabilities \citep[see][]{Luty:2003vm,Gorbunov:2005zk, Charmousis:2006pn}. In this model dark matter particles follow geodesics, implying that $\Xi = 0$ in \cref{eq:Perturb2}. The relation between $\Phi$ and $\delta$ is described by \cref{eq:poisson1}, with  \citep[see for example][]{Bose:2016qun,Bose:2018orj} 
\begin{align}
\mu(k,a) & =  1 + \frac{1}{3\,\beta(a)}  \ ,  \\ 
\gamma_2(\bfk_1,\bfk_2,a) & =  -\left[\frac{H_0}{H(a)}\right]^2\frac{1}{24\beta^3(a) \, \Omega_{\rm rc}} \left(\frac{\Omega_{\rm m,0}}{a^3}\right)^2 (1-\mu_{1,2}^2) \, , \\
\gamma_3(\bfk_1,\bfk_2,\bfk_3,a) & =   \left[\frac{H_0}{H(a)}\right]^2\frac{1}{144 \beta^5(a) \, \Omega_{\rm rc}^2} \left(\frac{\Omega_{\rm m,0}}{a^3}\right)^3 \nonumber \\ & \quad \times (1-\mu_{2,3}^2) (1-\mu_{1,23}^2) \, , 
\end{align}
where $H_0$ and $\Omega_{\rm m,0}$ are the Hubble parameter and total matter density fraction today,  
\begin{equation}
\beta(a) \coloneqq 1+\frac{H(a)}{H_0} \frac{1}{\sqrt{\Omega_{\rm rc}}}   \left(1+\frac{a \, H'(a)}{3H(a)}\right) \,.
\label{betadef}
\end{equation}
The third order kernel, $\gamma_3$, needs to be symmetrised and $\mu_{i,j} = \hat{\bfk}_i \cdot \hat{\bfk}_j$ is the cosine of the angle between $\bfk_i$ and $\bfk_j$ (recall $\bfk_{ij} = \bfk_i + \bfk_j$). Here $\Omega_{\rm rc} \coloneqq c^2/(4 r_c^2 H_0^2)$, where $r_{\rm c}$ represents the scale above which gravity deviates from GR, meaning $r_{\rm c}\to \infty$, or $\Omega_{\rm rc} \to 0$, is the GR-limit of the theory  \citep[see][]{Dvali:2000hr}. In this work we assume that the background expansion, $H$, is exactly the one of a flat $\Lambda$CDM model \citep{Schmidt:2009sv}.


\subsection{ \texorpdfstring{$f(R)$}{fR} gravity} \label{sec:nl-fofr}

In this work we consider the specific form for $f(R)$ proposed by \citet{Hu:2007nk}. This model is representative of the class of models exhibiting the Chameleon mechanism \citep{Khoury:2003rn}, which screens in a phenomenologically distinct way to Vainshtein screening.

As with DGP, in this theory we have $\Xi = 0$, and have the following modifications to the Poisson equation \citep{Koyama:2009me,Taruya:2014faa,Bose:2016qun}
 \begin{align}
\mu(k,a) & = 1 + \left(\frac{k}{a}\right)^2\frac{1}{3\Pi(k,a)}, 
\\ 
\gamma_2(\bfk_1,\bfk_2,a)  & = - \frac{3}{16}\left(\frac{k \, H_0}{a \, H(a)}\right)^2\left(\frac{\Omega_{\rm m,0}}{a^3}\right)^2  \frac{\Upsilon^5(a)}{f_0^2 \, (3\Omega_{\rm m,0}-4)^4} \nonumber \\ & \quad 
\times \frac{1}{\Pi(k,a)\, \Pi(k_1,a) \, \Pi(k_2,a)}, 
\label{frg2} 
\end{align}
and
\begin{align}
&\gamma_3(\bfk_1,\bfk_2,\bfk_3,a)    \nonumber \\ & = \frac{1}{32} \left(\frac{k \, H_0}{a \, H(a)}\right)^2   \left(\frac{\Omega_{\rm m,0}}{a^3}\right)^3 \frac{1}{\Pi(k,a)\, \Pi(k_1,a)\, \Pi(k_2,a)\, \Pi(k_3,a)}  
\nonumber 
\\ & \quad \times   \left\{-5\frac{\Upsilon^7(a)}{f_0^3 \, (3\Omega_{\rm m,0}-4)^6}  + \frac{9}{2}\frac{1}{\Pi(k_{23},a) }\left[ \frac{\Upsilon^5(a)}{ f_0^2 \, (3\Omega_{\rm m,0}-4)^4} \right]^2\right\},
\label{frg3} 
\end{align}
where  we have implicitly assumed the constraint coming from the 1-loop integral: $\bfk = \bfk_1+\bfk_2$ for $\gamma_2$ and $\bfk = \bfk_1+\bfk_2+\bfk_3$ for $\gamma_3$. Again, we note that $\gamma_3$ must be symmetrised. The functions $\Pi$ and $\Upsilon$ are given by
\begin{align}
\Pi(k,a) & = \left(\frac{k}{a}\right)^2+\frac{\Upsilon^3(a)}{2 f_0 \, (3\Omega_{\rm m,0}-4)^2},  \\   \Upsilon(a) & = \frac{\Omega_{\rm m,0}+4a^3\, (1-\Omega_{\rm m,0})}{ a^3} \label{eq:Xi} \, .
\end{align}
Here we set $f_0 = c^2 |{f}_{\rm R0}|/H_0^2$, where  $f_{\rm R0}$ is the value of $f_{\rm R} = \mathrm{d} f(R) / \mathrm{d}R $ today,  which gauges the strength of the modification to GR. The GR-limit is given by $f_{\rm R0} \to 0$. As with the DGP case, we also assume a $\Lambda$CDM background expansion, $H$.


\section{Nonlinear modelling}\label{sec:nonlinear}

When mapping the underlying theories described in the previous section to the observable quantity of interest, there are several choices one can make. This range of choices expands significantly when we wish to model scales where nonlinearities become important. In this section, we present a number of such choices which we test to varying degrees against $N$-body simulations in \cref{sec:results}.


\subsection{The quasi-nonlinear power spectrum}
Our goal is to calculate the power spectrum of the dark matter density and velocity fluctuations. This can be defined as 
\begin{equation}
    (2\pi)^3 \delta_{\rm D}(\bfk + \bfk') \, P^{ab} (k,a) \coloneqq \langle \varphi^a(\bfk,a) \, \varphi^b (\bfk',a) \rangle  \, ,
\end{equation}
where we have introduced the duplet $\varphi^a \coloneqq (\delta, \theta)$, and $\langle \cdot \rangle$ denotes an ensemble average. By expanding $\delta$ and $\theta$ using \cref{eq:ptexpansion}, we can perturbatively express the power spectrum as a leading-order term plus a next-to-leading-order (nlo) term
\begin{equation}
P^{ab} (k,a) = P_{11}^{ab}(k,a) +  P^{ab}_{\rm nlo}(k,a) + \ldots \,, \label{eq:1loopr} 
\end{equation}
where the leading term, $P_{11}^{ab}$, is the `tree-level' or linear power spectrum. The nlo or `1-loop' term can be expressed as the sum of three quantities,
\begin{equation}
P^{ab}_{\rm nlo}(k,a) = P_{22}^{ab}(k,a) + P_{13}^{ab}(k,a) + P_{31}^{ab}(k,a) \,, 
\end{equation}
where we have used the notation
\begin{equation}
    (2\pi)^3 \delta_{\rm D}(\bfk + \bfk') \, P^{ab}_{ij} (k,a) = \langle \varphi_i^a(\bfk,a) \, \varphi_j^b (\bfk',a) \rangle  \, .
\end{equation}
The ellipses in \cref{eq:1loopr} represent terms of higher-order, including contributions from 2-loop order and beyond. We define the `1-loop' power spectrum as 
\begin{equation}
    P_{\rm 1-loop}^{ab} \coloneqq  P_{11}^{ab}(k,a) +  P^{ab}_{\rm nlo}(k,a) \, .
    \label{eq:realpofk}
\end{equation}


\subsection{Redshift space}

So far we have worked in `real' space coordinates, $\bfr$, or its Fourier equivalent. But astronomical observations are composed of angular positions and redshifts, which contains all depth information. This means we must convert our theoretical predictions to redshift space coordinates, $\bfs$. This involves peculiar velocity information which combines with the Hubble velocity to give the total measured redshift. In terms of positional coordinates, this is expressed as 
\begin{equation}
    \bfs = \bfr + \frac{v_z(\bfr,a)}{a \, H(a)} \hat{\bfz} \, , 
\end{equation}
where we have taken the line-of-sight direction to be along the real space $z$-axis, and $v_z$ is the projection of the peculiar velocity along that axis. Conservation of mass then implies the translation 
\begin{equation}
    \delta^{s}(\bfs,a) = \left|\frac{\partial \bfs}{\partial \bfr} \right|^{-1} [1+\delta(\bfr,a)] -1 \,  ,
\end{equation}
where a superscript `$s$' denotes a redshift-space quantity. If we Fourier transform the redshift-space density field and take the ensemble average with itself at two points, we can arrive at the following expression for the redshift-space power spectrum at 1-loop order \citep{Matsubara:2007wj,Heavens:1998es,Taruya:2010mx},
\begin{align}
 P_{\rm 1-loop}^s(k,\mu) & =   P_{\rm 1-loop}^{\delta \delta} (k) + 2 \mu^2 P_{\rm 1-loop}^{\delta \theta}(k) +  \mu^4 P_{\rm 1-loop}^{\theta \theta} (k) \nonumber \\ & \quad +  A(k,\mu) + B(k,\mu) + C(k,\mu) \nonumber \\ & \quad - k^2 \mu^2 \, \tilde\sigma_{\rm v}^2 \left[ F_1(k)+ G_1(k)\, \mu^2 \right]^2 \, P_{\rm 11,i}^{\delta \delta}(k) 
 \, , \label{eq:rsdpofk}
\end{align}
 where we have dropped the explicit time dependence of all functions for compactness, except for $P_{\rm 11,i}^{\delta \delta}$ which is computed at some fixed early time. Here $\mu$ is the angle between the line of sight direction $\hat{\vec{z}}$ and the wave mode $\kv$.\footnote{Not to be confused with the linear Poisson equation modification, $\mu(k,a)$, which always appears with its arguments.} The linear theory estimate for the velocity dispersion, $\tilde\sigma_{\rm v}$, is given by
\begin{equation}
\tilde{\sigma}^2_{ \rm v} = \frac{1}{6\pi^2} \int \diff q \, G_1^2(q) \, P_{\rm 11,i}^{\delta \delta}(q) \, . 
\label{eq:veldisp}
\end{equation}
We refer the reader to \cite{Bose:2016qun,Taruya:2010mx} for expressions for $A(k,\mu)$, $B(k,\mu)$, and $C(k,\mu)$.
These terms come from higher-order interactions between the density and velocity perturbations. We give their explicit forms in terms of the kernels defined in \cref{eq:kernels} in \cref{app:rsdterms}.

\Cref{eq:rsdpofk} performs poorly when modelling nonlinear redshift-space effects. In particular, it does not capture the  fingers-of-god (FoG) effect, caused by the large positional distortions of measured objects towards the centres of potential wells, coming from their high peculiar velocities. Higher-order perturbation theory (higher than third order), has been shown to have poor convergence at the level of the power spectrum \citep[see, for example,][]{Carlson:2009it}, and is also very computationally expensive. To move further into the nonlinear regime, a number of proposals have been made in the past decade which modify \cref{eq:rsdpofk} by introducing new degrees of freedom, or nuisance parameters, quantifying our uncertainty on nonlinear effects. 

We consider two prominent such proposals, both of which have been applied in the biggest galaxy survey to date, the BOSS survey \citep{Beutler:2016arn,Ivanov:2019pdj,DAmico:2019fhj,Carrilho:2022mon}. Both of these can also be computed with a number of fast public codes, discussed next.

\subsubsection{Codes}

In this work we use a number of codes. In particular, we adopt:
\begin{itemize}
    \item 
The publicly available \texttt{MG-Copter} code \citep{Bose:2016qun}, recently absorbed by the \texttt{ReACT} code \footnote{Download \texttt{ReACT}: \url{https://github.com/nebblu/ACTio-ReACTio}} \citep{Bose:2020wch,Bose:2021mkz,Bose:2022vwi} which solves the first, second and third order continuity and Euler equations simultaneously and numerically for each external momentum; 
\item 
The publicly available \texttt{PyBird} code \footnote{Download \texttt{PyBird}: \url{https://github.com/pierrexyz/pybird}}~\citep{DAmico:2020kxu},  which uses a Fast Fourier Transform (FFT) decomposition and has been used in recent BOSS analyses
\citep[for instance see][for $\Lambda$CDM, $w$CDM, and nDGP respectively]{Zhang:2021yna,DAmico:2020kxu,Piga:2022mge}; 
\item 
The \texttt{PBJ} code \citep{Moretti:2023drg,Oddo:2020, Oddo:2021,2023JCAP...01..031R}, which implements the model of \cite{Ivanov:2019pdj} taking advantage of the \texttt{FAST-PT} algorithm \citep{mcewen2016, fang2017}, and has also been used in recent BOSS analyses for Dark Scattering \citep{Carrilho:2022mon}. This code is planned to be made public soon.
\end{itemize}
We summarise the codes in \cref{tab:codes} along with relevant implementations. \texttt{MG-Copter} is by far the slowest, needing to solve eight sets of \cref{eq:Perturb1,eq:Perturb2} for each external mode $k$ to find the perturbative kernels up to third order. More quantitatively, to compute the first two multipoles of \cref{eq:rsdpofk} in $\Lambda$CDM at fifty values of $k$ in the range $[0.01,0.3] \, h \, {\rm Mpc}^{-1}$, it takes roughly $45$ s on a MacBook Pro 2018 model. By comparison, \texttt{PBJ} and \texttt{PyBird} only need to compute a set of products on a fixed grid by using the FFT method, which allows them to produce the same result in roughly $20$ ms. Of course, this assumes a specific scale dependence of the perturbative kernels, which may not be known or even analytically available. This is not true for \texttt{MG-Copter} which only requires the specification of $H$, $\mu$, $\gamma_2$, $\gamma_3$ and $\Xi$. 

We have performed validation of these codes both in real (see \cref{eq:realpofk}) and redshift space (see \cref{eq:rsdpofk}) to sub-percent precision for $\Lambda$CDM, $w_0w_a$CDM, nDGP, and Dark Scattering (see \cref{app:validate} for some of these 
 comparisons). Additionally, \texttt{MG-Copter}, \texttt{PBJ} and \texttt{PyBird} have been validated independently against large $N$-body simulations in a number of papers: see \cite{Bose:2016qun,Bose:2017myh,Bose:2017jjx} for \texttt{MG-Copter}, \cite{Oddo:2021, Carrilho:2021hly, Tsedrik:2022cri} for \texttt{PBJ} and \cite{Nishimichi:2020, DAmico:2019fhj} for \texttt{PyBird}.

Within the \Euclid collaboration, both \texttt{PBJ} and \texttt{PyBird} are being used to fit measurements from the Flagship simulation \citep{Potter:2017}, with results to be presented in a series of papers. Finally, \texttt{PBJ} has been ported to the \Euclid likelihood code, named \texttt{CLOE}, and will be used to perform the official analysis of the spectroscopic sample for $\Lambda$CDM.

\begin{table*}
\caption{\label{t7}  Summary of codes used in this work.}
\centering
\begin{tabular}{| c | c | c | c | c | c |}
\hline  
 {\bf Name} & {\bf Solving Method} & {\bf Speed} [ms] & {\bf Beyond-$\Lambda$CDM Models}  & {\bf RSD Model} & {\bf Resummation} \\ \hline 
\texttt{PBJ} & FFT & $\mathcal{O}(10)$  & DGP${}_{\rm USA,EdS}$, DS${}_{\rm EdS}$, $w_0w_a$CDM${}_{\rm EdS}$  &  TNS, EFTofLSS & WnW  \\
\texttt{PyBird}  & FFT & $\mathcal{O}(10)$  &  DGP, $w_0w_a$CDM & EFTofLSS & Lagrangian \\ 
\texttt{MG-Copter}  & Solve \cref{eq:Perturb1,eq:Perturb2} per $k$ & $\mathcal{O}(10000)$ & $f(R)$, DGP, DS, $w_0w_a$CDM& TNS, EFTofLSS & WnW \\ 
 \hline 
\end{tabular}
\tablefoot{We show each code's method of solving for the SPT kernels (see main text for more details), the computational speed (in order of magnitude in ms) and available modelling. `EdS' and `USA` subscripts indicate only the Einstein--de Sitter and Unscreened approximations are available for that model (see \cref{sec:approximations}). Here `DS' stands for Dark Scattering and `WnW' stands for Wiggle-no-Wiggle (see \cref{sec:resummation}). We note that \texttt{MG-Copter} only supports an EdS implementation of the WnW - EFTofLSS model. }
\label{tab:codes}
\end{table*}


\subsubsection{Effective Field Theory of Large-Scale Structure} \label{sec:eftoflss}

The effective field theory of large-scale structure (EFTofLSS) prescription for the redshift-space matter power spectrum can be written as 
\citep{DAmico:2019fhj,Ivanov:2019pdj,Chudaykin:2020} 
\begin{align}
P^s_{\rm EFT}(k,\mu) &= P^s_{\rm 1-loop}(k,\mu) + \tilde{P}^s_{\rm ctr}(k,\mu)\,,   \label{eq:eftoflsspofk}
\end{align}
where the second term, $\tilde{P}^s_{\rm ctr}(k,\mu)$, contains `counterterms' which are used to model nonlinear effects, and contains a number of free constants that must be fit to the data. 

In \texttt{PBJ} the EFTofLSS counterterms are given by
\begin{align}
\tilde{P}^s_{\rm ctr}(k,\mu)  =  - 2 k^2 \, P_{11}^{\delta \delta}(k) \left( \tilde{c}_0 + \tilde{c}_2  f \, \mu^2 + \tilde{c}_4   f^2  \mu^4 \right)\, .
\label{eq:ct_PBJ}
\end{align}
This model has therefore a total of three nuisance parameters at the dark matter level, $\{\tilde{c}_0, \tilde{c}_2, \tilde{c}_4\}$, with dimensions $h^{-2} \, {\rm Mpc}^2$. 

In \texttt{PyBird} a slightly different, dimensionless, basis is used for the counterterms \citep[see][for example]{DAmico:2020kxu,Piga:2022mge}. One has
\begin{equation}
    P^s_{\rm ctr}(k,\mu) = 2 P_{11}^{\delta \delta}(k) \left(\frac{k}{k_{M}}\right)^2 (b_1 + f \, \mu^2)\left(c_{\rm ctr} + c_{r,1}  \mu^2 + c_{r,2}  \mu^4\right)\,,
    \label{eq:ct_PB}
\end{equation}
where $k_{M} =0.7 \, h \, {\rm Mpc}^{-1}$ parametrises the inverse spatial extension scale of galaxies. The total number of nuisance parameters is therefore the same: $\{ c_{\rm ctr}, c_{r,1}, c_{r,2} \}$. 

When we only consider the monopole and quadrupole of \cref{eq:eftoflsspofk}, as is done in the majority of this paper, we can set $\tilde{c}_4 = c_{r,2}=0$. 
This yields the following mapping between the two different counterterm bases 
\begin{equation}
    c_{\rm ctr} = -\frac{k_M^2 \left(35 b_1 \tilde{c}_0+30 \tilde{c}_0 f+3 \tilde{c}_2 f^2\right)}{35 b_1^2+30 b_1 f+3 f^2}\,,
\end{equation} 
\begin{equation}
    c_{r,1} = -\frac{35 k_M^2 (b_1 \tilde{c}_2-\tilde{c}_0)}{35 b_1^2+30 b_1 f+3 f^2}\,.
    \label{eq:map_cr1}
\end{equation}
This mapping was obtained after imposing that the monopole and the quadrupole of the two different expressions used for the counterterm, \cref{eq:ct_PBJ,eq:ct_PB},  have to be the same respectively. 

We remind the reader that in theories such as $f(R)$, the growth rate, $f = -G_1(k,a)/F_1(k,a)$, is both time and scale-dependent, and the linear power spectrum has additional scale dependencies coming from $F_1(k,a)$: $P_{\rm 11}^{\delta \delta} = F_1^2(k,a) P_{\rm 11,i}^{\delta \delta}$.


\subsubsection{Taruya--Nishimichi--Saito} \label{sec:tns}  

\noindent The TNS prescription is given by \citep{Taruya:2010mx}
\begin{align}
 P_{\rm TNS}(k,\mu) & = D_{\rm FoG}(\mu^2 k^2 \sigma_{\rm v}^2) \, \Big[P_{\rm 1-loop}^{\delta \delta} (k) + 2 \mu^2 P_{\rm 1-loop}^{\delta \theta}(k)  \nonumber \\ & \quad +  \mu^4 P_{\rm 1-loop}^{\theta \theta} (k)  +  A(k,\mu) + B(k,\mu) \Big] \, ,
\label{eq:tnspofk}
\end{align} 
where $A$ and $B$ terms are as in \cref{eq:rsdpofk}, and the $C$ term has been omitted as it can be largely absorbed in $D_{\rm FoG}$ \citep{Taruya:2010mx}. The damping factor $D_{\rm FoG}$ encodes a phenomenological description of the FoG damping. This is a function of the velocity dispersion, $\sigma_{\rm v}$, which is taken to be a free nuisance parameter. Here we assume this function takes a Lorentzian form so that
\begin{equation}
D_{\rm FoG}(k^2\mu^2 \sigma_{\rm v}^2) \coloneqq \frac{1}{1 + (k^2\mu^2 \sigma_{\rm v}^2)/2} \, .\label{DFoG}
\end{equation}
This form has been shown to be good at describing nonlinear effects~\citep[see for example][]{Bose:2019psj}. This model then has a single nuisance parameter at the dark matter level, $\{\sigma_{\rm v}\}$, with units $h^{-1} \, {\rm Mpc}$, which represents an effective velocity dispersion, not to be confused with the linear estimate given in \cref{eq:veldisp}.


\subsection{Tracer bias}\label{sec:nonlinear-bias}

The next step to connect theory with the spectroscopic observable is to translate the dark matter spectrum to that of galaxies, which are biased tracers of the CDM distribution. As a proxy, one can consider the CDM halo power spectrum. This can be modelled incorporating an appropriate bias model. 


\subsubsection{Eulerian bias expansion} 

A general Eulerian perturbative bias expansion can be used to relate the halo density field to the dark matter density field, with the (scale-independent) coefficients of each term being treated as free parameters to be fit to observations~\citep{McDonald:2009dh,Chan:2012jj,Assassi:2014fva,Senatore:2014eva,Mirbabayi:2014zca,Desjacques:2016bnm,Fujita:2016dne}. The terms in the bias expansion that are relevant for the calculation of the 1-loop power spectrum of halos are
\begin{equation}
\delta_{\rm halo} = b_1 \delta + \frac{b_2}{2} \delta^2 + b_{\mathcal{G}_2} \mathcal{G}_2 \paren{\tilde\Phi\, | \, \xv} + b_{\Gamma_3} \Gamma_3 + b_{\nabla^2 \delta} \nabla^2 \delta + {\rm noise} \, , \label{eq:biasexp}
\end{equation}
where $b_1$ and $b_2$ are the linear and quadratic bias parameters, $\nabla^2 \delta$ is a higher-derivative operator and `noise' denotes the stochastic contributions uncorrelated with $\delta$ and with zero means, encoding shot-noise.  The $\nabla^2 \delta$ term is completely degenerate with the effects of the first counterterm \citep[see for example][and \cref{eq:ct_PBJ}]{Perko:2016puo} and so we do not consider it further here, as was also done in previous EFTofLSS-based clustering data analyses \citep{Ivanov:2019pdj,DAmico:2019fhj}. $\mathcal{G}_2$ and $\Gamma_3$ are non-local operators which take into account the large-scale tidal fields at leading and next-to-leading order. In configuration space, they are defined as:
\begin{equation}
        \mathcal{G}_2\paren{\tilde\Phi\,|\,\xv} \coloneqq \brackets{\nabla_{ij}\,\tilde\Phi(\xv)}^{\,2}-\brackets{\nabla^{\,2}\,\tilde\Phi(\xv)}^{\,2},
        \label{eq:G2_operator_conf}
\end{equation}
and
\begin{equation}
    \Gamma_3(\xv)\coloneqq\mathcal{G}_2\paren{\tilde\Phi\,|\,\xv}-\mathcal{G}_2\paren{\tilde\Phi_{\rm v}\,|\,\xv},
        \label{eq:Gamma3_operator}
\end{equation}
where the potentials $\tilde \Phi$ and $\tilde \Phi_{\rm v}$ are defined as $\tilde \Phi=\nabla^{-2} \delta$,  $\tilde \Phi_{\rm v} =-\nabla^{-2} \left(\theta/f\right)$. We have also defined  $\nabla_{ij} \coloneqq \nabla_i \nabla_j$ and $\nabla^2 \coloneqq \nabla_i \nabla_i$. 

Moving to redshift space, we obtain the following halo redshift-space power spectrum \citep[see][for example]{Senatore:2014vja}
\begin{align}
P^s_g(k, \mu)  & =  D_{\rm FoG}(k^2 \mu^2 \sigma_{\rm v}^2) \, \left[P_{\rm g, 1-loop}^{s}(k, \mu)+P_{\rm g, noise} ^{s}(k, \mu) \right] \nonumber \\ 
& \quad +P_{\rm ctr}^{s}(k, \mu)\,.
\label{eq:rsdpofkbias}
\end{align}
The term in square brackets is what we get from taking the ensemble average of the redshift space halo density perturbation in \cref{eq:biasexp}.

We have also included the nonlinear corrections coming from the EFTofLSS and TNS. In the EFTofLSS, $D_{\rm FoG}(k^2 \mu^2 \sigma_{\rm v}^2) = 1$ and the counterterms $P_{\rm ctr}^{s}(k,\mu)$ are given by \cref{eq:ct_PBJ,eq:ct_PB} for \texttt{PBJ} and \texttt{PyBird} respectively. In TNS, $P_{\rm ctr}^{s}(k,\mu)=0$ while $D_{\rm FoG}(k^2 \mu^2 \sigma_{\rm v}^2)$ is given by \cref{DFoG}. Further, in the TNS case we do not include the last two terms of \cref{eq:rsdpofk}, in accordance with \cite{Taruya:2010mx}.

The power spectrum of the stochastic part of \cref{eq:biasexp}, and as appearing in \texttt{PBJ}, is given by  
\begin{align}
P_{ \rm g, noise}^{s}(k, \mu)  & = N + \frac{1}{\bar{n}} \left(\epsilon_{0} k^2 + \epsilon_{2} \mu^2 k^2 \right) \label{eq:sn_PBJA} \;,
\end{align}
where $\bar{n}$ is the galaxy number density in units of $h^3 \, {\rm Mpc}^{-3}$ and $N$, $\epsilon_0$, $\epsilon_2$ are constants with dimensions $h^{-3} \, {\rm Mpc}^3$, $h^{-2}\, {\rm Mpc}^2$ and $h^{-2}\, {\rm Mpc}^2$, respectively. \texttt{Pybird} uses the same modelling, with dimensionless parameters $c_{\epsilon,0} = N \bar{n}$, $c_{\epsilon,1} = \epsilon_0 k_M^2$ and $c_{\epsilon,2} = \epsilon_2 k_M^2/f$ to give 
\begin{align}
P_{ \rm g, noise}^{s}(k, \mu)  & =\frac{1}{\bar{n}} \left(  c_{\epsilon,0}  + c_{\epsilon,1} \frac{k^2}{k_M^2} + c_{\epsilon,2} f \, \mu^2 \frac{k^2}{k_M^2} \right) \label{eq:sn_PBJB} \, .
\end{align}
In this work we set $\epsilon_0 = \epsilon_2 =0$ and only consider the constant stochastic term, $N$, as done in the flagship BOSS analysis \citep{Beutler:2016arn} and in \cite{Ivanov:2019pdj}. We have checked that the addition of $\epsilon_0$ does not significantly improve the fits to the simulations. $\epsilon_2$ however was found to improve the modelling significantly and has been included in some EFTofLSS-BOSS analyses \citep[for example][]{DAmico:2019fhj}. We have indeed checked that the goodness-of-fit to selected simulations does improve by including $\epsilon_2$. Despite this, we choose to omit it in order to keep the complexity of the model minimal, which should not affect relative differences between $\Lambda$CDM and beyond-$\Lambda$CDM scenarios.

In the notation of \cref{eq:rsdpofk}, the 1-loop part can be expanded as 
\begin{align}
P_{\rm g, 1-loop}^{s}(k, \mu) & =   b_1^2 P_{\rm 1-loop}^{\delta \delta} (k) + 2 b_1 \mu^2 P_{\rm 1-loop}^{\delta \theta}(k) +  \mu^4 P_{\rm 1-loop}^{\theta \theta} (k) \nonumber \\ & \quad +  A(k,\mu, b_1) + B(k,\mu, b_1) +  C(k,\mu, b_1) \nonumber \\ & \quad- k^2\mu^2 \tilde\sigma_{\rm v}^2 \, \left[b_1 F_1(k)+ G_1(k)\, \mu^2 \right]^2 P_{\rm 11,i}^{\delta \delta}(k) \nonumber \\ 
& \quad + B_{\rm g,22}^{s}(k,\mu)+B_{\rm g,13}^{s}(k,\mu) \, ,
\label{eq:rsdpofk2}
\end{align}
where we now have included the linear bias contributions explicitly for the 1-loop spectra and implicitly for the $A$, $B$, and $C$ correction terms (see \cref{app:rsdterms}). 

The last two terms of \cref{eq:rsdpofk2} include the contributions from the higher-order bias. In this work we treat these terms under the Einstein--de Sitter approximation outlined in \cref{sec:approximations}. We make this approximation explicit in the relevant expressions to follow. Our reasoning is that any additional scale dependencies coming from beyond-$\Lambda$CDM physics can be absorbed by the higher-order bias coefficients, and indeed a full derivation of these terms for general cosmologies and theories of gravity has proven not to be necessary under the considerations of this paper (see \cref{sec:results}). 

Under this assumption, the higher-order bias terms can be expressed as \citep[see Appendix A of][for example]{Ivanov:2019pdj}
\begin{align}
\label{eq:loopintegrals}
B_{\rm g,22}^{s}(k,\mu) & =  2\,\int \mathrm{d}^3q \,P^{\delta \delta}_{11}(q)\,P^{\delta \delta}_{11}(|\kv-\qv|) \nonumber \\ 
& \quad \times Z_{b,2}(\qv,\kv-\qv) \left[  \frac{b_2}{2} + b_{\mathcal{G}_2} S^b(\bfq, \bfk-\bfq) \right] \, , \\
B_{g,13}^{s}(k,\mu) & =  6\,Z_{b,1} (\mu)\,P^{\delta \delta}_{11}(k)\int \mathrm{d}^3q\,Z_{b,3}^{\rm sym}(\qv,-\qv,\kv)\,P^{\delta \delta}_{11}(q)\, ,
\end{align}
where $S^b(\kv_1, \kv_2) = (\kv_1 \cdot \kv_2)^2 / k_1^2k_2^2 - 1$. We note here that we have included the linear growth implicitly in the linear spectra and have assumed it is calculated at the external wave mode, for example $P_{\rm 11}^{\delta \delta}(q) = F_1^2(k,a) \, P_{\rm 11, i}^{\delta \delta}(q)$. The $Z_{b,n}$ kernels can be written as \citep[see][]{Scoccimarro:1999ed,Bernardeau:2001qr} 
\begin{align}
Z_{b,1}(\mu) & = b_1 + f \, \mu^2 \,, \\
Z_{b,2}(\kv_1,\kv_2) & = 2 b_1 F_2^{\rm EdS}(\kv_1, \kv_2) + 2 f \mu^2 G_2^{\rm EdS}(\kv_1,\kv_2)  \nonumber \\ & \quad + f \, \mu \, k \, 
\left[ \frac{\mu_1}{k_1}(b_1+f\mu_2^2) + \frac{\mu_2}{k_2}(b_1+f\, \mu_1^2) \right] \nonumber \\ 
& \quad + \frac{b_2}{2} + b_{\mathcal{G}_2} S^b(\kv_1, \kv_2)  
\,, \\
Z_{b,3}(\kv_1,\kv_2,\kv_3) & =  \frac{b_2}{2} f \, \mu \, k
\frac{\mu_1}{k_1}
+\,b_{{\mathcal G}_2}
f \,  \mu \,  k \frac{\mu_1}{k_1} \,S^b(\kv_2,\kv_{3}) \, \nonumber \\
& \quad +  b_2 F_2^{\rm EdS}(\kv_1, \kv_2) + 2 b_{{\mathcal G}_2} S^b(\kv_1 ,\kv_{23}) \,  F_2^{\rm EdS}(\kv_2, \kv_3) \nonumber \\
& \quad + 2 b_{\Gamma_3}S^b(\kv_1, \kv_{23}) \left[F_2^{\rm EdS}(\kv_2, \kv_3) - G_2^{\rm EdS}(\kv_2, \kv_3) \right]  \, , 
\label{eq:zkernels}
\end{align}
where `EdS' means the kernel is an Einstein--de Sitter expression. We note that $f=-G_1(k,a)/F_1(k,a)$ is calculated exactly for the theory of gravity or dark energy under consideration at the external wave mode, $k$. Further, the $Z_{b,3}$ kernel needs to be symmetrised to get $Z_{b,3}^{\rm sym}$ and variables $k$ and $\mu$ are the amplitude of the sum of the momenta $\kv = \kv_1+ \kv_2+ \kv_3$ and cosine of angle between $\hat{k}$ and the line of sight $\hat{z}$.

Now  we report here the mapping among the different bias  bases used in the analyses of this work for both \texttt{PBJ} and \texttt{Pybird} codes. The bias basis used by \texttt{PBJ} is presented in \cref{eq:biasexp}, and uses $[b_1, b_2, b_{\mathcal{G}_2}, b_{\Gamma_3}]$, while \texttt{Pybird} adopts the basis presented in~\cite{DAmico:2019fhj}, that we indicate with $[\hat{b}_1,\hat{b}_2,\hat{b}_3,\hat{b}_4]$. There is a one-to-one correspondence between the two bases:
\begin{align}
    b_1 &= \hat{b}_1\,, \\
    b_2 &= -2 (\hat{b}_1-\hat{b}_2-\hat{b}_4)\,, \\
    b_{\mathcal{G}_2} &= \frac{2}{7}\left[\hat{b}_2 + \frac{7}{4}(a_\gamma - 2)\hat{b}_1\right] \,, \\
    b_{\Gamma_3} &= \left(\frac{a_\gamma^2}{2} - \frac{a_{\gamma a}}{2} +1\right)\hat{b}_1 + \frac{2}{7}a_\gamma \hat{b}_2 - \frac{2}{21}\hat{b}_3\,.
\end{align}
Here $a_\gamma$ and $a_{\gamma a}$ are the bootstrap time-dependent functions introduced in~\cite{DAmico:2021rdb} and are usually fixed to their EdS values $a_\gamma^{\rm EdS} = 10/7$ and $a_{\gamma a}^{\rm EdS} = 3/7$.
We fix $b_{\Gamma_3} = 0$ because it is degenerate with $b_{\mathcal{G}_2}$, as well as  $b_{\nabla^2 \delta} = 0 $ as it is degenerate with a counterterm. We vary all the other parameters.

Further, using the TNS model, we wish to test the Local Lagrangian relationship \citep{Sheth:2012fc,Baldauf:2012hs,Saito:2014qha} for beyond-$\Lambda$CDM cases, which is known to hold well for $\Lambda$CDM. Again, we adopt this choice for the TNS model only, as was applied in the BOSS survey \citep{Beutler:2016arn}. This leaves us with only the following bias parameters in the TNS case $\{b_1, b_2, N\}$, which are those used in \cite{Beutler:2016arn} and can be related to the basis used in \cite{McDonald:2009dh} through \cite{Saito:2014qha}. To map these to the basis used in \cite{Assassi:2014fva}, which is used in our EFTofLSS models, see \cite{Desjacques:2016bnm}.


\subsubsection{Scale-dependent Q-bias}
For the gravity and dark energy models that produce a scale-independent modification to the linear growth factor $F_1(a)$, one does not expect any significant additional scale-dependencies entering the bias expansion \citep[see, for example,][]{Valogiannis:2019nfz,Valogiannis:2019xed}. However, this is not the case in $f(R)$ which yields a scale-dependent growth factor $F_1(k,a)$ and growth rate $f(k,a)$. To address possible scale-dependent modifications to the bias parameters, we also tested the phenomenological `Q-bias' prescription for the linear bias of \cite{Song:2015oza},
\begin{equation}
b_1(k) = b_0 \frac{1+A_2 k^2}{1+A_1 k} \,, \label{eq:qbias}
\end{equation} 
which was shown to match $N$-body halo measurements well. 
This introduces two new free parameters $\{A_1,A_2\}$ with dimensions $h^{-1} \, {\rm Mpc}$ and $h^{-2} \, {\rm Mpc}^2$ respectively, characterising possible scale-dependencies of the linear bias. To keep the degrees of freedom comparable to the Eulerian bias expansion, when adopting this model, we set all higher-order bias terms in \cref{eq:rsdpofk2} to zero, and simply replace $b_1$ with \cref{eq:qbias}.


\subsection{Infrared-Resummation}\label{sec:resummation}

A further complication comes from the baryonic acoustic oscillations whose imprint on the power spectrum should be treated properly to avoid percent-level oscillatory inaccuracies \citep{Carrasco:2013mua,Blas:2016sfa}. To treat this insufficient damping of the baryon acoustic oscillations, various resummation methods have been proposed. This is particularly relevant for the EFTofLSS and we do not adopt an infrared-resummation scheme in the TNS case. We describe the method implemented by each EFTofLSS code below.

\subsubsection{ Wiggle-no-Wiggle decomposition}\label{sec:wignowig}

\texttt{PBJ} and \texttt{MG-Copter} adopt a \enquote{\textit{Wiggle-no-Wiggle}} (WnW) decomposition approach, both under the Einstein--de Sitter approximation (see \cref{sec:approximations}). This resummation method is based on a splitting of the linear power spectrum into a \enquote{wiggle} part (containing the baryon acoustic oscillations features) and a \enquote{no-wiggle}, broadband part \citep{Baldauf:2015xfa,Vlah:2015sea, delaBella:2017qjy}:
 \begin{equation}
      P_{11}^{\delta \delta} = P_{11}^{\rm w} + P_{11}^{\rm nw} \, . 
  \end{equation}
The loop integrals of \cref{eq:rsdpofk2} are then computed separately on the two components, and resummed with a damping applied to the wiggle component. Specifically, \texttt{PBJ} computes the smooth component by convolving the Eisenstein and Hu fit for the broadband linear spectrum \citep{Eisenstein:1997jh} with a Gaussian filter:
\begin{equation}
    P_{11}^{\rm nw}(k) = P_{\rm EH}(k) \,  \left[ \mathcal{F} \ast R \right] (k) \, ,
\end{equation}
where $P_{\rm EH}$ is the Eisenstein and Hu prescription for the linear matter power spectrum, $R = P^{\delta \delta}_{11}(k) / P_{\rm EH}(k)$ and $\mathcal{F}$ is a 1D Gaussian filter, so that the convolution reads
\begin{equation}
    \left[\mathcal{F} \ast R \right](k)  =  \frac{1}{\sqrt{2 \pi}\lambda} \int_{q_{\rm min}}^{q_{\rm max}}  \frac{\diff q}{q}~R(q)\, \exp{ \left\{ - \frac{(\ln k/q)^2}{2\lambda^2} \right\}} \, ,
\end{equation}
with $\lambda$ being the dimensionless width of the Gaussian filter. 

We then compute the loop integrals using the linear ($P_{\rm 1-loop}^{s}$) and smooth power spectra ($P_{\rm 1-loop}^{s,{\rm nw}}$), with their difference being the loop corrections to the wiggle component ($P_{\rm 1-loop}^{s,{\rm w}} = P_{\rm 1-loop}^{s} - P_{\rm 1-loop}^{s,{\rm nw}}$). Finally, we construct the infrared-resummed 1-loop power spectrum as 
\begin{align}
    P_{\rm NL}^{s,{\rm IR-resum}}(k) & =  (b_1 + f~\mu^2)^2 \left\{ P^{\rm nw}_{11}(k)  \right. \nonumber \\ & \quad \left. + {\rm e}^{-k^2 \Sigma^2(\vec{k},\vec{\ell}_{\rm osc})}~P^{\rm w}_{11}(k) \left[ 1 + k^2 \Sigma^2(\vec{k},\vec{\ell}_{\rm osc}) \right] \right\}  \nonumber \\ & \quad + P^{s, {\rm nw}}_{\rm 1-loop}(k) + {\rm e}^{-k^2 \Sigma^2(\vec{k},\vec{\ell}_{\rm osc})}~P^{s, {\rm w}}_{\rm 1-loop}(k) \, ,
\label{Pres}
\end{align}
where $\Sigma^2(\vec{k},\vec{\ell}_{\rm osc})$ is the RSD damping function computed as 
\begin{align}
\Sigma^2(\vec{k}, \vec{\ell}) & = \frac{1}{2} \left\{ \Xi_0(\ell) \left(1 + 2 f \, \mu^2 + f^2 \mu^2\right) \right. \nonumber \\ 
    & \quad + \left. \Xi_2(\ell) \left[ (\hat{\vec{k}} \cdot \hat{\vec{\ell}})^2 + 2 f \mu \, \mu_\ell \, (\hat{\vec{k}} \cdot \hat{\vec{\ell}}) + f^2 \mu^2 \mu_\ell^2\right] \right\} \, , 
 \label{eq:sigmas}
\end{align}
and we have defined 
\begin{align}
    & \Xi_0(\ell) = \frac{2}{3} \int_0^\infty \frac{\mathrm{d} p}{2 \pi^2} \, {\rm e}^{-p^2/\Lambda_{\rm IR}^2} P^{\delta \delta}_{11}(p) \left[ 1 - j_0(p \,  \ell) - j_2(p \, \ell) \right] \, ,
    \label{eq:Xi0} \\
    & \Xi_2(\ell) = 2 \int_0^\infty \frac{\mathrm{d} p}{2 \pi^2} \, {\mathrm e}^{-p^2/\Lambda_{\mathrm IR}^2} P^{\delta \delta}_{11}(p) j_2(p \, \ell) \, . \label{eq:Xi2}
\end{align}
Here $j_n$ are the spherical Bessel functions of order $n$ and $\mu_\ell$ is the angle between $\vec{\ell}$ and the line of sight. $\Lambda_{\rm IR}$ is a cutoff scale. In \texttt{PBJ} we do not include the exponential cutoff but rather truncate the upper bound of the integral to  $k_{\rm s}$. \texttt{PyBird} also uses these functions (see \cref{sec:lagrangianresum}), but instead of a cutoff, it sets $\Lambda_{\rm IR}=0.2 \,h\,{\rm Mpc}^{-1}$. $\ell_{\rm osc} = | \vec{\ell}_{\rm osc}|$ is set to be the baryon acoustic oscillations scale. 

For \texttt{PBJ}, we  adopt the following choice of parameters: $q_{\rm min} = k \, {\rm e}^{-4 \lambda}$, $q_{\rm max}= k\,{\rm e}^{4\lambda}$, $k_s=0.2\,h\,{\rm Mpc}^{-1}$, $\lambda=0.25$ and $\ell_{\rm osc}=102.707\,h^{-1}\,{\rm Mpc}$ with $\hat{\vec{\ell}}_{\rm osc}$ and $\hat{\vec{k}}$ having the same orientation. We note that $\ell_{\rm osc}$ is in principle degenerate with cosmology, and in a full cosmological analysis it should also be varied. This being said, the impact of varying this parameter in such an analysis has been checked internally within the Euclid Collaboration and has been found to be minimal.


\subsubsection{Lagrangian resummation}\label{sec:lagrangianresum}
\texttt{Pybird} adopts a different approach to resummation, explicitly developed in~\cite{Senatore:2014vja}.
The details of the numerical implementation are explained in~\cite{DAmico:2020kxu}. 

The Lagrangian resummation starts from the expression of the overdensity as a functional of the displacement field $\vec{\psi} (\vec{y}, t) \coloneqq \vec{l}(\vec{y},t) - \vec{y}$, where $\vec{l}(\vec{y},t)$ is the final position of the particle and $\vec{y}$ is the initial position. The overdensity is given as 
\begin{equation}
\begin{split}
    1 + \delta(\vec x, t) & = \int \mathrm{d}^3 y \,  \delta_\mathrm{D}^{(3)}[\vec{x} - \vec{y} - \vec{\psi}(\vec{y}, t)] \\
    &= \int \mathrm{d}^3 y \int \frac{\mathrm{d}^3 k}{(2 \pi)^3} \,  {\rm e}^{{\rm i} \, \vec{k} \cdot [\vec{x} - \vec{y} - \vec{\psi}(\vec{y}, t)]} \, .
\end{split}
\end{equation}
In redshift space, one needs to separate the components perpendicular and parallel to the line of sight 
\begin{align}
     1 + \delta^s(\vec s, t) &= \int \mathrm{d}^3 y\, \delta_\mathrm{D}^{(2)}[\vec{s}_\perp - \vec{y}_\perp - \vec{\psi}^s_\perp(\vec{y}, t)]  \nonumber \\ 
     &\quad \times \delta_\mathrm{D}^{(1)}[s_\parallel - y_\parallel - \psi^s_\parallel(\vec{y}, t) - \dot{\psi}^s_\parallel(\vec{y}, t)/H ]  \nonumber \\ 
    &= \int \mathrm{d}^3 y\int \frac{\mathrm{d}^2 k_\perp}{(2 \pi)^2} \int \frac{\mathrm{d} k_\parallel}{2 \pi} \, {\rm e}^{{\rm i} \, \vec{k}_\perp \cdot \left[\vec{s}_\perp - \vec{y}_\perp - \vec{\psi}^s_\perp(\vec{y}, t) \right]} \nonumber \\ 
    &\quad \times {\rm e}^{{\rm i} \, k_\parallel \left[s_\parallel - y_\parallel - \psi^s_\parallel(\vec{y}, t) - \dot{\psi}^s_\parallel(\vec{y}, t)/H \right] } \, . \label{eq:redshiftspacedelta}
\end{align}
Here $\vec{\psi}^s_\parallel(\vec{y}, t)  =  (\vec{\psi}^s(\vec{y},t) \cdot \hat{z} ) \, \hat{z}$ and $\vec{\psi}^s_\perp(\vec{y}, t) = \vec{\psi}^s(\vec{y},t) - \vec{\psi}^s_\parallel(\vec{y}, t) $. From \cref{eq:redshiftspacedelta},  it is easy to derive the power spectrum
\begin{equation}
   P(\vec{k}_\perp, k_\parallel) = \int \mathrm{d}^3 y \,  {\rm e}^{-{\rm i} \, \vec{k} \cdot \vec{y}} K_r(\vec{k}, \vec{y}) \, ,
\end{equation}
where
\begin{equation}
\begin{split}
  K_r(\vec{k}, \vec{y}) &=  \ave{ {\rm e}^{{\rm i} \, \vec{k}_\perp \cdot \left[\vec{\psi}^s_\perp(\vec{y}) - \vec{\psi}^s_\perp(\vec{0}) \right]} 
   {\rm e}^{{\rm i} \, k_\parallel \left[\psi^s_\parallel(\vec{y}) - \psi^s_\parallel(\vec{0}) + \frac{\dot{\psi}^s_\parallel(\vec{y}, t) - \dot{\psi}^s_\parallel(\vec{0}, t)}{H} \right]} } \\
   &= \exp{\sum_{N=1}^{\infty} \frac{1}{N!}  \ave{ \left\{ {\rm i} \,\vec{k}_\perp \cdot \vec{\Delta}_\perp(\vec{y}) + i \, k_\parallel \left[ \Delta_\parallel(\vec{y}) + \dot{\Delta}_{\parallel}(\vec{y}) /H \right] \right\}^N } } \, , 
\end{split}
\end{equation}
having defined $\vec{\Delta}(\vec{y}) \coloneqq \vec{\psi}(\vec{y}) - \vec{\psi}(\vec{0})$.
Now, one only resums the linear displacement field on large scales, while expanding perturbatively the short-scale displacements and the density field.
In practice, one cuts off the displacement integrals in the exponential, as done in \cref{eq:Xi0,eq:Xi2}.
After some lengthy algebra, one arrives at the formula
\begin{equation}
    \left. P^\ell(k) \right|_N = \sum_{j=0}^N \sum_{\ell'} 4 \pi (-{\rm i})^{\ell'} \int_0^\infty \mathrm{d} r  \, r^2 Q_{||N-j}^{\ell \ell'}(k, r) \, \xi_j^{\ell'}(r) \, , 
    \label{eq:Presum}
\end{equation}
where  $\left.P^\ell(k) \right|_N$ is the resummed ($\ell$-th multipole of the) power spectrum up to order $N$, $\xi_j^{\ell}(r)$ is the $j$-th loop order term in Eulerian perturbation theory of the ($\ell$-th multipole of the) correlation function, and $Q_{||N-j}^{\ell \ell'}(k, r)$ is given by
\begin{align}
     Q_{||N-j}^{\ell \ell'}(k, r) & = {\rm i}^{\ell'} \frac{2 \ell+1}{2}  \nonumber \\
    & \quad   \times \int_{-1}^1 \mathrm{d} \mu_k \int \frac{\mathrm{d}^2 \hat{r}}{4 \pi} {\rm e}^{-{\rm i} \vec{k} \cdot \vec{r}} F||_{N-j}(\vec{k}, \vec{r}) \, \mathcal{L}_\ell(\mu_k)\, \mathcal{L}_{\ell'}(\mu_r) \, , \\
     F||_{N-j}(\vec{k}, \vec{r}) &= T_{0,r}(\vec{k}, \vec{r})  \, T_{0,r}^{-1}||_{N-j}(\vec{k}, \vec{r}) \, , \\
     T_{0, r}(\vec{k}, \vec{r}) & = \exp \bigg\{- k^2 \Sigma^2(\vec{k},\vec{r})\bigg\} \, , 
\end{align}
where $\diff^2 \hat{r}$ is an angular integration and $\mu=\mu_k$ is now an integration variable. Here $T_{0,r}^{-1}||_{N-j}(\vec{k}, \vec{y})$ means that we need to perturbatively expand (in powers of $P^{\delta \delta}_{11}$) $T_{0,r}^{-1}$ up to loop order $N-j$.

If in \texttt{PyBird} the option $\textsc{optiresum}$ is chosen, the correlation function is split in a smooth part and a baryon acoustic oscillations peak, and \cref{eq:Presum} is only applied to the baryon acoustic oscillations peak. The result is then added to the smooth power spectrum. For all \texttt{PyBird} results shown in this paper we have used $\textsc{optiresum}$.

We find that $\textsc{optiresum}$ and the WnW decomposition are highly consistent, but are both significantly discrepant with the Lagrangian resummation predictions. The reader is directed to \cref{app:validate} for more details on these comparisons. 


\subsubsection{Resummation in scale-dependent models}\label{scaledep}
We discuss now the extension of the infrared-resummation procedure to models where the linear growth is scale-dependent, as in the $f(R)$ model of \cref{sec:nl-fofr} or when massive neutrinos are considered (see next section).  Both infrared-resummation schemes discussed above consider contributions of the form (we discuss only real space, for simplicity)
\begin{align}
& - \frac{k^2}{3}\int \frac{\mathrm{d}^3 q}{(2\pi)^3} {\rm e}^{-q^2/\Lambda_{\rm IR}^2} \frac{P^{\delta \delta}_{11}(q)}{q^2}\left[P(k)-P(|\bfk-\bfq|)\right]\,,
\label{eqIR}
\end{align}
where the power spectra inside square brackets are computed at a given order in perturbation theory. The cutoff ensures that $q\ll k$ in most of the integration domain. 
 If $P(k)$ contains a new scale, as is the case for $\ell_{\rm osc}$ in $P^{\rm w}(k)$, the difference in parentheses is either of order $P_{11}^{\rm w}(k)\, q^2 \ell_{\rm osc}^2$ (for $q \,\ell_{\rm osc} \ll 1$) or  $P_{11}^{\rm w}(k)$ (for $q \,\ell_{\rm osc} \gg 1$, as the oscillations in the second term inside parentheses average out in this limit). Both regimes give a contribution to \cref{eqIR} which is $k^2$ enhanced with respect to other perturbative contributions of  the same order, and therefore should be resummed. 
 
 If, on the other hand, no oscillatory feature is present in $P(k)$,  the difference inside parentheses can be approximated as \begin{align}
&\left[P(k)-P(|\bfk-\bfq|)\right]\simeq -\frac{1}{2}\frac{\diff^2 P(k)}{\diff k^2} q^2 x^2 \,,
\label{2der}
\end{align}
with
\begin{equation}
 x=\frac{\bfk\cdot \bfq}{k \, q}.  
\end{equation}
 For an approximately scale-invariant power spectrum, the second derivative gives a contribution of order $P(k)/k^2$ and, consequently, \cref{eqIR} is not  $k^2$-enhanced. Therefore, in resummation schemes which focus on the wiggly part of the power spectrum, as in WnW or in  the  $\textsc{optiresum}$ option of \texttt{PyBird}, only leading contributions are resummed. In other schemes, as the full Lagrangian one, also subleading contributions are included, The difference between the two methods is of order 2-loop terms. The wiggle-no wiggle method has been derived as an approximation of the Lagrangian resummation in \cite{Lewandowski:2018ywf}.

In scale-dependent models the separation between leading and next-to-leading contributions could be, potentially, complicated by the presence of a new scale. However, considering for definiteness $f(R)$, we can show that it is not the case. Indeed, even assuming that the whole scale-dependence of the function $\mu(k,a)$ of \cref{frg2} is inherited by the power spectrum would imply a contribution to \cref{2der} proportional to 
\begin{equation}
\frac{\partial^2 \ln \mu(k,a)}{\partial k^2}  \ll  \ell_{\rm osc}^2 \,,
\end{equation}
where the inequality holds in the whole relevant $k$-range for the $f(R)$ models considered in this work. We conclude that the extra-scale-dependence induced in $f(R)$ is always subdominant with respect to the one induced by baryon acoustic oscillations, and  does not require any modification of the infrared-resummation schemes with respect to those for scale-independent models. \texttt{Pybird} implements this scheme, and computes the linear power spectrum in \cref{eq:Xi0,eq:Xi2}  by taking the scale-dependent linear growth into account.

Since the scale-dependence in $f(R)$ is much stronger than that induced by massive neutrinos, the above conclusions extend also to models with nonvanishing neutrino masses.


\subsection{Massive neutrinos} \label{sec:massive-neutrinos}

With the measurements of flavour oscillations \citep{Super-Kamiokande:1998qwk,SNO:2003bmh}, massive neutrinos entered the realm of standard physics. These have been shown to have a significant effect on cosmological observables \citep{Lesgourgues:2006nd,Wright:2019qhf,Bose:2021mkz}. We look to include these effects in \cref{eq:rsdpofkbias}.

Since galaxies are biased tracers of the CDM plus baryon field only \citep[see][for example]{Villaescusa-Navarro:2013pva,Castorina:2013wga,Costanzi:2013bha}, we only consider the effects of massive neutrinos on the CDM+baryon (cb) spectrum, and opt to include them only through the linear cb power spectrum $P_{\rm 11}^{\rm (cb)}(k, a)$.  We do not include higher order massive neutrino effects in the kernels $F_n^{\rm app}$ and $G_n^{\rm app}$ (`app' stands for approximate), which are calculated assuming a single CDM matter fluid. This differs slightly from the prescription of \cite{Wright:2019qhf}, which was shown to work very well by comparing to simulations. In particular, we do not include the effects of massive neutrinos on the growth rate $f$. We have checked that this effect is sub-percent for neutrino masses considered here. 

In practice, we use the following initial linear spectrum in the 1-loop integrals
\begin{equation}
    P_{11,{\rm i}}^{\rm (cb)}(k) = \frac{P_{11}^{\rm (cb)}(k,a)}{F^{\rm (m) 2}_1(k,a)} \,, \label{eq:mnurescale}
\end{equation}
where $F_1^{\rm (m)}(k,a)$ assumes no massive neutrino contribution but the same total matter contribution to the background evolution and Poisson equation, $\Omega_{\rm m}$. We have checked that using $\Omega_{\rm cb}$ instead makes negligible difference on the final RSD power spectrum predictions. On the other hand, $P_{11}^{\rm (cb)}(k,a)$  has the full massive neutrino and modified gravity or dark energy dependence, which we calculate using the Boltzmann code \texttt{MGCAMB} \citep{Zucca:2019xhg}. The goal is to check for any deterioration of this approximation between the $\Lambda$CDM and beyond-$\Lambda$CDM comparisons with the simulation measurements.


\subsection{Approximations}\label{sec:approximations}

The efficient calculation of the 1-loop power spectrum can be challenging, particularly when the modifications include some sort of scale-dependence, for example in $f(R)$ gravity. The numerical approach, as adopted by the \texttt{MG-Copter} code, is both time consuming and prone to numerical inaccuracies making it ill-suited for fast and comprehensive statistical analyses. On the other hand, the FFT approach, as adopted by \texttt{PyBird} and \texttt{PBJ} is highly computationally efficient and free from numerical instabilities. The challenge is extending the FFT to non-trivial kernels as introduced by modifications to gravity. Alternatively, one can apply various approximations to the computation which are easily implemented in current FFT frameworks. 

Motivated by these issues, we examine various approximations that can alleviate computational cost while maintaining the desired degree of accuracy for the halo power spectrum in redshift space. These approximations, relevant for the loop integrals implicit in the first two lines of \cref{eq:rsdpofk2}, are computed by rescaling the approximate second and third order kernels with the appropriate linear growth factors as follows,
\begin{align} 
F_2(\bfk_1,\bfk_2) & = \frac{F_1(k_1) \, F_1(k_2)}{F_{\rm 1}^{\rm app}(k_1) \, F_{\rm 1}^{\rm app}(k_2)}  F_{\rm 2}^{\rm app}(\bfk_1,\bfk_2) \, , \nonumber \\ 
G_2(\bfk_1,\bfk_2) & = \frac{f(k_{12}) \, F_1(k_1) \,F_1(k_2)}{f^{\rm app}(k_{12})\, F_{\rm 1}^{\rm app}(k_1) \,F_{\rm 1}^{\rm app}(k_2) } G_{\rm 2}^{\rm app}(\bfk_1,\bfk_2)  \, , \nonumber \\ 
F_3(\bfk_1,\bfk_2, \bfk_3) & = \frac{F_1(k_1) \, F_1(k_2) \,F_1(k_3)}{F_1^{\rm app}(k_{1}) \,F_1^{\rm app}(k_2) \,F_1^{\rm app}(k_3)} F_{\rm 3}^{\rm app}(\bfk_1,\bfk_2,\bfk_3) \, , \nonumber \\ 
G_3(\bfk_1,\bfk_2, \bfk_3) & = \frac{f(k_{123}) \, F_1(k_1) \,F_1(k_2) \, F_1(k_3)}{f^{\rm app}(k_{123})\, F_{\rm 1}^{\rm app}(k_1)\, F_{\rm 1}^{\rm app}(k_2)\, F_{\rm 1}^{\rm app}(k_3) } \nonumber \\ & \quad \times G_{3}^{\rm app}(\bfk_1,\bfk_2,\bfk_3) \, , \label{eq:rescaling}
\end{align}
where we have dropped the time dependence for compactness and we remind the reader that $f(k) = -G_1(k)/F_1(k)$ is the growth rate. Recall that `app' stands for approximate. The approximations we consider for the kernels are 
\begin{enumerate}
    \item 
    {\bf The Einstein--de Sitter approximation (EdS)}:  $F_1^{\rm app}(k,a) = F_1(k,a)$ and $G_1^{\rm app}(k,a) = G_1(k,a) $, but the higher-order kernel's scale dependence are given by the standard Einstein--de Sitter universe ($\Omega_{\rm m}(a)=1$) expressions, $F_n^{\rm EdS}$ and $G_n^{
    \rm EdS}$ \citep[see][]{Bernardeau:2001qr}. The time dependence is then given by the appropriate factors of $F_1$ and $G_1$, for example $F_2^{\rm app}(\bfk_1,\bfk_2) = F_1(k_1,a) \, F_1(k_2,a) \, F_2^{\rm EdS}(\bfk_1,\bfk_2)$.
    \item 
    {\bf The unscreened approximation (USA)}: $F_1^{\rm app}(k,a) = F_1(k,a)$ and $G_1^{\rm app}(k,a) = G_1(k,a)$ and the higher-order kernels 
    are computed by solving the perturbation evolution equations using $S(\bfk)=0$ (see ~\cref{eq:poisson1}), that is we ignore the effects of higher-order mode-coupling terms responsible for screening. 
    \item 
    {\bf The $\Lambda$CDM-screened approximation ($\Lambda$CDM-scr)}: $F_1^{\rm app}(k)$ and $G_1^{\rm app}(k)$ are the $\Lambda$CDM growth factors which are scale-independent to a very good approximation: ~$F_1^{\Lambda {\rm CDM}}(k,a) \approx F_1^{\Lambda {\rm CDM}}(a)$ and $G_1^{\Lambda {\rm CDM}}(k,a) \approx G_1^{\Lambda {\rm CDM}}(a)$. The higher-order kernels are computed using the approximate expression for $S(\bfk)$ as given in \cref{app:lcdmscr}.  This expression does not depend on the integrated momentum mode and hence allows it to be decomposed using a FFT approach.
\end{enumerate}

\section{Simulations} \label{sec:sims}

Before discussing our results, we list the various simulations we consider in \cref{tab:sims}. We describe each of these in detail below. 
\begin{table*}
\caption{\label{t7} Simulations and associated cosmological or gravitational models considered in this work.}
\centering
\begin{tabular}{| c | c | c | | c | c | c | c | | c | c | c | c |  }
\hline  
  \multicolumn{3}{| c ||}{\texttt{ELEPHANT}} & \multicolumn{4}{  c || }{\texttt{DAKAR} } & \multicolumn{4}{  c | }{\texttt{DEMNUni}}
  \\ 
    \multicolumn{3}{| c ||}{$V = 5 \times 1 \,  h^{-3} \, {\rm Gpc}^3$} & \multicolumn{4}{  c || }{$V = 1\,  h^{-3} \, {\rm Gpc}^3$} & 
   \multicolumn{4}{  c | }{$V = 8\, h^{-3} \, {\rm Gpc}^3$} \\ \hline \hline 
 {\bf Code} & $\mathbf{|f_{\rm R0}|}$ & $\mathbf{\Omega_{\rm rc}}$ & {\bf Code}  &  $\mathbf{w_0}$ & $\mathbf{w_a}$ & $\mathbf{\xi}$ [bn $\si{GeV}^{-1}$] &  {\bf Code}  &  $\mathbf{w_0}$ & $\mathbf{w_a}$ & $\mathbf{\sum m_\nu}$ [eV]  \\ \hline 
 F5 & $10^{-5}$ & - & w09 & $-0.9$ & 0 & 10 &    CPL3 & $-1.1$ & $-0.3$ & 0.00 \\ \hline 
 F6 & $10^{-6}$ & - & w11 &  $-1.1$ & 0 & 10 & CPL3-16 & $-1.1$ & $-0.3$ & 0.16 \\ \hline 
 N1 & - & 0.25 & CPL2 & $-1.1$ & 0.3 & 50 &  
  CPL4 & $-0.9$ & 0.3 & 0.00   \\ \hline 
 N5 & - & 0.01 &  -  & - & - &  - &  CPL4-16 &$-0.9$ & 0.3 & 0.16 \\ \hline 
\end{tabular}
\tablefoot{We show the volume of each simulation box, noting that for \texttt{ELEPHANT} we have five realisations. Further, each simulation set has a $\Lambda$CDM simulation with the same initial seeds as the beyond-$\Lambda$CDM simulations.}
\label{tab:sims}
\end{table*}


\subsection{The \texttt{ELEPHANT} simulations} \label{sec:ELEPHANT}
The Extended LEnsing PHysics using ANalaytic ray Tracing ({\texttt{ELEPHANT}) simulations \citep{Cautun:2017tkc,Hernandez-Aguayo:2018oxg} are a suite of five independent realisations of the GR ($\Lambda$CDM), F6 ($f_{\rm R0}=-10^{-6}$), F5 ($f_{\rm R0}=-10^{-5}$), N5 ($H_0r_{\rm c}/c=5$) and N1 ($H_0r_{\rm c}/c=1$) models, where F stands for an $f(R)$ model and N for a DGP model. The simulations were run with the \texttt{ECOSMOG} adaptive mesh refinement code \citep{Li:2011vk,Li:2013nua} and they follow the evolution of $1024^3$ dark-matter particles in a cubical box of length $1024\, \si{\hMpc}$, giving a mass resolution of $m_p = 7.78\times 10^{10}\, h^{-1} \, \si{\solarmass}$. Their initial conditions were generated at $z=49$ using the Zel'dovich approximation with the \texttt{MPgraphic} code \citep{Prunet:2008fv} and the cosmological parameters are consistent with those of the \WMAP (WMAP) data release nine collaboration \citep{Hinshaw:2012aka},
$$\{\Omega_{\rm b,0}, \Omega_{\rm m,0}, h, n_{\rm s}, A_{\rm s}\} = \{0.046,0.281,0.697,0.971,2.297\times 10^{-9}\}\, , $$
where $A_{\rm s}$ is the amplitude of primordial scalar perturbations, $n_{\rm s}$ is the scalar spectral index and $\Omega_{\rm b,0}$ is the baryon density fraction today.
The halo catalogues were constructed with the \texttt{rockstar} halo finder \citep{Behroozi:2011ju}, where we chose the $M_{200c}$ halo mass definition, which is the mass enclosed within a sphere of radius $r_{200c}$ with $200$ times the critical density of the universe.

\Refr{These simulations offer one of the largest suites of modified gravity simulations available. There are larger volume simulations \citep[see for example][]{Arnold:2018nmv}, but single realisation, and so the effective volume the \texttt{ELEPHANT} simulations provide is deemed optimal. The unavailability of simulations with a volume comparable to \Euclid largely stems from the fact that running large volume modified gravity simulations is currently extremely computationally challenging \citep[see][and references therein]{Arnold:2021xtm}.}


\subsection{The \texttt{DAKAR} simulations}
\label{sec:DS_sims}

We made use of the \texttt{DAKAR} simulations presented in \cite{Baldi:2016zom} which were run using a modified version of the \texttt{GADGET-2} $N$-body code \citep{Springel:2005mi} which consistently implements the effects of the momentum exchange within the dark sector. The simulations consisted of $1024^3$ dark matter particles in a periodic cosmological box of length $1 \,h^{-1} \,  {\rm Gpc} $, evolved from a starting redshift of $z_{\rm i}=99$. The resulting CDM particle mass is $m_{\rm c} = 8\times 10^{10} \, h^{-1} \, \si{\solarmass}$ and the spatial resolution is $\epsilon = 24 \, h^{-1} \, {\rm kpc}$. The cosmological parameters are 
$$\{\Omega_{\rm b,0}, \Omega_{\rm m,0}, h, n_{\rm s}, A_{\rm s}\} = \{0.048,0.308, 0.678,0.966, 2.115\times 10^{-9}\}.$$
We refer the interested reader to \cite{Baldi:2016zom} for a more extended description of the simulations and of the modified $N$-body code. 

\Refr{We note that these simulations have a very limited volume, far smaller than that to be probed by \Euclid. As will be seen in \cref{sec:apptest} and \cref{sec:rspdm}, the approximations considered for the Dark Scattering model are extremely good even without introducing bias degrees of freedom, and more so when considering recent lensing and clustering constraints on this model \citep{Carrilho:2022mon,Carrion:2024itc}.}


\subsection{The \texttt{DEMNUni} simulations} 
\label{sec:DEMNUni_sims}
The \enquote{Dark Energy and Massive Neutrino Universe} (\texttt{DEMNUni}) simulations~\citep{DEMNUni_simulations,Parimbelli2022} have been produced with the aim of investigating large-scale structures in the presence of massive neutrinos and dynamical dark energy, and they were conceived for the nonlinear analysis and modelling of different probes, including dark matter, halo, and galaxy clustering \citep[see][and Carella et al., in prep.]{DEMNUni1,Moresco2017,Zennaro2018,Ruggeri2018,Bel2019,Parimbelli2021,Parimbelli2022, Guidi_2022, Baratta_2022, Gouyou_Beauchamps_2023}, weak lensing, CMB lensing, SZ and ISW effects \citep{Roncarelli2015,DEMNUni_simulations,fabbian2018,Hernandez-Molinero:2023jes}, cosmic void statistics \citep{Kreisch2019,Schuster2019,verza_2019,Verza_2022a,Verza_2022b, Vielzeuf2023}, and cross-correlations among these probes \citep{Cuozzo2022}.
The \texttt{DEMNUni} simulations combine a good mass resolution with a large volume to include perturbations both at large and small scales. They are characterised by a softening length $\varepsilon=20\, h^{-1} \, {\rm kpc}$, a comoving volume of $8 \, h^{-3} \, \mathrm{Gpc}^3$ filled with $2048^3$ dark matter particles and, when present, $2048^3$ neutrino particles. The simulations are initialised at $z_{\rm i}=99$ with Zeldovich initial conditions. The initial power spectrum is rescaled to the initial redshift via the rescaling method developed in~\cite{zennaro_2017}. Initial conditions are then generated with a modified version of the \texttt{N-GenIC} software, assuming Rayleigh random amplitudes and uniform random phases.
The \texttt{DEMNUni} simulations were run using the tree particle mesh-smoothed particle hydrodynamics (TreePM-SPH) code \texttt{P-Gadget3} \citep{Springel:2005mi}, specifically modified as in \cite{Viel_2010} to account for the presence of massive neutrinos. This modified version of \texttt{P-Gadget3} follows the evolution of CDM and neutrino particles, treating them as two separated collisionless components. 

The reference cosmological parameters are chosen to be close to the baseline Planck 2013 cosmology \citep{Ade:2013zuv}
$$\{\Omega_{\rm b,0}, \Omega_{\rm m,0}, h, n_{\rm s}, A_{\rm s} \} = \{0.05, 0.32, 0.67, 0.96, 2.127 \times 10^{-9} \}.$$
Given these values, the reference (i.e., the massless neutrino case) CDM-particle mass resolution is $m^{\rm p}_{\rm CDM} = 8.27\times 10^{10} \, h^{-1} \, \si{\solarmass}$ and is decreased according to the mass of neutrino particles, in order to keep the same $\Omega_{\rm m,0}$ among all the \texttt{DEMNUni} simulations. In fact, massive neutrinos are assumed to come as a particle component in a three mass-degenerate scenario. Therefore, to keep $\Omega_{\rm m,0}$ fixed, an increase in the massive neutrino density fraction yields a decrease in the CDM density fraction.

\section{Results} \label{sec:results}


\subsection{Setup} \label{sec:setup}
We tested the various beyond-$\Lambda$CDM modelling approaches against measurements from $N$-body simulations, with the $\Lambda$CDM case serving as a performance benchmark. We considered the redshift space power spectrum monopole and quadrupole as they contain most of the cosmological and gravitational information, and only considered $z=1$, as this will be one of the lowest targeted redshifts of the \Euclid clustering probe \citep[see, for example,][]{Euclid:2019clj,Amendola:2016saw,EUCLID:2011zbd}, with the higher redshifts relying less on the accuracy of nonlinear modelling, and where modified gravity or dark energy effects on the power spectra are less pronounced. 

The analyses we performed are limited by the speed of \texttt{MG-Copter} which is capable of providing the spectra predictions without employing any of the approximations outlined in \cref{sec:approximations}. Because of this, we restricted our analysis to the calculation of the minimum $\chi^2$ statistic between the theory predictions and simulation measurements, comparing this statistic between the various modelling choices. This gave us a basic measure of the applicability of each choice without having to perform more expensive Bayesian parameter inference analyses. Our results aim at informing and optimising such future parameter posterior analyses (for example, D'Amico et al., in prep.).

We performed two separate analyses, one for CDM and one for CDM halos. The first tested the validity of the approximations outlined in \cref{sec:approximations} as at the level of CDM we have a very reduced nuisance parameter set with which to fit the simulations. The second analysis tested the robustness of all the modelling approaches detailed in \cref{sec:nonlinear}; the various kernel approximations of \cref{sec:approximations}, the two RSD models, the two bias schemes, and massive neutrino modelling as described in \cref{sec:nonlinear}.

The $\chi^2$ statistic is given by 
\begin{align}
\chi^2(k_{\rm max})  = & \frac{1}{N_{\rm dof}}\sum_{k=k_{\rm min}}^{k_{\rm max}} \sum_{\ell,\ell'=0,2} \left[P^{s}_{\ell,{\rm data}}(k)-P^{s}_{\ell,{\rm model}}(k)\right] \nonumber \\ & \times \mbox{Cov}^{-1}_{\ell,\ell'} (k) \left[P^{s}_{\ell',{\rm data}}(k)-P^{s}_{\ell',{\rm model}}(k)\right],
\label{covarianceeqn}
\end{align}
where $P^s_{\ell}$ is the $\ell^{\rm th}$ multipole of the redshift space CDM or halo power spectrum and $\mbox{Cov}_{\ell,\ell'}$ is the Gaussian covariance matrix between the different multipoles. In the case of a Gaussian covariance, the number of degrees of freedom are given by $N_{\rm dof} = 2\, N_{\rm bins} - N_{\rm params}$, where $N_{\rm bins}$ is the number of $k-$bins summed over  and $N_{\rm params}$ is the number of free parameters in the theoretical model. In this case, $N_{\rm params}$ is just be the number of nuisance parameters (bias and RSD) as we fixed the cosmological ones to the fiducial values.

We used a Gaussian, linear covariance between the multipoles \citep[see Appendix C of][for details]{Taruya:2010mx}. This has been shown to reproduce $N$-body results up to $k\leq 0.3 \,h\,{\rm Mpc}^{-1}$ at $z=1$ \citep{Taruya:2010mx}. We used the simulation volume (see \cref{tab:sims}) to calculate the covariance, and in the case of halos, we assumed $b_1$ as measured from the simulations. No supersample covariance was included. The effects of this was shown to be minimal in past surveys \citep{Wadekar:2020hax}, and will be investigated in the context of \Euclid in an upcoming \Euclid paper. As we mostly use simulations with a finite size, no supersample contribution should be included in these cases.  

Further, no shot noise term was considered. We have found that including a shot noise term based on the simulation-measured halo number density yields unreasonably good fits by our theoretical prescriptions to the simulations. More specifically, we found $\chi^2 \approx 1$ for $k_{\rm max} \leq 0.6~h \, {\rm Mpc}^{-1}$ and beyond in some cases. This is likely because the models are efficient at fitting shot noise above a certain scale. \Refr{We have checked that the relative behaviour of the $\chi^2$ between $\Lambda$CDM and beyond-$\Lambda$CDM models is unchanged around $\chi^2\sim 1$ with and without the shot noise contribution. This gives us confidence that the $\chi^2\sim 1$ criteria we used is still meaningful to compare the relative goodness-of-fit between the various approximations considered.}

On this note, a few things should be mentioned. The actual measurements from \Euclid will encode a larger number density of tracers than the simulations considered making us more sensitive to inaccuracies in the modelling. For example, the \texttt{ELEPHANT} halo measurements at $z=1$ have $\bar{n} \approx 3.1 \times 10^{-4}~h^3 \, {\rm Mpc}^{-3} $ while \Euclid is expected to have over twice this density of tracers \citep{Euclid:2019clj}. Further, \Euclid will measure galaxies not halos, which come with additional bias modelling considerations. Lastly, despite having fixed the cosmology and gravity to their fiducial values, such goodness-of-fit comparisons do not ensure an absence of bias in the recovered parameters in a full posterior analysis, which can emerge from complex degeneracies between nuisance and cosmological parameters, and the shape of the full parameter space posterior distribution. 

The simulations also all have smaller volumes than the expected \Euclid survey volume, $V(0.9\leq z \leq 1.1) = 7.94 \, h^{-3} \, {\rm Gpc}^3$ \citep{Euclid:2019clj}, except for \texttt{DEMNUni}, which has a slightly larger volume. This limits our capacity to make robust statements for \Euclid  based on the simulation measurements alone. 

To address these issues, we performed some analyses on mock data produced using the theoretical prescriptions. In this case we could employ a larger volume in the covariance as well as the \Euclid-estimated shot noise contribution, $\bar{n} = 6.86 \times 10^{-4} \, h^3 \, {\rm Mpc}^{-3}$ \citep{Euclid:2019clj}. In what follows, this number density and  $V = 8.8 \, h^{-3} \, {\rm Gpc}^3$ is assumed in the covariance whenever we refer to a \Euclid-like setting.  \Refr{This volume is larger than the expected volume of \Euclid at this redshift bin, which accounts partially for any additional information coming from higher redshift bin observations.}

\Refr{On this point, we remark that in all our analyses we only vary the redshift space and bias model nuisance parameters, which will be fit at every redshift bin in the \Euclid spectroscopic analysis. This means our results will not likely degrade with the inclusion of other redshift bins. It is also true that all considered beyond-$\Lambda$CDM effects are diminished at higher redshifts, and so the tested approximations will perform better at all other \Euclid redshift bins \citep[see Table.~1 of][for the specific binning]{Euclid:2019clj}. It should however be noted that the lessening of modifications to $\Lambda$CDM can be compensated by the larger volumes probed at higher redshifts. } 

We use the largest value of $k_{\rm max}$, the maximum applicable wave mode, at which theory gives a good fit to the simulation measurements, as our performance metric for testing the approximations and modelling approaches. To determine this, we followed the procedure of \cite{Bose:2019psj}:   
\begin{enumerate}
    \item
    We performed a least-squares fit to the $N$-body data by varying the set of nuisance parameters but fixing all cosmological and gravitational parameters to their fiducial values. We repeated this for $k$-bins within $0.125 \,h\,{\rm Mpc}^{-1} \leq k_{\rm max} \leq 0.300 \,h\,{\rm Mpc}^{-1}$.
    \item 
    We then calculated the $1\sigma$ confidence intervals ($\Delta \chi^2$) on a $\chi^2$ distribution using $N_{\rm dof}$ degrees of freedom. 
    \item 
     The model validity scale $k_{\rm max}$ was then chosen as the largest value of $k_{\rm max}$ which has $[\chi^2(k_{\rm max}) - \Delta \chi^2(k_{\rm max})] \leq 1$. A $\chi^2$ value of one indicates a good fit to the data. 
\end{enumerate}
This procedure provides a fair estimate of the range of validity of the models in which we can recover the fiducial cosmology with a $1\sigma$ criterion. This procedure was validated in $\Lambda$CDM in \cite{Bose:2019psj,Markovic:2019sva} using a Markov Chain Monte Carlo analysis. 

We imposed the physically motivated priors $b_1, \sigma_{\rm v} > 0 $ in the fits. Otherwise all fits assume an extremely wide prior on all other nuisance parameters, representing minimal prior information. We made this choice, as it is now known from the results of \cite{Carrilho:2022mon, Simon:2022lde}, that the choice of narrow priors on nuisance parameters can lead to differences in the inferred marginal posteriors of cosmological parameters. While also shown in the aforementioned references that prior volume effects are important and increase with the size of the priors,\footnote{Solutions to this issue are currently being investigated in the literature, including using Jeffreys priors~\citep{Donald-McCann:2023kpx} or profile likelihoods~\citep{Moretti:2023drg, Holm:2023laa}.} our results are not affected, since we are maximising the posterior, which is this case of large priors is equivalent to a maximum likelihood estimation of the parameters.


\subsection{Approximation selection}\label{sec:apptest}

We begun by performing direct spectra comparisons for the approximations outlined in \cref{sec:approximations} in order to evaluate the effectiveness of each approximation. To do this, we compared the monopole and quadrupole as predicted using the TNS model (see \cref{eq:tnspofk}) under the various approximations against the full, most rigorous numerical calculation. We considered the TNS as opposed to the 1-loop SPT prediction (see \cref{eq:rsdpofk}) as it can probe smaller scales while not invoking too many degrees of freedom which may be degenerate with the approximation's effects. We fixed the free parameter, $\sigma_{\rm v}$, for all the predictions, to the exact value which is given in the figures. \Refr{This value was chosen to be the best-fit value found by comparing to the respective dark matter $P_0$ and $P_2$ simulation measurements, where $\chi^2(k_{\rm max}) \sim 1$ (see \cref{covarianceeqn})}. The hexadecapole comparisons are also shown to check the approximation's accuracy at modelling even greater nonlinear RSD effects. 

We note that for $f(R)$ and DGP, the EdS and USA approximations have been examined in real space, and to a lesser extent in redshift space, in \cite{Bose:2016qun,Bose:2017myh,Bose:2018orj,Aviles:2020wme}. On the other hand, the $\Lambda$CDM-scr approximation has not been tested at all, and is presented for the first time in this work. For Dark Scattering and $w_0w_a$CDM, the EdS approximation was mildly tested at the level of the redshift space power spectrum in \cite{Bose:2017jjx,Carrilho:2021hly}. Given that \texttt{PyBird} has recently been extended to include the exact calculation for both DGP \citep{Piga:2022mge} and $w_0w_a$CDM \citep{DAmico:2020kxu}, we restrict ourselves to $f(R)$ and Dark Scattering in this section.

Despite Dark Scattering having been tested independently in other works, we also performed a limited test of the EdS approximation. We show this in \cref{fig:dsapp} along with multipole errors coming from a Gaussian covariance with our \Euclid-like volume at $z=1$. In particular we considered the largest modification to $\Lambda$CDM, the CPL2 case (see \cref{tab:sims}). We see that the EdS deviates by more than $2\%$ in the quadrupole at small scales, which may signal that this approximation is inadequate for \Euclid. We have checked that lowering the coupling parameter, $\xi$, produces better consistency but, more importantly, changing the RSD degree of freedom can account for the entire deviation down to less than $0.5\%$ for monopole and quadrupole, and brings all multipole ratios well within the monopole error assuming the \Euclid-like volume at this redshift bin. Given this, we find the EdS to be a good approximation for low interaction strengths within the Dark Scattering model. We investigate this further in the next section. Further, we note that in the limit $\xi \rightarrow 0$ we recover $w_0w_a$CDM and so we confirm that for mild $w_0w_a$CDM modifications, the EdS approximation is very good.

In \cref{fig:elfrapp} we show the $f(R)$ case. Here, scale dependencies in the linear growth factor and rate cause the EdS and USA approximations to break down completely, with deviations significantly larger than the \Euclid-like measurement errors. This has already been shown in the literature \citep[see, for example,][]{Aviles:2020wme}. Remarkably though, the $\Lambda$CDM-scr approximation is sub-percent consistent with the exact kernel calculation for all scales considered and well within the \Euclid-like error bands}. This provides a promising implementation for analyses pipelines which can use the FFT approach with this approximation. We investigate this further in the next section. 

To summarise, the EdS approximation works very well for models inducing a scale-independent modification to the linear growth of structure, such as DGP, $w_0 w_{\rm a}$CDM or Dark Scattering. In the case of the former, the absence of a scalar field mass term and any environment dependence of the modification, leaves its impact on scales below the crossover scale $r_{\rm c}$ uniform. Similarly, the latter two models give no additional scale dependencies in the Poisson equation nor the Euler or continuity equations \cref{eq:Perturb1,eq:Perturb2,eq:Perturb3}. This ensures deviations to the scale dependencies of the standard EdS kernels are minimal. On the other hand, $f(R)$ gravity comes with a scalar field potential term and associated mass, which induces a scale-dependent modification of the linear growth of structure. It also has the more complex, environment-dependent, Chameleon screening mechanism. This breaks the EdS approximation as well as makes the omission of screening more noticeable on the resulting power spectrum. The $\Lambda$CDM-scr approximation attempts to match the large and small scale behaviour of $\gamma_2$ and $\gamma_3$ without introducing additional scale dependencies from integrated wave modes, allowing the perturbative kernels to remain accurate (see \cref{app:lcdmscr} for more details).
\begin{figure*}
    \centering
    \includegraphics[width=\textwidth]{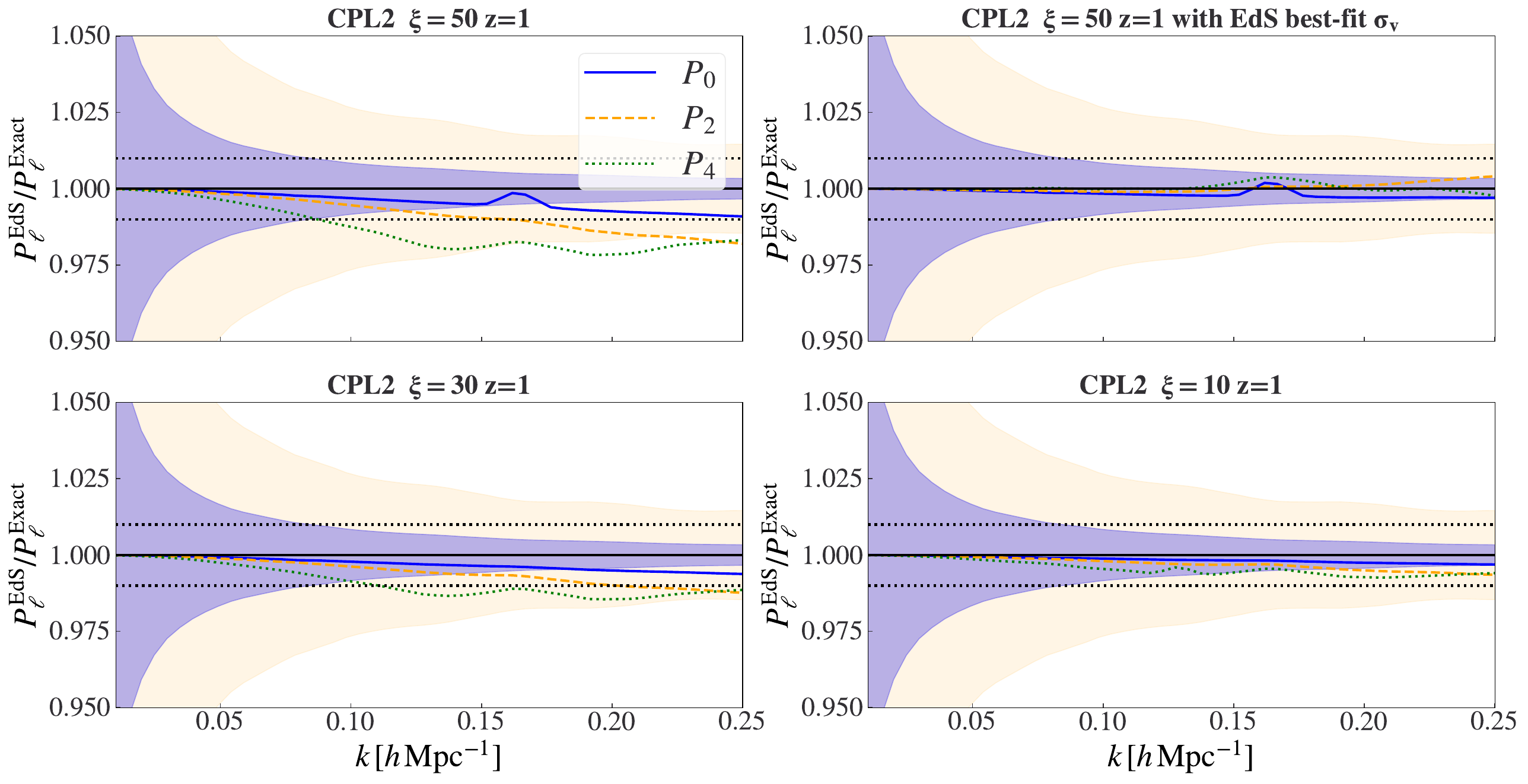}
    \caption{Ratio of the EdS approximation to the exact calculation of the redshift space matter power spectrum multipoles, computed with the TNS prescription, for the Dark Scattering model with equation of state of dark energy described by the CPL parametrization (see \cref{eq:cplpar}) with $w_0=-1.1$ and $w_a=0.3$ (denoted  CPL2 in the text, see \cref{tab:sims}), and at redshift $z=1$. Blue solid, orange dashed and green dotted lines denote the monopole, the quadrupole and the hexadecapole ratios, respectively. In the top-left, bottom-left and bottom-right panels, we use $\sigma_{\rm v}=5.2 \, h^{-1} \, {\rm Mpc}$ and $\xi=50 \, \text{bn} \, \si{GeV}^{-1}$, $\xi=30 \, \text{bn} \,  \si{GeV}^{-1}$ and $\xi=10 \, \text{bn} \, \si{GeV}^{-1}$, respectively. In the top-right panel, we use $\sigma_{\rm v}=5.14\, h^{-1}\, {\rm Mpc} $ in the EdS prediction, which is the value obtained by refitting to the exact calculation which uses $\sigma_{\rm v}=5.2 \, h^{-1} \, {\rm Mpc}$. Blue and beige bands indicate errors on the monopole and quadrupole assuming a $\Lambda$CDM Gaussian covariance with $V = 8.8 \, h^{-3} \, {\rm Gpc}^3$ and no shot noise contribution (we compare dark matter multipoles). We note that the hexadecapole error fills the plot and so we have omitted it.}
    \label{fig:dsapp}
\end{figure*}
\begin{figure*}
    \centering
    \includegraphics[width=0.49\textwidth]{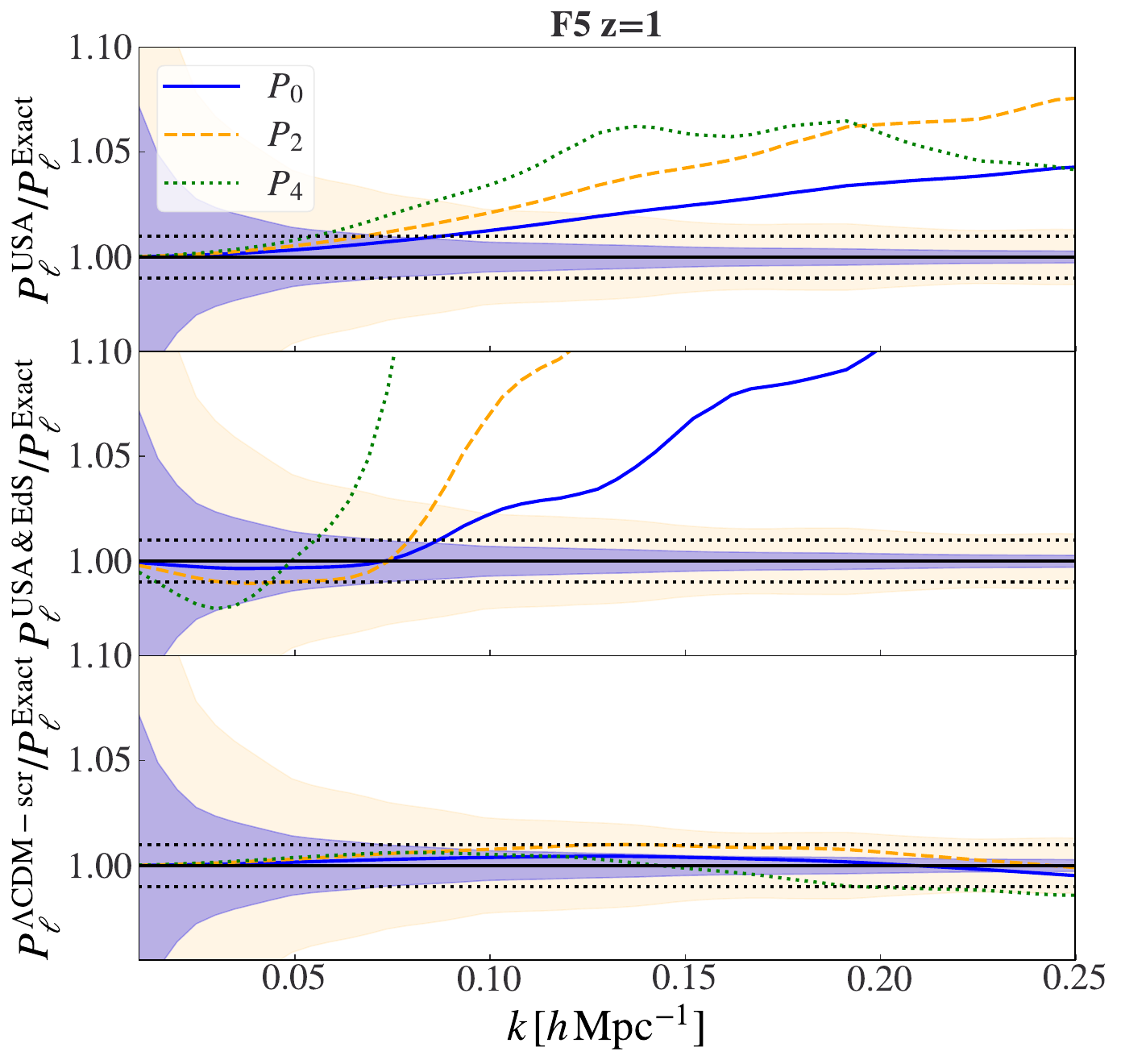}
    \includegraphics[width=0.49\textwidth]{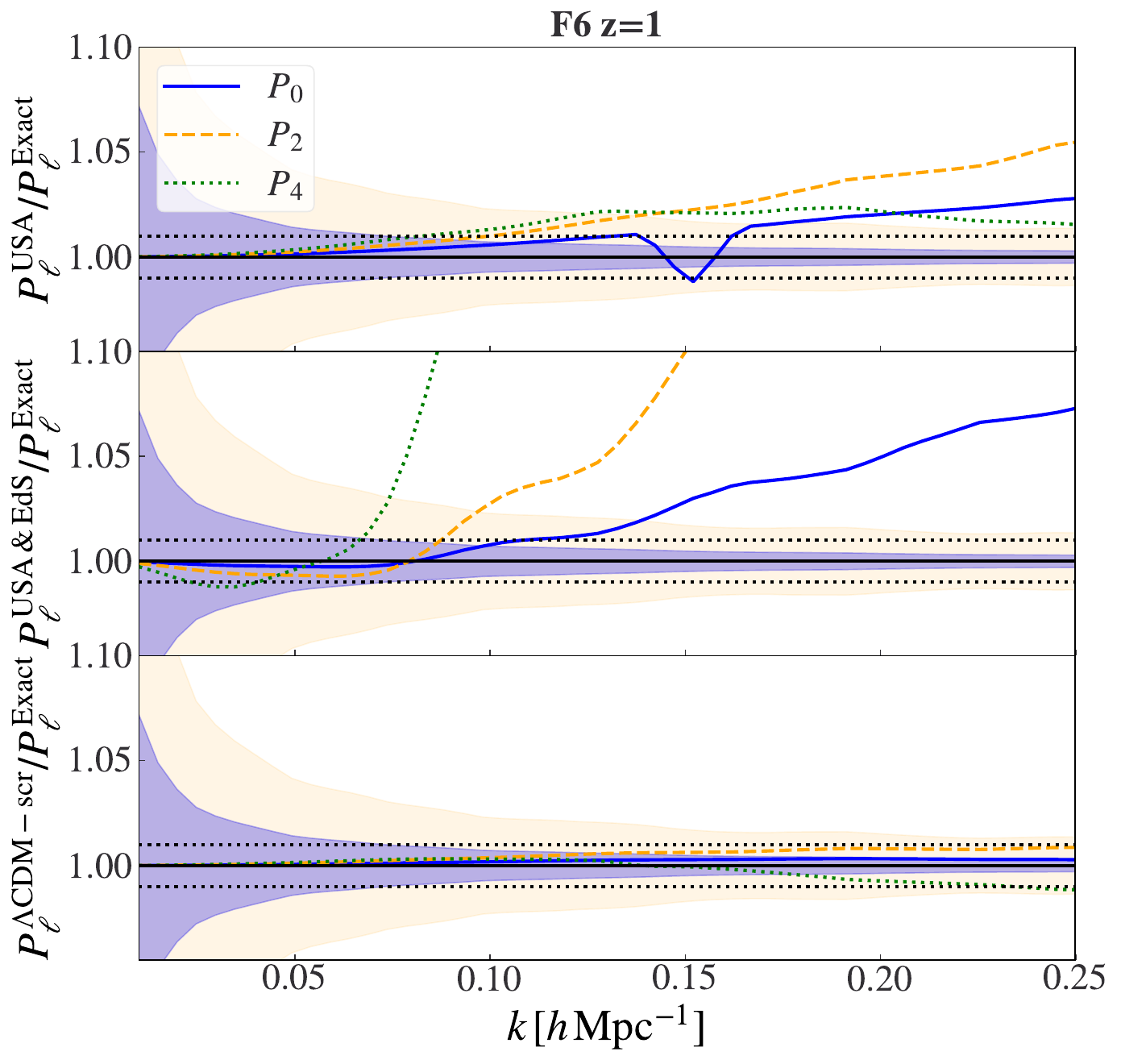} 
    \caption{Ratio of the three approximations considered in this paper to the exact calculation of the redshift space matter power spectrum multipoles, computed with the TNS prescription, for two $f(R)$ models: $|f_{\rm R0}| = 10^{-5}$ and $\sigma_{\rm v}=5.8\, h^{-1}\, {\rm Mpc} $ (left panels), and  $|f_{\rm R0}| = 10^{-6}$ and $\sigma_{\rm v}=5.6\, h^{-1}\, {\rm Mpc} $ (right panels), at redshift $z=1$. The top, middle and bottom panels show the USA, the USA and EdS, and the $\Lambda$CDM-scr approximations, respectively. Blue and beige bands indicate errors on the monopole and quadrupole assuming a $\Lambda$CDM Gaussian covariance with $V = 8.8 \, h^{-3} \, {\rm Gpc}^3$ and no shot noise contribution. We note that the hexadecapole error fills the plot and so we have omitted it. }
    \label{fig:elfrapp}
\end{figure*}


\subsection{Dark matter \texorpdfstring{$\chi^2$}{chi2} tests}\label{sec:rspdm}

This analysis validated the necessary approximations currently needed by the FFT-based fast codes described in \cref{tab:codes}. In particular, we tested the EdS approximation for Dark Scattering, the EdS and USA for DGP and $f(R)$, and the $\Lambda$CDM-scr approximation for $f(R)$. We restricted all these tests to the TNS model of RSD as only this implementation in \texttt{MG-Copter} uses the exact kernel calculations. If the approximation does sufficiently well using the TNS, we expect that it should do equally well or better in the EFTofLSS case which makes use of more RSD fitting parameters. Further, our ultimate goal is to test these approximations at the level of CDM halos, which we do in the next section. As we are only considering CDM, to compute the minimum $\chi^2$ as a function of $k_{\rm max}$, we only vary the TNS RSD nuisance parameter, $\sigma_{\rm v}$, in order to minimise the $\chi^2$ statistic given in \cref{covarianceeqn}. 

\begin{figure*}
    \centering
    \includegraphics[width=\textwidth]{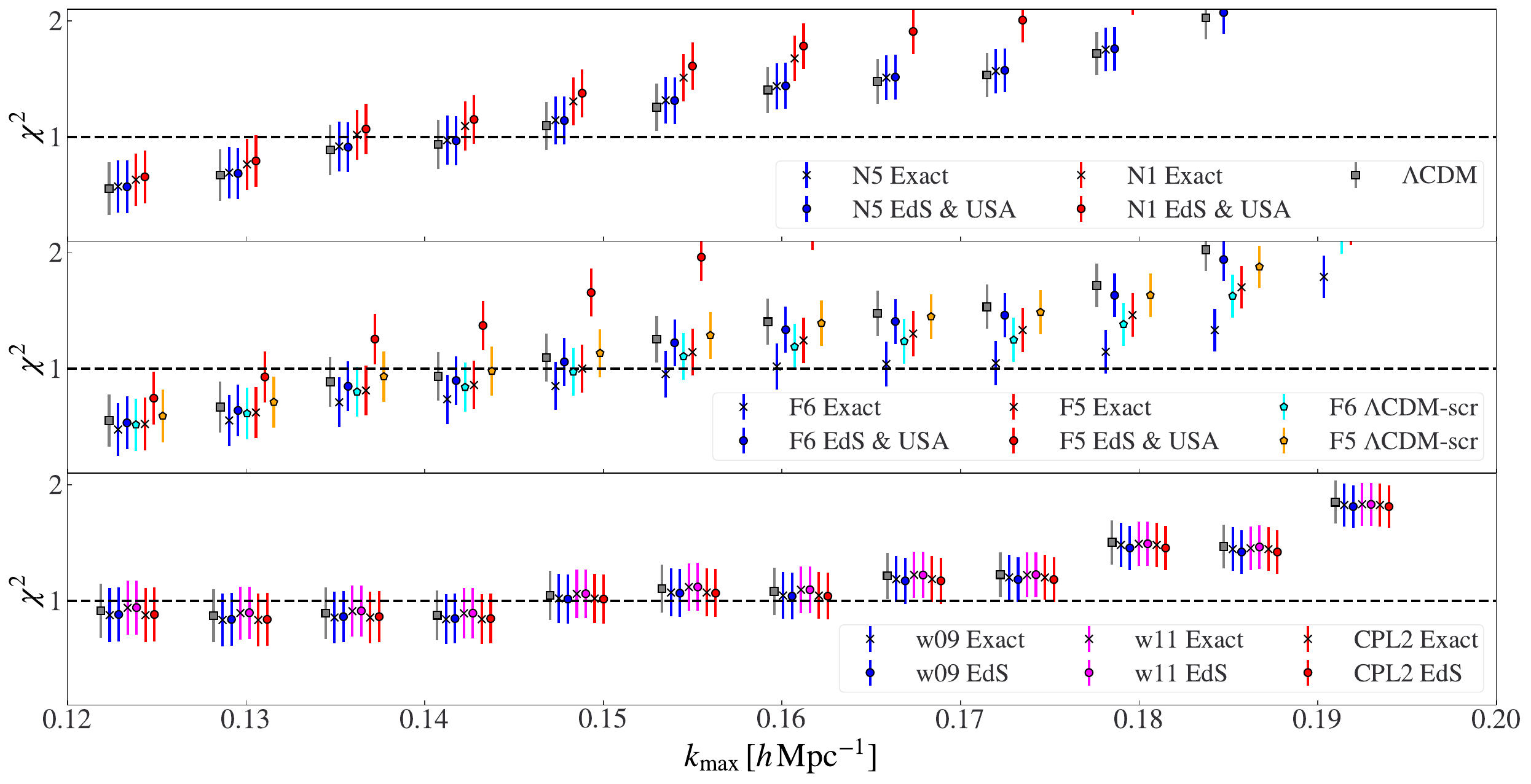} 
    \caption{Reduced $\chi^2$ for the fit of the redshift space matter power spectrum multipoles, computed with the TNS prescription, to the one from numerical simulations, as a function of $k_{\rm max}$ and at redshift $z=1$. In the top panel we consider the DGP \texttt{ELEPHANT} simulations, in the middle panel the $f(R)$ \texttt{ELEPHANT} simulations, and in the  bottom panel the Dark Scattering \texttt{DAKAR} simulations. Crosses indicate the exact computation while circles indicate the EdS and USA approximations. For the $f(R)$ models, the $\Lambda$CDM-scr approximation is shown as pentagons. The $\Lambda$CDM case is shown as grey squares. The error bars are the $1\sigma$ errors on the  $\chi^2$ statistic with $N_{\rm dof}=2 N_k -1$ degrees of freedom, where $N_k$ are the number of wave modes used in calculating $\chi^2$. We subtract one because the TNS model has one degree of freedom.} 
    \label{fig:dm_chi}
\end{figure*}

In \cref{fig:dm_chi} we show the results for all cases. The top panel show the \texttt{ELEPHANT} simulation DGP model results.  In this case, the low modification (N5, exact) does equally well as the $\Lambda$CDM case. The high modification (N1, exact) does slightly worse, which is expected as nonlinearities are amplified by the significantly enhanced growth, causing SPT to break down at smaller $k$. In both cases, the EdS combined with the USA approximation does very well, following the goodness of fit of the exact prediction.

In the middle panel, we show the \texttt{ELEPHANT} simulation $f(R)$ results. Here the high modification case (F5, exact) does equally well as $\Lambda$CDM, while we see a much better fit in the low modification case (F6, exact). This is in part due to the degeneracy between damping effects induced by modified gravity and the TNS damping parameter $\sigma_{\rm v}$. The added damping of the quadrupole due to $f(R)$ increases the effectiveness of $\sigma_{\rm v}$ in damping both multipoles appropriately. This only works to an extent as illustrated by the F5 case.  As the damping from $f(R)$ is also redshift dependent, we do not expect this result to hold for all redshifts.

Reassuringly, the $\Lambda$CDM-scr approximation does only slightly worse than the exact computation. On the other hand, the EdS combined with the USA approximation does significantly worse than the exact computation in the F5 case. This was expected given \cref{fig:elfrapp}. Despite this poor performance, it is yet to be seen if this approximation is sufficient (or necessary) at the level of CDM halos, when we introduce a number of bias parameters. 
 
Finally, the bottom panel shows the Dark Scattering simulation results. In this case, we find the EdS approximation is a good approximation in all cases, even in the high interaction case (CPL2), with all cases following the $\Lambda$CDM goodness of fit very closely. This was expected given the top right panel of \cref{fig:dsapp}. Based on this, we do not consider Dark Scattering in the next section. 

Finally, we comment on the particularly small $\chi^2$ in the \texttt{ELEPHANT} simulations at small $k_{\rm max}$. This is likely due to the very small scatter in these measurements as they are the averages of five realisations, yielding small $\chi^2$ despite the smaller error expressed in the covariance. In the absence of this scatter at linear scales, the nonlinear models applied here, with their additional degree of freedom, will over-fit the data as seen. Moving to nonlinear scales, the size of the errors (larger in \texttt{DAKAR}), begins to play a bigger role as the measurements from both simulations are very low in scatter. This is also seen to some extent in the next section.


\subsection{Halo \texorpdfstring{$\chi^2$}{chi2} tests}\label{sec:haloresults}

Having validated the various approximations in the TNS case at the level of CDM, we move to compare the goodness of fit in the beyond-$\Lambda$CDM models to the $\Lambda$CDM case at the level of CDM halos. These are biased tracers of the CDM distribution and so serve as a proxy for galaxies, and a means to test the modelling prescriptions in the presence of bias degrees of freedom. 

In the TNS case, we primarily adopted the model used in the flagship BOSS survey analysis \citep{Beutler:2016arn}, which has two bias degrees of freedom $b_1$ and $b_2$, as well as a shot noise parameter $N$.  We also considered a Q-bias model independently for the $f(R)$ cases (see \cref{eq:qbias}) which has three degrees of freedom, on top of which we also include the shot-noise parameter $N$. These are in addition to the TNS RSD parameter, $\sigma_{\rm v}$. 

In the  EFTofLSS analyses for biased tracers we adopted a model with three perturbative biases, $b_1, b_2 $ and $b_{\mathcal{G}_2}$, one shot-noise parameter, $N$, and two counterterms, $\tilde{c}_0$ and $\tilde{c}_2$. Adding more shot-noise or counterterm parameters gradually improves the reached $k_{\rm max}$ but, at the same time, could erase the information about the specific model analysed, since many of these EFTofLSS parameters may be degenerate with the beyond-$\Lambda$CDM ones. This will be investigated in an upcoming \Euclid paper (D'Amico et al., in prep.). Our main goal here is to compare the $\Lambda$CDM to beyond-$\Lambda$CDM scenarios, which should not depend on the inclusion of the additional shot-noise parameters. Since in EFTofLSS analyses these parameters are fixed and/or marginalised analytically, we fixed all the higher-order nuisance parameters to zero.

We summarise the best-fit results for all models with Eulerian bias in \cref{tab:chi2halos}. The TNS model with the Q-bias  prescription fits are shown in \cref{tab:qbias}.


\subsubsection{\texorpdfstring{$f(R)$}{fR} and DGP}

\begin{figure*}
    \centering
    \includegraphics[width=\textwidth]{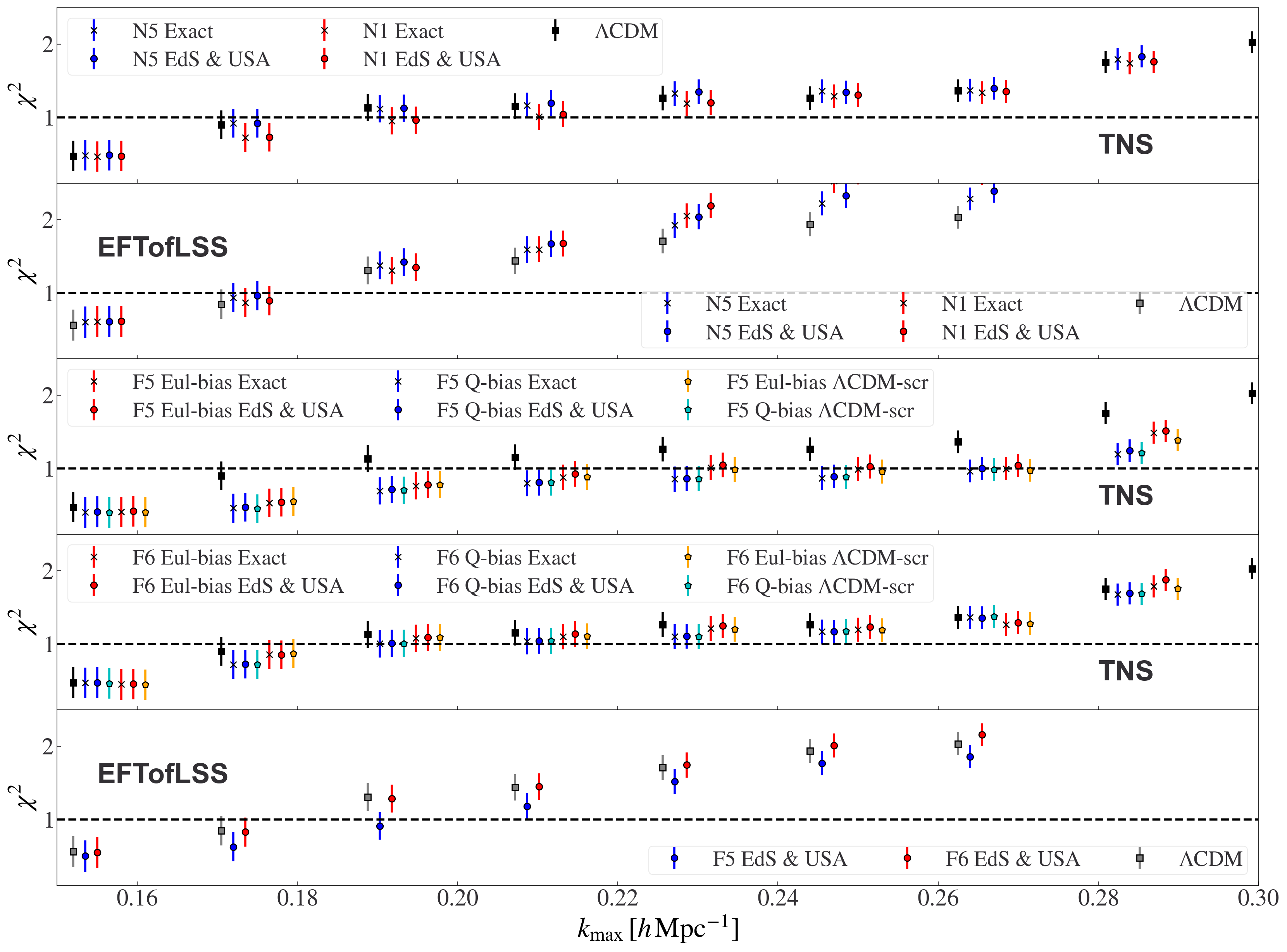}
    \caption{Reduced $\chi^2$ for the fit of the redshift space halo power spectrum multipoles, computed with the TNS  and the EFTofLSS prescriptions, to the measured multipoles from the DGP and $f(R)$ \texttt{ELEPHANT} simulations,  as a function of $k_{\rm max}$ and at redshift $z=1$. From top to bottom, TNS fit to DGP, EFTofLSS fit to DGP, TNS fit to $f(R)$ with $|f_{\rm R0}| = 10^{-5}$, TNS fit to $f(R)$ with $|f_{\rm R0}| = 10^{-6}$, and EFTofLSS fit to $f(R)$. Crosses, circles and pentagons  indicate the exact computation,  the EdS and USA approximation and the $\Lambda$CDM-scr approximation, respectively. The $\Lambda$CDM cases are shown as black (TNS) and grey (EFTofLSS) squares. The error bars are the $1\sigma$ errors on the $\chi^2$ statistic with $N_{\rm dof}=2 N_k - N_{\rm x}$ degrees of freedom, where $N_k$ are the number of wave modes used in calculating $\chi^2$. We use $N_{\rm x} = 4$ for the TNS model using Eulerian bias and $N_{\rm x} = 5$ for EFTofLSS and the TNS model with Q-bias. All EFTofLSS exact and $f(R)$ calculations are performed using \texttt{PyBird}, while the DGP EdS and USA calculations are performed using \texttt{PBJ}. The $\Lambda$CDM predictions are made using the exact kernel calculations. } 
    \label{fig:EL_halo_chi_tns}
\end{figure*}

We present the dependence on the $\chi^2$ as a function of $k_{\rm max}$ for the \texttt{ELEPHANT} simulations in \cref{fig:EL_halo_chi_tns}. Here we show both EFTofLSS and TNS model predictions for DGP (top two panels) and $f(R)$ (bottom three panels). We find that the Eulerian bias expansion, keeping scale-independent bias coefficients, is effective for the DGP case. Both levels of modification follow the $\Lambda$CDM trend very closely, with the high modification doing slightly better at the mid-range of scales presented. Further, the EdS and USA are very good approximations, following the exact computation almost exactly as was seen in \cref{fig:dm_chi}. This holds for both TNS and EFTofLSS cases. 

In the $f(R)$ case, we tested two different bias models and both EdS and USA as well as the $\Lambda$CDM-scr approximations. The bias models we examine are the Q-bias and Eulerian bias (see \cref{sec:nonlinear-bias}). For the Eulerian bias, we assume EdS kernels for the higher-order bias terms and scale-independent bias coefficients. Interestingly, when the bias degrees of freedom were introduced, the EdS and USA approximations seem to become much more applicable, with the inaccuracy it incurs likely being absorbed by the bias degrees of freedom. We also note that there is little difference between the TNS Eulerian bias prescription and the Q-bias model, despite the Q-bias having one additional parameter. 

It is also interesting to note that all $f(R)$ cases do better in terms of their fit to the simulations (a lower $\chi^2$) than the $\Lambda$CDM fits. This could be because of an increased efficiency of the bias and RSD degrees of freedom to damp both multipoles appropriately in the presence of $f(R)$ effects, as was noted in the CDM case. We also see this enhanced fit in the EFTofLSS case which applies the EdS and USA approximation. \Refr{This improved efficiency of the RSD models to capture the full shape of the power spectrum when $f(R)$ effects are present does not necessarily mean we will gain more information as it also suggests there are no distinct $f(R)$ features at these scales, but rather just an overall damping of the spectrum, highly degenerate with $\sigma_v$ or counterterms. 

This being said, we do note that an improved fit to the full shape does not mean there are no additional features in the multipoles coming from $f(R)$, specifically ones that are not pure damping in nature. These effects may become relevant with smaller error bars. Moreover, both the damping and these possible additional features are crucially dependent on the modification strength, which we can see clearly in \cref{fig:EL_halo_chi_tns} with the F6 case fitting the data marginally better than the $\Lambda$CDM case, while the F5 case fitting the multipole measurements significantly better.}  

An important point here is that we have fit the simulation data using a linear covariance that assumes the simulation volume, which is much less than that which \Euclid will be probing. We do however omit shot noise which would improve the fits at small scales. We have found that including shot noise contributions, based on number densities from the halo catalogue measurements, enables the TNS model to fit the data to well beyond $k>0.5 \, h \, {\rm Mpc}^{-1}$. This is likely due to the model simply fitting the shot noise which it can do efficiently. 

\begin{figure}
    \centering
    \includegraphics[width=0.49\textwidth]{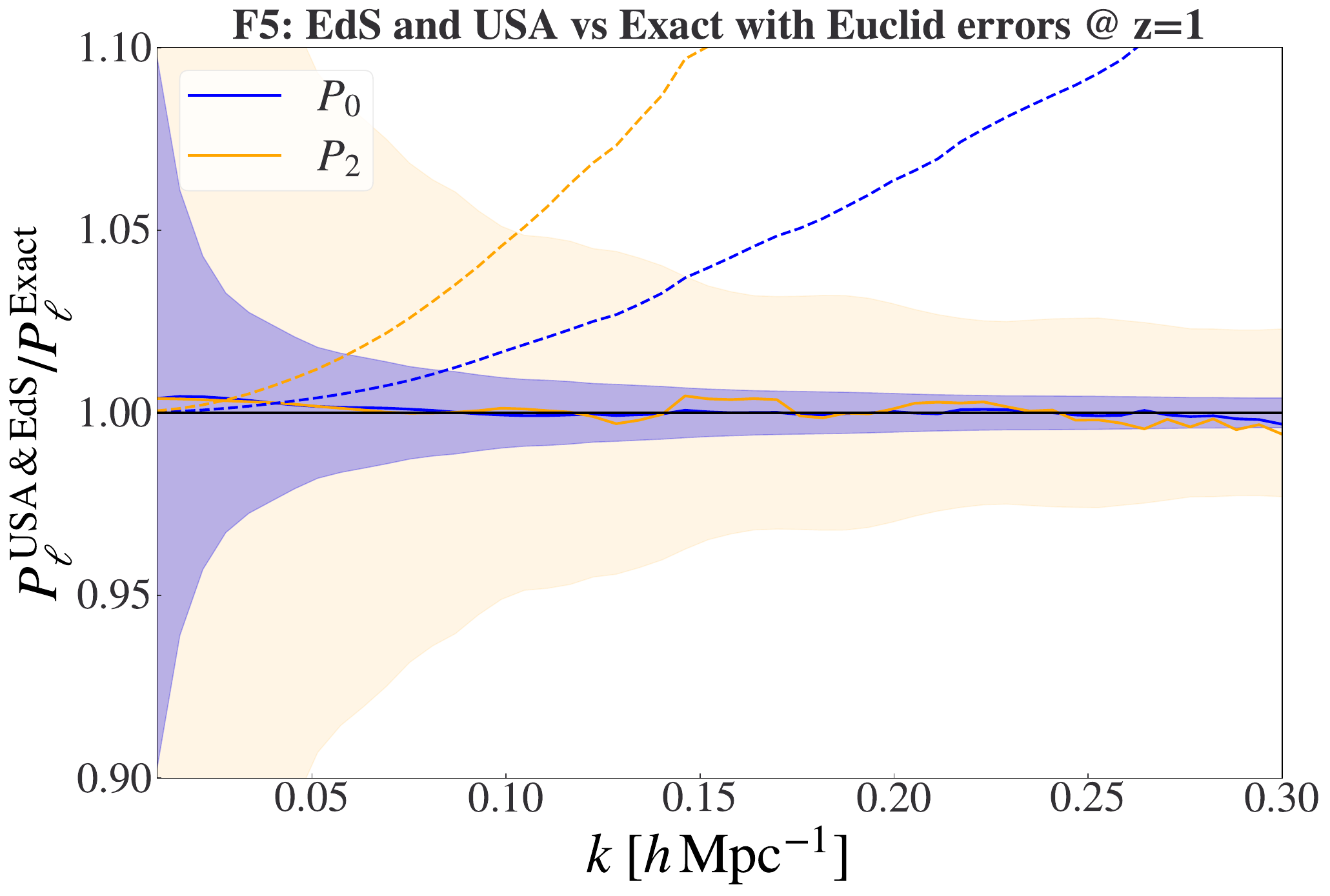}
    \caption{\Refr{Ratio of the USA and EdS approximation to the exact calculation of the redshift space matter power spectrum multipoles, computed with the TNS prescription, for the  $|f_{\rm R0}| = 10^{-5}$ (F5) model at redshift $z=1$. The models assume the fits as found in \cref{tab:chi2halos}. The dotted lines show a USA and EdS prediction where the nuisance parameters are not refit to the simulation data, and take the same values as the exact computation. Blue and beige bands indicate errors on the monopole and quadrupole assuming a \Euclid-like Gaussian covariance with $V = 8.8 \, h^{-3} \, {\rm Gpc}^3$ and $\bar{n} = 6.86 \times 10^{-4} \, h^3 \, {\rm Mpc}^{-3}$.}} 
    \label{fig:eucliderrf5}
\end{figure}

\Refr{To check whether or not the approximations are good enough for \Euclid, we checked the multipoles using the best-fit parameters determined from the simulations. In particular, we investigated the case which is expected to perform worst: the USA combined with EdS for the F5 case, which we saw in \cref{fig:elfrapp} performed terribly for dark matter. In \cref{fig:eucliderrf5} we show the ratio of the monopole and quadrupole for the TNS EdS and USA approximation for F5 to the exact solution with the \texttt{ELEPHANT}-based simulation best-fit parameters found in \cref{tab:chi2halos}. We also overlay as bands the expected cosmic variance and shot noise errors expected from the \Euclid survey at the $z=1$ redshift bin \citep{Euclid:2019clj}. As dotted lines we show the EdS and USA approximation prediction but without refitting the nuisance parameters and using the same fits as the exact solution. This highlights how the nuisance parameters  are effectively accounting for the approximation, even to within \Euclid precision. Without this freedom the approximation fails, even before the shot noise begins to dominate. } 

\begin{figure*}
    \centering
   \includegraphics[width=\textwidth]{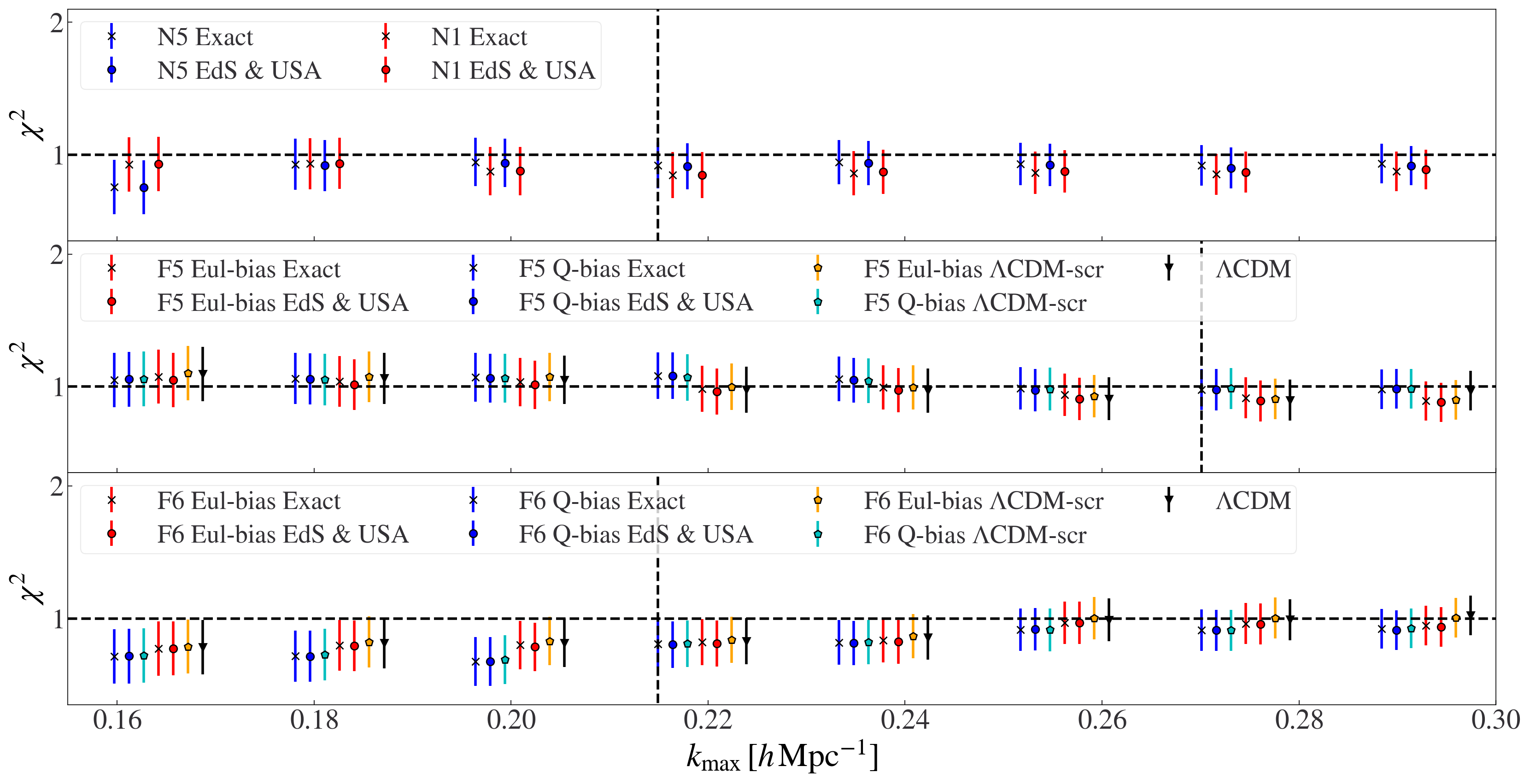}
    \caption{Same as the TNS panels of \cref{fig:EL_halo_chi_tns} but fitting to a fiducial SPT-mock data vector using the exact prediction for the SPT kernels and the Eulerian bias (red crosses, and blue crosses for the N5 model). The model parameters are those found in \cref{tab:chi2halos}. The mock data vector is given scatter using a \Euclid-like Gaussian covariance, using $V = 8.8 \, h^{-3} \, {\rm Gpc}^3 $ and $\bar{n} = 6.86 \times 10^{-4} \, h^3 \, {\rm Mpc}^{-3}$ and the same linear bias as the mock data. The fits are also made using the same covariance. We delimit the $k_{\rm max}$ which was used to choose the best-fit bias and RSD nuisance parameters when comparing to the \texttt{ELEPHANT} simulations, roughly corresponding to the values found in \cref{tab:chi2halos}. The black triangles indicate a modelling where no $f(R)$ effects are accounted for, i.e. a pure $\Lambda$CDM modelling.} 
    \label{fig:EL_SPT_halo_chi}
\end{figure*}

To further investigate whether or not the approximations applied here would be valid in a \Euclid-like setting, we prepared a mock data vector using the TNS model and exact SPT kernel predictions,\footnote{We used the values of the nuisance parameters found at the $k_{\rm max}$ given in \cref{tab:chi2halos}.} and attached to it a covariance computed with \Euclid-like shot noise and volume ($V = 8.8 \, h^{-3} \, {\rm Gpc}^3 $ and $\bar{n} = 6.86 \times 10^{-4} \, h^3 \, {\rm Mpc}^{-3}$, see \cref{sec:setup}). We further added scatter to the mock data vector using this covariance. Then, we fitted the theoretical prescriptions using approximations to this data vector and checked the goodness of fit as a function of $k_{\rm max}$. The fits are shown in \cref{fig:EL_SPT_halo_chi}. All approximations do as well as the exact calculation (which the mock data vector was produced with), and give a $\chi^2 \approx 1$ for a large range of $k_{\rm max}$, exceeding that determined by fitting the simulations. 

We also fitted this mock data vector in the $f(R)$ cases using a pure $\Lambda$CDM modelling ($\mu(k,a)=1$ and $S(\bfk)=0$), which also shows an excellent fit to the mock data. This suggests that the nuisance parameters are completely degenerate with $f(R)$ effects, at least for $|f_{\rm R0}|\leq 10^{-5}$ and over the range of scales that our models are valid within, even in a \Euclid-like setup. 

One way of checking this hypothesis is to observe the behaviour of the nuisance parameters as a function of $k_{\rm max}$. We find, in the TNS case, that the best-fit linear bias $b_1$ and $\sigma_{\rm v}$ are $\sim 5\%$ and $\sim 10\%$ consistent between all approximations and the exact results over the full range of $k_{\rm max}$, with $b_1$ also being consistent with the simulation measurement. Interestingly, the best-fit values of $\sigma_{\rm v}$ in the $\Lambda$CDM-scr approximation match extremely well with the exact solution, which confirms it being a very good approximation for the SPT kernels. Differing values of $\sigma_{\rm v}$ for the other approximations and Q-bias model indicate that the effects of $f(R)$ are being absorbed by a change in this parameter. Similarly, $b_2$ and $N$ also vary significantly, both as a function of $k_{\rm max}$ and between the various approximations. Such degeneracies will likely be able to be broken by using a combination of redshift bins, giving far more information about the evolution of structure, as well as the bispectrum for example, which will yield more information on bias parameters, particularly if the modelling prescription for both statistics is consistent \citep{Hashimoto:2017klo,Philcox:2022frc,Tsedrik:2022cri,Ivanov:2021kcd}. We leave further investigation of this to future work which should employ large-volume simulations.  


\subsubsection{\texorpdfstring{$w_0w_a$}{w0wa}CDM and massive neutrinos}

\begin{figure*}
    \centering
    \includegraphics[width=\textwidth]{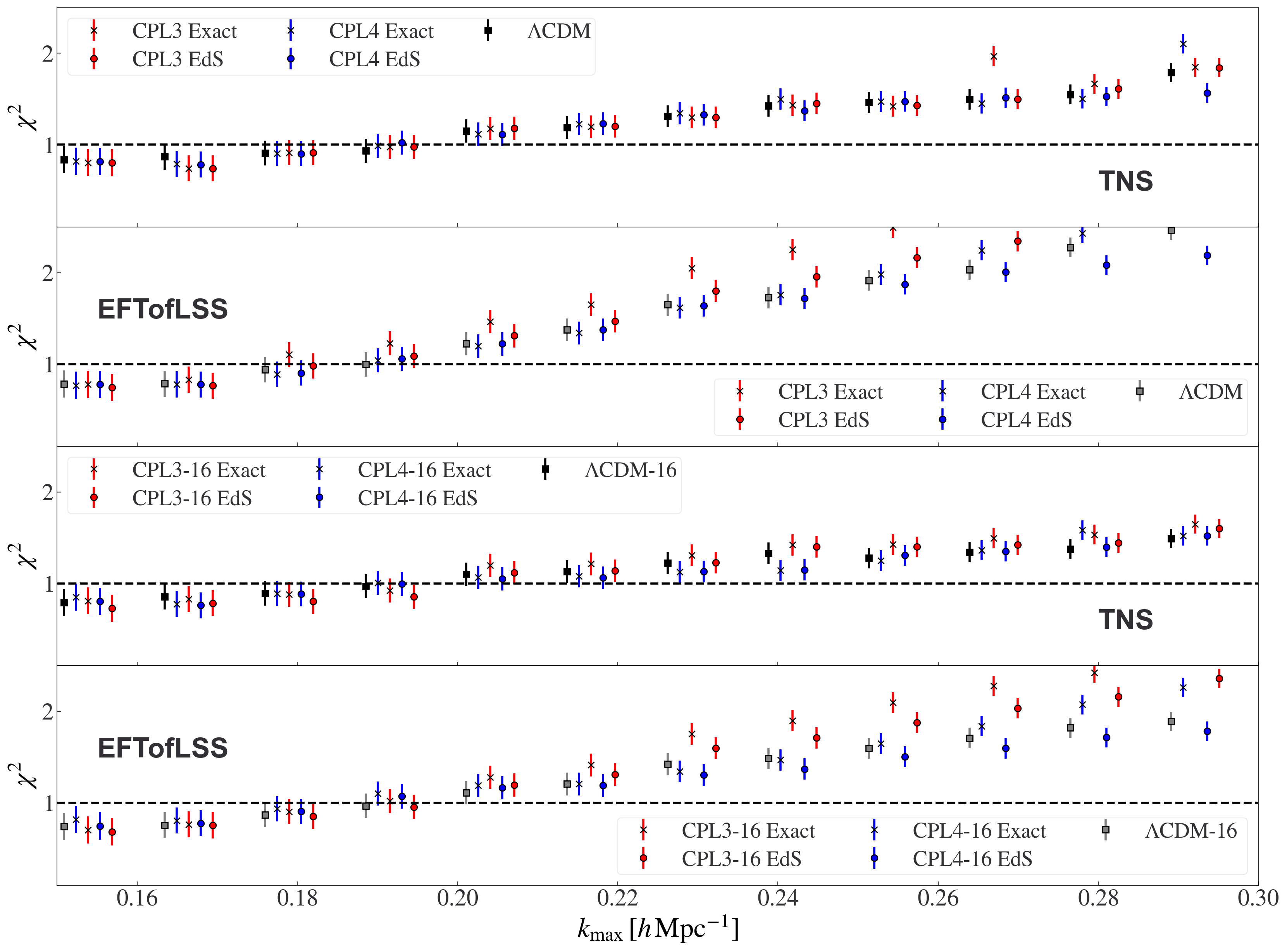} 
    \caption{Same as\cref{fig:EL_halo_chi_tns} but for the \texttt{DEMNUni} simulation models. We show the cases without massive neutrinos (top two panels: TNS upper, EFTofLSS lower) and with massive neutrinos (bottom two panels: TNS upper, EFTofLSS lower).} 
    \label{fig:DEM_halo_chi}
\end{figure*}

In \cref{fig:DEM_halo_chi} we show the $\chi^2$ as a function of $k_{\rm max}$ for the \texttt{DEMNUni} simulations (see \cref{tab:sims}). The TNS and EFTofLSS model predictions for the massless neutrino cases are shown in the top two panels, while the bottom two panels show the massive neutrino cases ($\sum m_\nu = 0.16~{\rm eV}$). We find that the EdS approximation does equally well to the exact kernel calculations for all cases, at least until the $\chi^2 \approx 1$. Further, all beyond-$\Lambda$CDM cases follow very similar trends to the $\Lambda$CDM case. This suggests that the EdS approximation is sufficiently accurate for $w_0w_a$CDM cases, and our linear-only treatment of massive neutrinos (see \cref{sec:massive-neutrinos}) is equally good in all cases. 

As with the DGP and $f(R)$ cases, we find that the TNS model shows slightly lower $\chi^2$ over the full range of $k_{\rm max}$ considered than EFTofLSS, but both RSD models have roughly a $k_{\rm max} (\chi^2 \approx 1) \approx 0.19 \, h \, {\rm Mpc}^{-1}$. This is similar to the $k_{\rm max}$ for most of the \texttt{ELEPHANT} simulations (except F5), despite the \texttt{DEMNUni} simulations having a larger volume. This is likely compensated by the slightly lower $\sigma_8$ of these simulations, and hence slightly lower levels of nonlinearity, leading to a better performance of SPT. 

\begin{figure*}
    \centering
    \includegraphics[width=\textwidth]{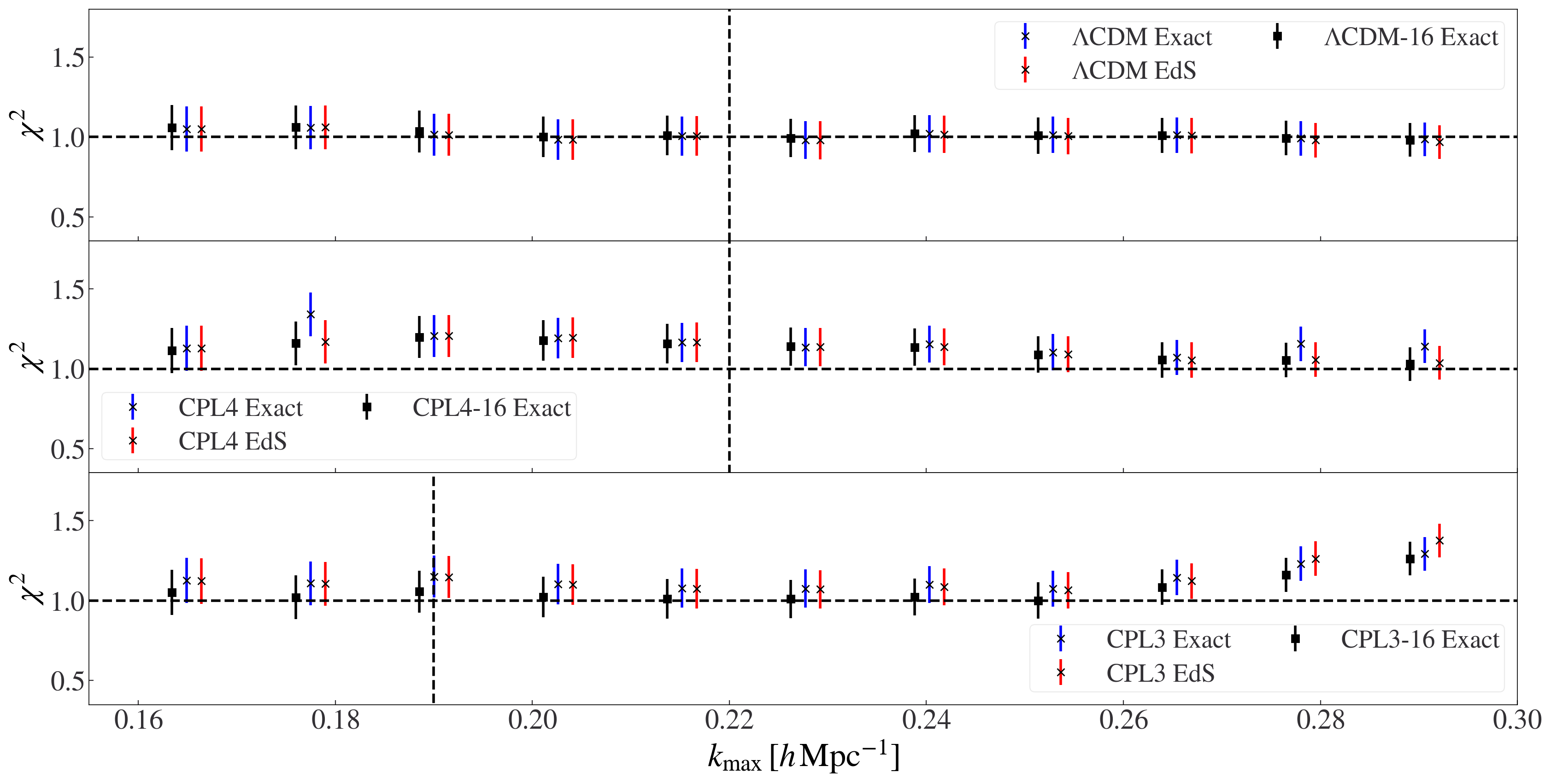}
    \caption{Same as \cref{fig:EL_SPT_halo_chi} but for the $w_0w_a$CDM \texttt{DEMNUni} cosmologies (see \cref{tab:sims}). The mock data vector in this case includes massive neutrinos ($\sum m_\nu = 0.16~{\rm eV}$), and the black squares use the same modelling as the mock data. The crosses assume $\sum m_\nu = 0.0~{\rm eV}$ in the modelling. } 
    \label{fig:DEM_SPT_halo_chi}
\end{figure*}

The volume of the \texttt{DEMNUni} simulations is comparable to the volume \Euclid will probe at $z=1$ and so we did not perform any additional tests of the EdS approximation against SPT mock data. On the other hand, we did test the importance of including massive neutrinos. We created a mock data vector using the exact calculation of the SPT kernels and including massive neutrino effects assuming $\sum m_\nu = 0.16 \, {\rm eV}$ in the linear  $P_{\rm cb}$ power spectrum (see \cref{sec:massive-neutrinos}), again using the best-fit nuisance parameters found in \cref{tab:chi2halos}. We then created a covariance with the \Euclid-like volume and tracer number density ($V = 8.8 \,  h^{-3} \, {\rm Gpc}^3 $ and $\bar{n} = 6.86 \times 10^{-4} \,  h^3 \, {\rm Mpc}^{-3}$) and a linear bias as given in \cref{tab:chi2halos}. Using this covariance, we added scatter to the mock data. 
We then fitted this mock data vector using the massless neutrino modelling. The fits of the various theoretical predictions to the mock data vectors are shown in \cref{fig:DEM_SPT_halo_chi}. 

Surprisingly, we find all massless neutrino fits, both in $\Lambda$CDM and $w_0w_a$CDM, follow the massive neutrino modelling fits almost perfectly, with $\chi^2 \approx 1$ at all $k_{\rm max}$ considered. This further suggests that the effects of massive neutrinos with masses $\sum m_\nu \leq 0.16~{\rm eV}$ are degenerate with nuisance parameters, at least at a fixed redshift bin and only with the clustering power spectrum. This \Refr{is} expected as the FoG damping effect, controlled by the free parameter $\sigma_{\rm v}$ in the TNS model or counterterms in EFTofLSS, are likely highly degenerate with the effects of massive neutrinos, which act to damp power on small scales.

We await further tests to be performed in D'Amico et al., in prep. to further explore these results. Unlike screening, the inclusion of massive neutrinos at the linear level is not significantly computationally expensive and so no approximation needs to be made in principle. 
\begin{table*}
\caption{\label{t7} Summary of TNS and EFTofLSS fits to the simulations.}
\centering
\begin{tabular}{| c | c | c | c | c ||  c | c | c | }
\hline  
   & & \multicolumn{3}{ | c || }{{\bf TNS}} &    \multicolumn{3}{  c | }{{\bf EFTofLSS}}   \\ \hline \hline 
 {\bf Model } &  {\bf Approx.} & $\mathbf{k_{\rm max}}$  & $ \mathbf{\chi^2}$ & $\{ b_1, b_2, N, \mathbf{\sigma_{\rm v}} \}$ &   $\mathbf{k_{\rm max}}$  & $ \mathbf{\chi^2}$ & $\{ b_1, b_2, b_{\mathcal{G}_2}, N, \tilde{c}_0, \tilde{c}_2  \}$ \\ \hline 
 \multirow{2}{*}{{\bf $\Lambda$CDM}} & Exact & 0.21 & 1.14 & $\{ 2.22 , 0.94 , -627 , 4.34 \} $ & 
 0.18 & 1.04 & $ \{ 2.20 , 0.11 , 0.05 , 733 , 5.39 , 33.6 \} $ \\ 
 &EdS &  0.21 & 1.16 & $\{ 2.22 , 0.77 , -535 , 4.32 \} $
& 0.18 & 1.08 & $\{ 2.21 , 0.12 , 0.03 , 654 , 4.98 , 39.92 \} $ \\ \hline 
 \multirow{2}{*}{{\bf N5}} & Exact &  0.19 & 1.12 & $\{ 2.17 , -0.32 , 306 , 4.20 \} $ &  0.18 & 1.09 & $ \{ 2.19 , -0.02 , 0.01 , 400 , 1.47 , 34.11 \} $ \\ 
& EdS and USA & 0.19 & 1.12 & $\{ 2.17 , -0.39 , 365 , 4.18 \} $ & 0.18 & 1.13 & $\{ 2.19 , -0.02 , -0.01 , 309 , 0.92 , 40.57 \} $
\\ \hline 
    \multirow{2}{*}{{\bf N1}} & Exact &  0.22 & 1.17 & $\{ 2.12 , 0.56 , -553 , 4.84 \} $ & 0.18 & 0.95 & $ \{ 2.12 , -0.10 , -0.03 ,  0, -2.21 , 38.16 \} $ \\ 
& EdS and USA & 0.22 & 1.19 & $\{ 2.12 , 0.39 , -443 , 4.84 \} $ & 0.18 & 0.98 & $\{ 2.12 , -0.01 , -0.05 , -297 , -4.38 , 46.82 \} $ \\ \hline     
    \multirow{3}{*}{{\bf F6}} & Exact &  0.21 & 1.12 & $\{ 2.20 , 1.00 , -637 , 4.27 \} $ & - & -  & -  \\ 
    & EdS and USA & 0.21 & 1.15 & $\{ 2.21 , 0.96 , -692 , 4.49 \} $  & 0.18 & 1.00 & $ \{ 2.19 , -0.04 , 0.04 , 0.16 , -0.64 , -5.45 \} $  \\ 
   & $\Lambda$CDM-scr &  0.21 & 1.12 & $\{ 2.20 , 0.83 , -566 , 4.33 \} $ & - & - & - \\ \hline 
    \multirow{3}{*}{{\bf F5}} & Exact &  0.26 & 0.99 & $\{ 2.11 , 1.07 , -769 , 4.62 \} $ & - & - & - \\ 
   & EdS and USA & 0.26 & 1.04 & $\{ 2.12 , 1.18 , -1030 , 5.21 \} $ &  0.21 & 1.18 & $ \{ 2.07 , 0.37 , -0.14 , 0.17 , -4.21 , -7.54 \} $
 \\ 
   & $\Lambda$CDM-scr &  0.26 & 0.97 & $\{ 2.10 , 0.90 , -672 , 4.62 \} $ & - & - & - \\ 
   \hline \hline
 \multirow{2}{*}{{\bf $\Lambda$CDM}} & Exact &  0.20 & 1.12 & $\{ 2.12 , -1.38 , 1118 , 3.12 \} $ & 0.19 & 0.98 & $\{ 2.17 , -0.09 , -0.41 , 1666 , -4.85 , 32.91 \} $ \\ 
   & EdS &   0.20 & 1.12 & $\{ 2.12 , -1.36 , 1108 , 3.12 \} $ & 0.19 & 1.02 & $\{ 2.17 , -0.07 , -0.43 , 1592 , -4.87 , 39.18 \} $
\\ \hline
 \multirow{2}{*}{{\bf CPL3}} & Exact & 0.19 & 0.97 & $\{ 2.08 , -0.92 , 747 , 3.53 \} $ & 0.19 & 1.02 & $\{ 2.16 , -0.93 , -0.82 , 2200, -0.65 , 13.17 \} $
 \\ 
& EdS & 0.19 & 0.97 & $\{ 2.08 , -0.93 , 751 , 3.53 \} $ & 0.19 & 1.07 & $\{ 2.11 , -0.22 , -0.46 , 1403 , -5.86 , 41.34 \} $
 \\ \hline
 \multirow{2}{*}{{\bf CPL4}} & Exact &  0.20 & 1.11 & $\{ 2.31 , 2.96 , -603 , 3.09 \} $ & 0.19 & 1.02 & $\{ 2.16 , -0.32 , -0.66 , 1262, -4.53 , 16.55 \} $ \\ 
& EdS & 0.20 & 1.11 & $\{ 2.31 , 3.00 , -597 , 3.06 \} $ & 0.19 & 1.04 & $\{ 2.31 , 0.20 , -0.54 , 1720 , -2.99 , 39.77 \} $ \\ \hline 
 \multirow{2}{*}{{\bf $\Lambda$CDM-16}} & Exact &  0.22 & 1.10 & $\{ 2.21 , -1.95 , 1326 , 3.00 \} $ & 0.21 & 1.11 & $\{ 2.23 , 0.98 , -0.46 , 1433 , -2.65 , 50.17 \} $
\\ 
& EdS & 0.22 & 1.10 & $\{ 2.21 , -1.95 , 1326 , 3.00 \} $ & 0.20 & 1.09 & $\{ 2.23 , 0.58 , -0.44 , 1578 , -3.03 , 51.02 \} $
\\ \hline 
 \multirow{2}{*}{{\bf CPL3-16}} & Exact &  0.19 & 0.97 & $\{ 2.20 , -0.53 , 522 , 3.45 \} $ & 0.19 & 1.00 & $\{ 2.22 , -1.83 , -0.55 , 3182 , -14.14 , -2.22 \} $ \\ 
& EdS & 0.19 & 0.85 & $\{ 2.23 , 1.75 , -560 , 3.61 \} $ & 0.19 & 0.94 & $\{ 2.18 , 0.16 , -0.35 , 1366 , -10.98 , 45.87 \} $
 \\ \hline 
 \multirow{2}{*}{{\bf CPL4-16}} & Exact &  0.22 & 1.07 & $\{ 2.37 , -1.94 , 1622 , 2.42 \} $ & 0.19 & 1.10 & $\{ 2.24 , -0.31 , -0.49 , 1239 , -15.73 , 13.38 \} $ \\ 
& EdS & 0.22 & 1.10 & $\{ 2.39 , 3.82 , -146 , 2.61 \} $ & 0.19 & 1.06 & $\{ 2.38 , 0.01 , -0.34 , 2283 , -7.25 , 35.05 \} $
\\ \hline
\end{tabular}
\tablefoot{We show the approximations employed, the highest $k_{\rm max}$ with its associated value of the reduced $\chi^2$ and best-fitting model parameter. The upper table shows the \texttt{ELEPHANT} fits, the lower table the \texttt{DEMNUni} fits. We note that $b_2$ and $N$ are normalised differently in the different models, so they are not expected to have similar values. $b_1$, $b_2$ and $b_{\mathcal{G}_2}$ are dimensionless. $N$ has dimensions $h^{-3} \, {\rm Mpc}^3$. $\sigma_v$ and $\{ \tilde{c}_0, \tilde{c}_2\}$ have dimensions $h^{-1} \, {\rm Mpc}$ and $h^{-2} \, {\rm Mpc}^2$ respectively.} 
\label{tab:chi2halos} 
\end{table*}
\begin{table*}
\caption{\label{t7} Q-bias TNS fits for $f(R)$ {\tt ELEPHANT} models.}
\centering
\begin{tabular}{| c | c | c | c | c | }
\hline  
 {\bf Model } &  {\bf Approximation} & $\mathbf{k_{\rm max}}$  & $ \mathbf{\chi^2}$ & $\{ b_0, A_1, A_2, N, \mathbf{\sigma_{\rm v}} \}$  \\ \hline 
    \multirow{3}{*}{{\bf F6}} & Exact &  0.23 & 1.10 & $\{2.17 , 0.02 , -2.00 , 912 , 3.48\}$    \\ 
   & EdS and USA &  0.23 & 1.10 & $\{2.18 , 0.04 , -2.00 , 886 , 3.71\}$  \\ 
   & $\Lambda$CDM-scr &  0.23 & 1.10 & $\{2.18 , 0.06 , -2.00 , 940 , 3.52\}$  \\ \hline
    \multirow{3}{*}{{\bf F5}} & Exact &  0.26 & 0.96 & $\{2.07 , -0.12 , -1.40 , 322 , 4.10\}$    \\ 
   & EdS and USA &  0.26 & 1.00 & $\{2.08 , -0.11 , -1.15 , 137 , 4.74\}$ \\ 
   & $\Lambda$CDM-scr &  0.26 & 0.98 & $\{2.07 , -0.13 , -1.04 , 196 , 4.26\}$  \\ 
  \hline  
\end{tabular}
\tablefoot{$A_1$ and $A_2$ have units $h^{-1} \, {\rm Mpc}$ and $h^{-2} \, {\rm Mpc}^2$ respectively. $N$ has dimensions $h^{-3} \, {\rm Mpc}^3$ and $\sigma_v$ has units $h^{-1} \, {\rm Mpc}$.}
\label{tab:qbias} 
\end{table*}

\section{Conclusions} \label{sec:conclusions}

In this paper we have investigated various theoretical approximations in beyond-$\Lambda$CDM scenarios, necessary for the computational demands of forthcoming \Euclid galaxy clustering analyses, as well as different RSD models. Before listing our results, as a reference for the reader, we provide a summary of previous related works and their main conclusions.
\\
\\
\subsection{Previous work} 

The Einstein--de Sitter (EdS) approximation has been investigated in a number of works. \cite{Bose:2018orj} compare the EdS approximation to simulations and the exact 1-loop real space matter power spectrum in DGP using the EFTofLSS. They find this is a very good approximation, especially given the additional degrees of freedom of the EFTofLSS. \cite{Bose:2017myh} perform Markov Chain Monte Carlo analyses on DGP simulations using the EdS TNS redshift space dark matter multipoles with and without screening effects, finding a $2\sigma$ bias is incurred on the growth rate, $f$, when screening is omitted at $z=1$ for a survey of $V_{\rm s}= 20~h^{-3} {\rm Gpc}^3$, but not for $V_{\rm s}= 10~h^{-3} {\rm Gpc}^3$. Further, \cite{Carrilho:2021hly} check the EdS approximation for the Dark Scattering real space spectra against the exact solutions and find sub-percent agreement for $z=1$, for scales less than $k\approx 0.25~h \, {\rm Mpc}^{-1}$ and for $\xi$ comparable to the values considered in this paper. 

Regarding tracer bias schemes, \cite{DAmico:2021rdb} show that the scale-independent bias parameters are likely a good approximation for models of gravity inducing a scale-independent growth factor and rate of structure formation, without explicitly comparing to simulations. A number of other works have developed exact calculations for redshift space correlations for dark matter and biased tracers, but have not explicitly investigated the need for this by comparing, directly or through best-fit analyses, to simulations or mock data \citep{Aviles:2017aor,Aviles:2020wme,Valogiannis:2019nfz,Valogiannis:2019xed,Bose:2016qun,Bose:2017dtl,Bose:2017jjx}, specifically when higher order nuisance parameters  are included. We do however note that \cite{Aviles:2020wme} find the local Lagrangian approximation to be valid for $f(R)$ gravity.

\subsection{This work}

In this work we have considered the TNS and EFTofLSS models for RSD with an Eulerian bias expansion. We further considered the local Lagrangian bias relation, known to hold well for $\Lambda$CDM and \Refr{$f(R)$ \citep{Aviles:2020wme}}. For the scale-dependent theory, $f(R)$, we also investigate the phenomenological Q-bias  model (see ~\cref{eq:qbias}). 

These approximations were checked by comparing the monopole and quadrupole of both dark matter and halo clustering to high quality numerical simulations as well as standard perturbation theory (SPT) based mock data generated with a minimal amount of approximations. For all comparisons, we have fixed cosmology and only varied the model nuisance parameters related to RSD, nonlinear clustering and tracer bias. Our main conclusions are summarised below.

\subsubsection{Dark Matter}

The EdS approximation without screening predictions gives an equally good $\chi^2$ fit to the simulation dark matter multipoles as the predictions without these approximations in DGP. The EdS approximation is generally excellent for DGP, $w_0w_a$CDM, and Dark Scattering models, both at the level of dark matter and halos. It does however give a significantly worse fit for $f(R)$. On the other hand, the $\Lambda$CDM-screened approximation (see \cref{app:lcdmscr}) for $f(R)$ gravity, developed in this work, gives an equally good fit to the simulation dark matter multipoles as the exact SPT predictions. This approximation is suitable for a fast Fourier transform implementation.
 
\subsubsection{Halos}
The EdS approximation without screening predictions gives an equally good $\chi^2$ fit to the simulation halo multipoles as the predictions without these approximations, both in $f(R)$ gravity and DGP. Further, both Eulerian and Q-bias prescriptions give equally good fits to the simulation data, with the local Lagrangian bias relation seeming to be valid for all beyond-$\Lambda$CDM scenarios, with the $\chi^2$ of these scenarios being comparable to that of $\Lambda$CDM. 

We found that not modelling any $f(R)$ effects, both in the bias terms and in the perturbation theory kernels, gives an equally good fit to the SPT-$f(R)$ mock data as the exact kernel computation. This suggests such effects can efficiently be absorbed into higher-order bias and RSD nuisance parameters. In fact, we have noted a significant change in the RSD and higher order bias parameters from the fiducial mock values in these fits. Particularly prominent shifts were noted for the largest approximations, for example using the Einstein--de Sitter and unscreened approximations for $f(R)$ gravity. These shifts also grew with the maximum scale, $k_{\rm max}$, included in the fit.

Similarly, not modelling any massive neutrino effects gives an equally good fit to the SPT-$\Lambda$CDM and -$w_0w_a$CDM mock data as the modelling including massive neutrinos, for $\sum m_\nu \leq 0.16~{\rm eV}$. This suggests such effects can also be efficiently absorbed into higher-order bias and RSD nuisance parameters. We note that changes in these parameters away from the fiducial mock values was also noted in the fits, in support of this claim.

Lastly, TNS and EFTofLSS both give a similar value of $k_{\rm max}$ such that $|\chi^2(k_{\rm max})-1| \lesssim 0.15$ when fit to the simulations, with the TNS yielding a slightly larger $k_{\rm max}$ systematically over all beyond-$\Lambda$CDM scenarios. \footnote{Note that we have not included the EFTofLSS $\sim \mu^2 k^2$ shot-noise term, which improves the fit.} This range is roughly $0.19 \, h \, {\rm Mpc}^{-1} \leq k_{\rm max} \leq 0.21\, h \, {\rm Mpc}^{-1}$, with the exception being the strong $f(R)$ modification where we find the TNS model can fit the data well up to $k_{\rm max} \approx 0.26\, h \, {\rm Mpc}^{-1}$. This is likely due to an enhanced efficiency of the phenomenological fingers-of-god RSD damping parameter $\sigma_{\rm v}$ when strong $f(R)$ effects are at play. 
\\
\\

We note that we do not vary cosmological parameters, and so these results, in particular the use of the various approximations considered for the perturbative kernels and bias, does not ensure an unbiased estimation of cosmology. They do however strongly suggest large degeneracies between screening and massive neutrino effects and higher-order bias as well as RSD effects. If validated, this would mean that a power spectrum monopole and quadrupole analysis at a single redshift will offer no significant information on modified gravity. This result was known in the case of massive neutrinos due to their degeneracies with bias parameters \citep[see][for example]{Marulli:2011he,Villaescusa-Navarro:2017mfx,Hahn:2019zob}, but not in the case of modified gravity, which is a key finding of this work.

This result may change when including higher-order multipoles or statistics \citep{Garcia-Farieta:2021hda}, such as the hexadecapole or bispectrum, galaxy lensing, or informative priors on nuisance parameters, which will help to break clear degeneracies between beyond-$\Lambda$CDM effects and bias parameters \citep{Chan:2016ehg,Bose:2018zpk,Bose:2019psj,Markovic:2019sva,Bose:2019ywu,Tsedrik:2022cri,Kacprzak:2022oit}. Such informative priors must be chosen carefully though, as they can lead to biases on the inferred cosmology or gravitational model \citep{Carrilho:2022mon, Simon:2022lde}. If such priors can however be identified, or if we can identify a more efficient nuisance parameter set, through for example principal component analyses \citep[see for example][]{Eifler:2014iva} or machine learning methods \citep{Piras:2023lid}, then the theoretical approximations used, particularly in the case of $f(R)$ or scale-dependent models, must be revised. In this case a more accurate prescription, such as the $\Lambda$CDM-screened approximation, can be applied. Conversely, we can use the methods presented here to gain information on the appropriate width of nuisance parameter priors. For example, we find that $\sigma_v$ changes by up to $10\%$ when employing approximate perturbative kernels. So, any prior must be at least so wide about some central value if we choose to employ such an approximation in a real analysis.

In light of these findings, we do not advocate additional effort in either improving computational efficiency to calculate the exact beyond-$\Lambda$CDM SPT kernels or theoretically developing new, more accurate kernel approximations. We also deem the Eulerian bias expansion, with constant bias coefficients to be sufficient for the considered beyond-$\Lambda$CDM cases. The RSD models perform similarly and so we do not advocate one over the other. This being said, tools and theoretical prescriptions are already available which can compute the clustering multipoles highly accurately and efficiently for DGP, Dark Scattering, $w_0w_a$CDM, and massive neutrinos \citep{DAmico:2020kxu,Noriega:2022nhf,Piga:2022mge,Carrilho:2022mon}. A forthcoming \Euclid paper (D'Amico et al., in prep.) will provide a more detailed Markov Chain Monte Carlo-based analyses of both RSD models. 

\Refr{In summary, the main contribution of this work to \Euclid is to have identified useful approximations for determining scale cuts in beyond-$\Lambda$CDM scenarios, which will require both high-volume simulation measurements and computationally expensive Markov Chain Monte Carlo analyses. On this note, this work has instructed the minimal extension of current internal pipelines for beyond-$\Lambda$CDM theories, in particular for the TNS model. We have also developed a useful theoretical approximation ($\Lambda$CDM-screened) in the case where standard approximations in $f(R)$-gravity break under the combined power of the \Euclid spectroscopic probe. This approximation can be integrated easily into codes employing the Fast Fourier transform method such as \texttt{PyBird} and \texttt{PBJ}.}

Finally, we have also validated the clustering predictions from various SPT based codes being employed within the \Euclid consortium. In particular, we have found the predictions from \texttt{PBJ}, \texttt{Pybird} and \texttt{MG-Copter} (now part of \texttt{ReACT}) (see \cref{tab:codes}) to be in sub-percent agreement for $k\leq 0.5\, h \, {\rm Mpc}^{-1}$ (see \cref{app:validate}). 

\begin{acknowledgements}
\AckEC
BB was supported by a UK Research and Innovation Stephen Hawking Fellowship (EP/W005654/2).
AP is a UK Research and Innovation Future Leaders Fellow [grant MR/S016066/2]. 
CM and PC's research for this project was supported by a UK Research and Innovation Future Leaders Fellowship [grant MR/S016066/2].  CM's work is supported by the Fondazione ICSC, Spoke 3 Astrophysics and Cosmos Observations, National Recovery and Resilience Plan (Piano Nazionale di Ripresa e
Resilienza, PNRR) Project ID CN\_00000013 ``Italian Research Center on High-Performance Computing, Big Data and Quantum Computing'' funded by
MUR Missione 4 Componente 2 Investimento 1.4: Potenziamento strutture di ricerca e creazione di "campioni nazionali di R\&S (M4C2-19 )" - Next Generation EU (NGEU).
MP acknowledges support by the MIUR `Progetti di Ricerca di Rilevante Interesse Nazionale' (PRIN) Bando 2022 - grant 20228RMX4A.
FV acknowledges partial support by the ANR Project COLSS (ANR-21-CE31-0029).
NF is supported by the Italian Ministry of University and Research (MUR) through the Rita Levi Montalcini project \enquote{Tests of gravity on cosmic scales} with reference PGR19ILFGP and she also acknowledges the FCT project with ref.\ number PTDC/FIS-AST/0054/2021.
CC acknowledges a generous CPU and storage allocation by the Italian Super-Computing Resource Allocation (ISCRA) as well as from the coordination of the ``Accordo Quadro MoU per lo svolgimento di attività congiunta di ricerca Nuove frontiere in Astrofisica: HPC e Data Exploration di nuova generazione'', together with storage from INFN-CNAF and INAF-IA2.
FP acknowledges partial support from the INFN grant InDark, the Departments of Excellence grant L.232/2016 of the Italian Ministry of Education, University and Research (MUR) and from the European Union -- Next Generation EU. FP also acknowledges the FCT project with ref.\ number PTDC/FIS-AST/0054/2021.
We acknowledge the hospitality of the Institute for Fundamental Physics of the Universe (IFPU) of Trieste for the group meeting held there in December 2022.
For the purpose of open access, the author(s) has applied a Creative Commons Attribution (CC BY) licence to any Author Accepted Manuscript version arising.
\end{acknowledgements}

\bibliographystyle{aa}
\bibliography{biblio}


\begin{appendix}


\section{Redshift space correction terms}\label{app:rsdterms}

In this appendix we present the general forms for the $A$, $B$ and $C$ corrections terms in \cref{eq:rsdpofk} as well as their linear bias dependence. These can be matched to the EdS expressions in, for example, \cite{Scoccimarro:1999ed,Bernardeau:2001qr} by expanding \cref{eq:rsdpofk2}. We note that these are the forms that are also found in \texttt{MG-Copter}, which can be found \href{https://github.com/nebblu/ACTio-ReACTio/blob/master/reactions/src/SPT.cpp}{here}. Following \cite{Bose:2016qun} we have 
\begin{align}
A(k,\mu) & = -(k \, \mu) \int \mathrm{d}^3 q \,  \left[ \frac{\mu_q}{q}  B_\sigma (\bfq, \bfk-\bfq, -\bfk) \right. \nonumber \\ 
 & \quad  \left. + \frac{k \, \mu - q \, \mu_q}{|\bfk - \bfq|^2} B_\sigma (\bfk-\bfq,\bfq,  -\bfk) \right] \, , 
\end{align}
where $\mu_q = \hat{\bfq}\cdot\hat{\vec{z}}$ and the cross bispectrum is given in terms of the general kernels and linear bias as 
\begin{align}
& B_\sigma (\bfk_1, \bfk_2, \bfk_3)  = 2  \nonumber \\
& \times \left\{ \left[ b_1 F_1(k_2) - \mu_{k_2}^2 G_1(k_2)\right] \left[ b_1 F_1(k_3) - \mu_{k_3}^2  G_1(k_3)\right] G_2(\bfk_2,\bfk_3) \right. \nonumber \\ 
& \times P_{\rm 11,i}(k_2) \,  P_{\rm 11,i}(k_3)  \nonumber \\ 
& +  \left[ b_1 F_1(k_3) - \mu_{k_3}^2  G_1(k_3)\right] \left[ b_1 F_2(\bfk_1,\bfk_3) - \mu_{k_2}^2  G_2(\bfk_1,\bfk_3)\right] G_1(k_1) \nonumber \\ 
& \times P_{\rm 11,i}(k_1) \, P_{\rm 11,i}(k_3)  \nonumber \\ 
& +  \left[ b_1 F_1(k_2) - \mu_{k_2}^2  G_1(k_2)\right] \left[ b_1 F_2(\bfk_1,\bfk_2) - \mu_{k_3}^2  G_2(\bfk_1,\bfk_2)\right] G_1(k_1) \nonumber \\ 
& \times P_{\rm 11,i}(k_1) \,  P_{\rm 11,i}(k_2) \Bigr\} \, . 
\end{align}
The $B$ term is given as
\begin{equation}
    B(k,\mu) = (k \, \mu)^2 \int \mathrm{d}^3 q \, F(\bfq) \, F(\bfk-\bfq) \, , 
\end{equation}
with 
\begin{equation}
    F(\bfq) = \frac{\mu_q}{q} G_1(q) \left[ b_1 F_1(q)- \mu_q^2 G_1(q)\right]P_{\rm 11,i}(q) \, . 
\end{equation}
Finally, the $C$ term is given as 
\begin{align}
    C(k,\mu) & =   (k \, \mu)^2 \int \mathrm{d}^3 p \,  \mathrm{d}^3 q \, \delta_{\rm D}(\bfk-\bfp -\bfq) \, P_{\rm 11,i}(p) \, P_{\rm 11,i}(q) \nonumber \\ 
    & \quad  \times \frac{\mu_p^2}{p^2} G_1^2(p) \left[b_1 F_1(q) - \mu_q^2 G_1(q)\right]^2 \, .
\end{align}

\section{\texorpdfstring{$\Lambda$}{L}CDM-screened for \texorpdfstring{$f(R)$}{fR}} \label{app:lcdmscr}

In this approximation we have the following modifications to the Poisson equation 
\begin{align}
\mu(k,a) & =  1 + \left(\frac{k}{a}\right)^2\frac{1}{3\Pi(k,a)} \, ,\\
\gamma_2^{\rm app}(\bfk-\bfq,\bfq,a)  & =   \gamma_2(\bfk,0,a)  \,\frac{\Pi(0,a_{\rm f})}{\Pi(q,a_{\rm f})} \left\{ \frac{\Pi(0,a_{\rm f})}{\Pi(q,a_{\rm f})} \right. \nonumber \\ &  \quad \left. + \right.  \left. \frac{ \left[\Pi(k,a_{\rm f}) - \Pi(0,a_{\rm f}) \right]\, \left[\Pi(q,a_{\rm f}) - \Pi(0,a_{\rm f}) \right]}{\Pi^2(q,a_{\rm f})}  \right\} \, ,\nonumber\\
\gamma_2^{\rm app}(\bfk,-\bfq,a)  & =   \gamma_2(\bfk,0,a) \, \frac{\Pi(0,a_{\rm f})}{\Pi(q,a_{\rm f})}  \, ,\nonumber\\
\gamma_2^{\rm app}(\bfk,\bfq-\bfk,a)  & =  \frac{q^2}{k^2}\, \gamma_2^{\rm app}(\bfk-\bfq,\bfq,a)   \, ,
\end{align}
where $a_{\rm f}$ is the final scale factor at which the power spectrum is computed. For $\gamma_3$ we use the following expressions for their cyclic permutations
\begin{align}
 &   \gamma_3^{\rm app}(\bfq, \bfk, -\bfq ,a) = \gamma_3^{\rm app}(\bfq, -\bfq, \bfk ,a)  = \gamma_3(0,\bfk,0,a) \frac{\Pi^2(0,a_{\rm f})}{\Pi^2(q,a_{\rm f})} \, , \nonumber\\
 &  \gamma_3^{\rm app}(\bfk, \bfq, -\bfq ,a)  =\gamma_3(\bfk,0,0,a) \frac{\Pi^2(0,a_{\rm f})}{\Pi^2(q,a_{\rm f})} \,. 
\end{align}
The terms depending on the loop momentum $\bfq$ in these expressions are evaluated at a fixed value for the scale factor, $a=a_{\rm f}$, allowing a factorization of scale and time integrations. Moreover, the momentum dependence is only through powers of 
\begin{equation}
    a_{\rm f}^2 \,  \Pi^2(q,a_{\rm f}) = q^2 + m^2(a_{\rm f})\,,
\end{equation}
where, from \cref{eq:Xi},
\begin{equation}
    m^2(a_{\rm f})\coloneqq \frac{a_{\rm f}^2\,\Upsilon^3(a_{\rm f})}{2f_0 \, (3\Omega_{\rm m,0}-4)^2}\,,
\end{equation}
which are well suited for an extension of the FFT approach for the fast evaluation of loop integrals.

\section{Code and implementation validation}\label{app:validate}

We have performed some basic validation tests between the three perturbation theory codes used in this work. In particular, we have compared predictions for the 1-loop SPT dark matter monopole, quadrupole and hexadecapole (see \cref{eq:rsdpofk}) which involves no free nuisance parameters. We have done this for $\Lambda$CDM and DGP at $z=0.5$ where nonlinearities are larger compared to $z=1$, providing a better validation test. We have also compared the different resummation implementations described in \cref{sec:resummation}. Additionally, we note that the $\chi^2$ obtained from the different codes when comparing the 1-loop SPT prediction to the $\Lambda$CDM \texttt{ELEPHANT} simulations at $z=1$ are extremely similar over a large range of $k_{\rm max}$ (see \cref{covarianceeqn}).

\subsection{Code validation}
\begin{figure*}
    \centering
    \includegraphics[width=0.49\textwidth]{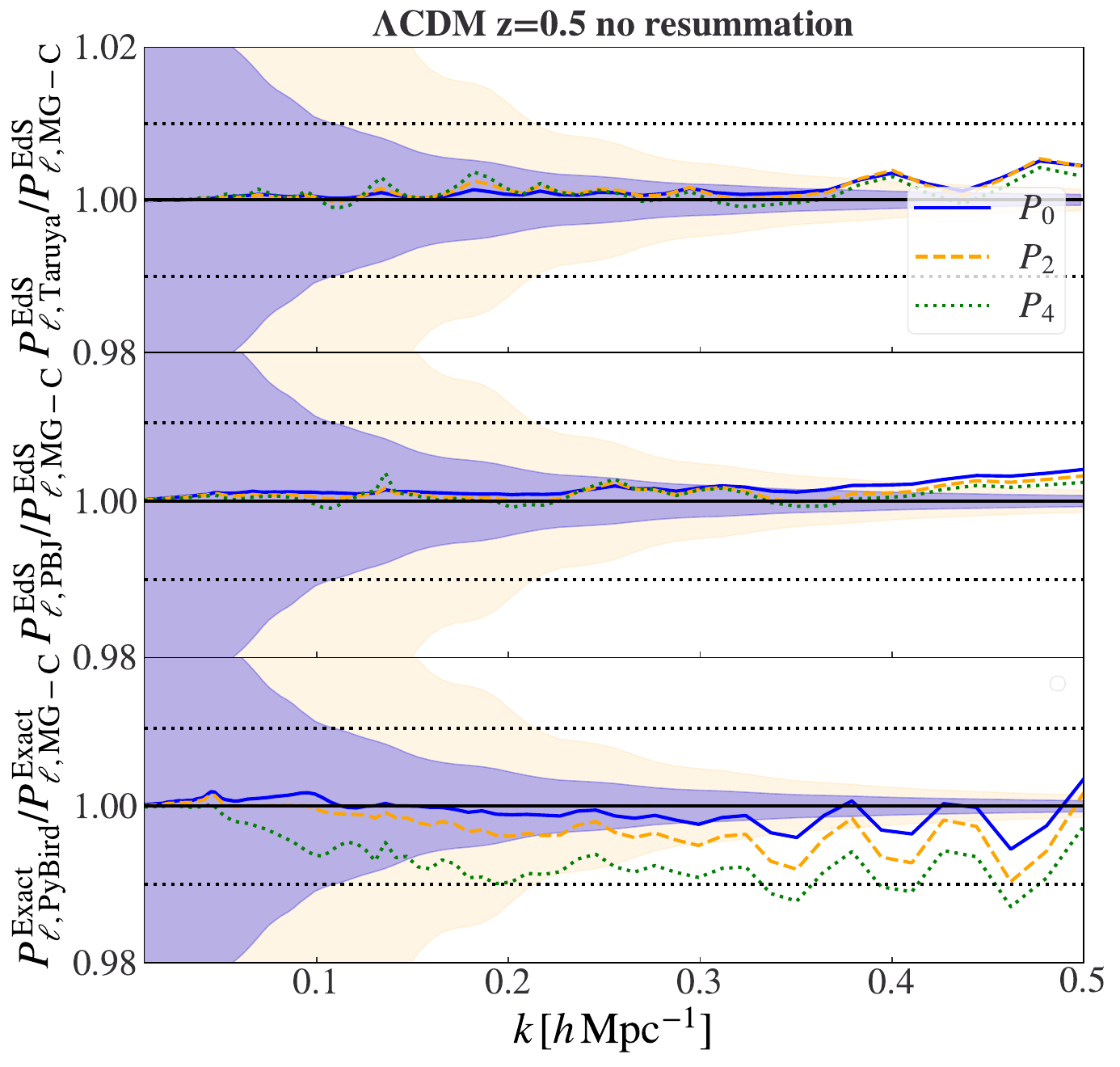}
    \includegraphics[width=0.49\textwidth]{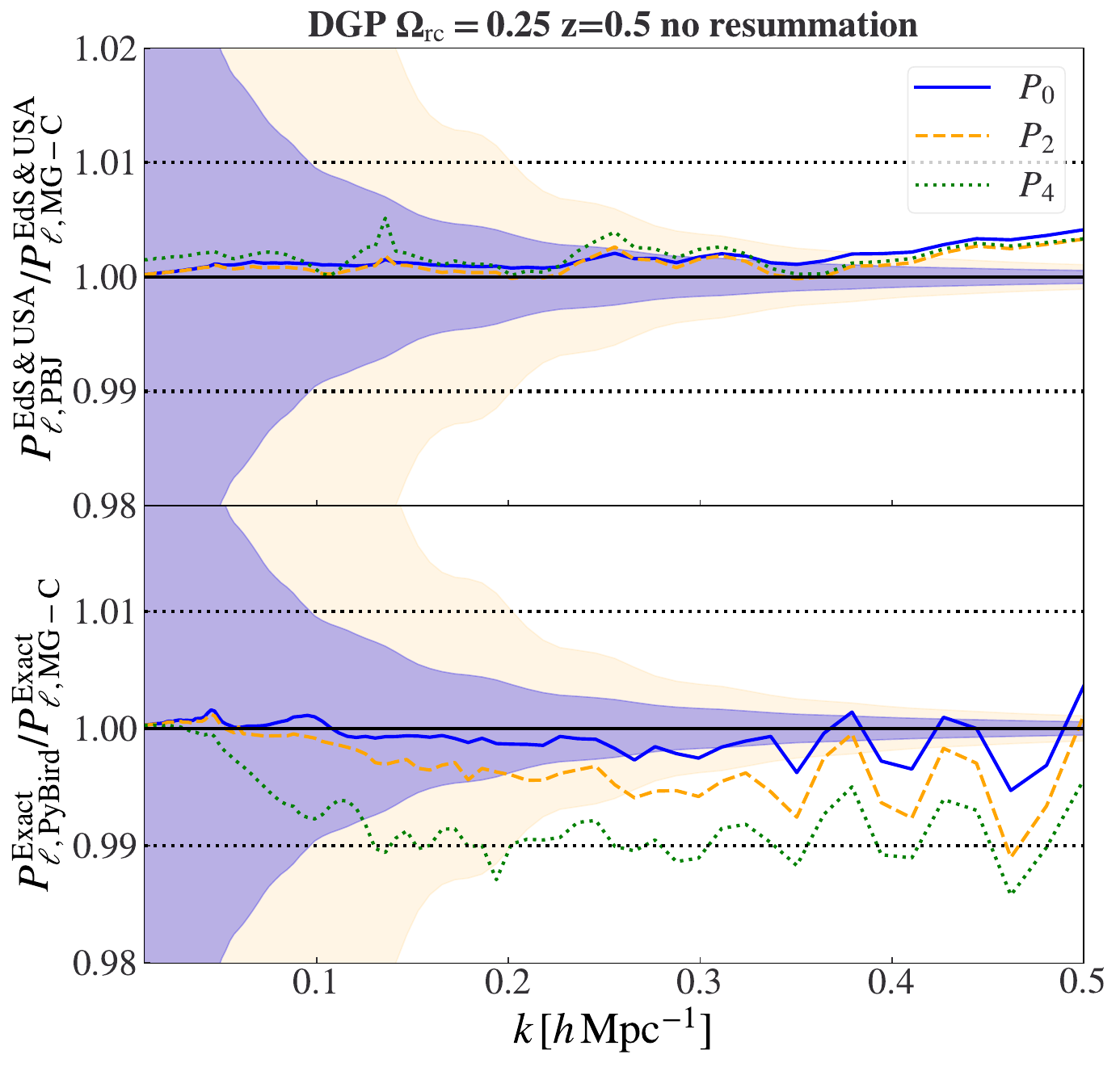}
    \caption{The ratio of the SPT 1-loop monopole (blue solid), quadrupole (orange dashed) and hexadecapole (green dotted) between different codes for $\Lambda$CDM (left) and DGP (right) at $z=0.5$ without infrared-resummation applied. The \texttt{MG-Copter} prediction is taken as the reference in all plots. The {top left} shows the prediction by Atsushi Taruya's code, the {middle left} and {top right} plots show the \texttt{PBJ} predictions and the {bottom left} and {bottom right} show the \texttt{Pybird} predictions. Atsushi Taruya's code and \texttt{PBJ} predictions are computed using the EdS approximation and without screening in the DGP case. We also adopt these approximations in their respective reference spectra from \texttt{MG-Copter}. \texttt{Pybird} predictions are exact and so are the respective \texttt{MG-Copter} reference spectra. The DGP parameter is set to $\Omega_{\rm rc} = 0.25$. Blue and beige bands indicate errors on the monopole and quadrupole assuming a $\Lambda$CDM Gaussian covariance with $V = 3.8 \, h^{-3} \, {\rm Gpc}^3$ and no shot noise contribution. We note that the hexadecapole error fills the plot and so we have omitted it.}
    \label{fig:validation1}
\end{figure*}

In \cref{fig:validation1} we show the ratio of the 1-loop SPT multipoles between the various codes for $\Lambda$CDM (left) and DGP (right) for $k\leq0.5\, h \, {\rm Mpc}^{-1}$ at $z=0.5$, without infrared-resummation applied.  For all cases we used the \texttt{MG-Copter} prediction as our reference spectrum. We have also included error bands from a $\Lambda$CDM Gaussian covariance assuming $V = 3.8 \, h^{-3} \, {\rm Gpc}^3$ at z=0.5, estimated using \cite{EUCLID:2011zbd}.\footnote{Note the difference in volume taken here and in the main text for a \Euclid-like setup comes from the different redshifts considered.} In the $\Lambda$CDM case, we find sub-0.5\% agreement between \texttt{MG-Copter}, \texttt{PBJ} and an additional code by Atsushi Taruya (AT).\footnote{Download this code: \url{http://www2.yukawa.kyoto-u.ac.jp/~atsushi.taruya/cpt_pack.html}.} We note that \texttt{PBJ} and AT's code both implement the expression for the 1-loop redshift space power spectrum as given by \cite{Matsubara:2007wj} while \texttt{MG-Copter} uses the expression given in \cref{eq:rsdpofk} \citep[see also Eq.~23 of][]{Taruya:2010mx}. These comparisons highlight the equivalence of these expressions. Further, the oscillations seen in the top left panel are caused by the interpolation over $k$. These three predictions have all made use of the Einstein--de Sitter (EdS) approximation.

In the bottom left panel we compare the exact kernel predictions between \texttt{MG-Copter} and \texttt{Pybird}. We find sub-1\% agreement in this case, with the monopole being sub-0.5\%. Interestingly, we observe almost the same discrepancy between these two codes when applying the EdS approximation. We have found that these predictions can be brought into better agreement by retuning the Fast Fourier Transform (FFT) bias parameter in \texttt{Pybird}. We note that the spike observed in the monopole is due to numerical instabilities in \texttt{MG-Copter}.

We observe similar agreement in the DGP scenario. In this case we compare \texttt{MG-copter} to \texttt{PBJ}, both using the EdS and unscreened approximations as this is the only current implementation in \texttt{PBJ}. Again we observe sub-0.5\% agreement. For the comparison with \texttt{Pybird} we use the exact time and screening predictions for both codes and observe sub-1\% agreement between the predictions, with some oscillations caused by interpolation. 

Finally, we find all comparisons are within the estimated measurement errors down to scales well beyond the expected validity regime of the predictions, which at this lower redshift will be lower than the rough estimates found in \cref{tab:chi2halos} ($k_{\rm max} \leq 0.2~h \, {\rm Mpc}^{-1}$). 

\subsection{Resummation}

\begin{figure*}
    \centering
\includegraphics[width=0.49\textwidth]{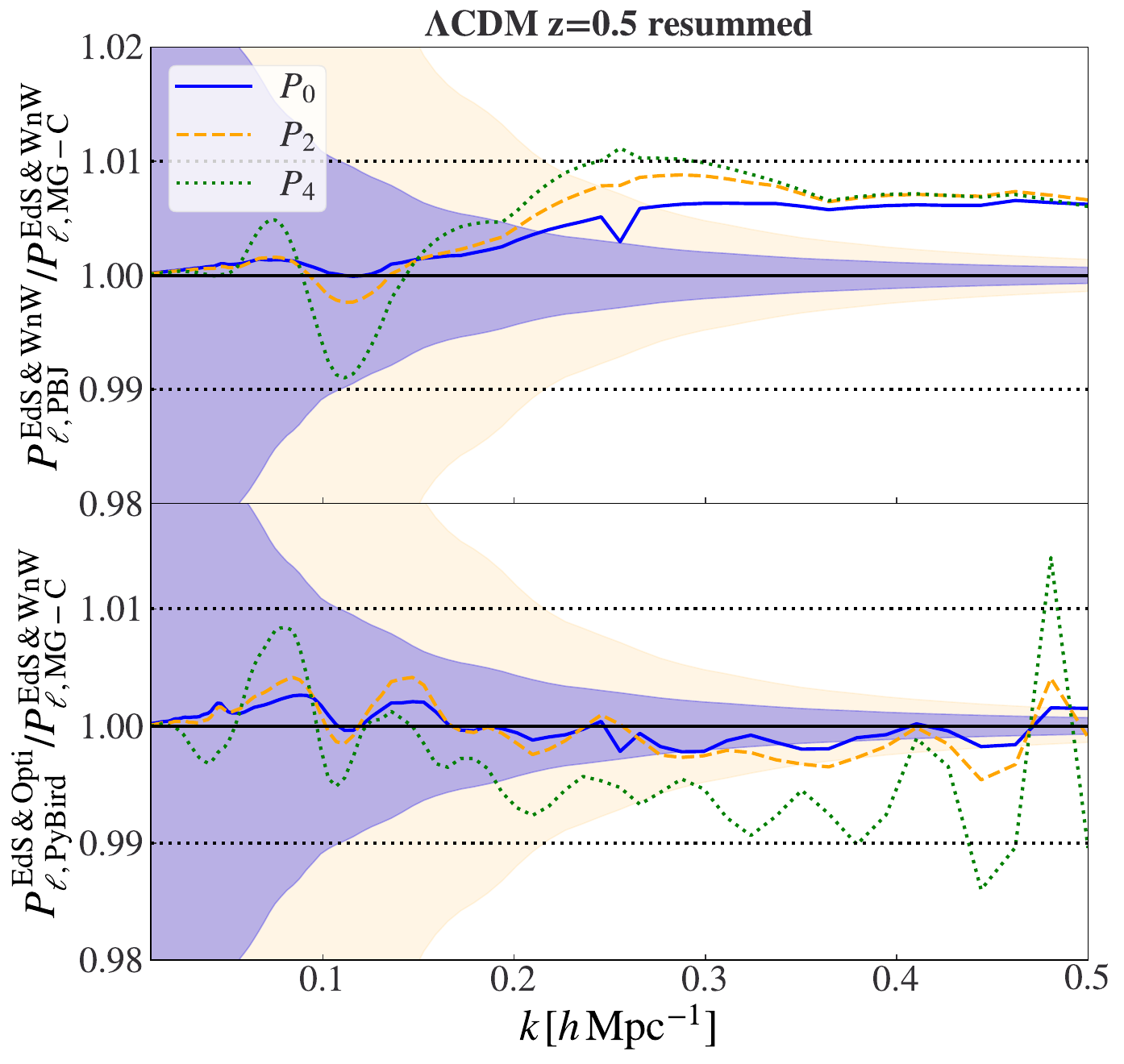}
\includegraphics[width=0.49\textwidth]{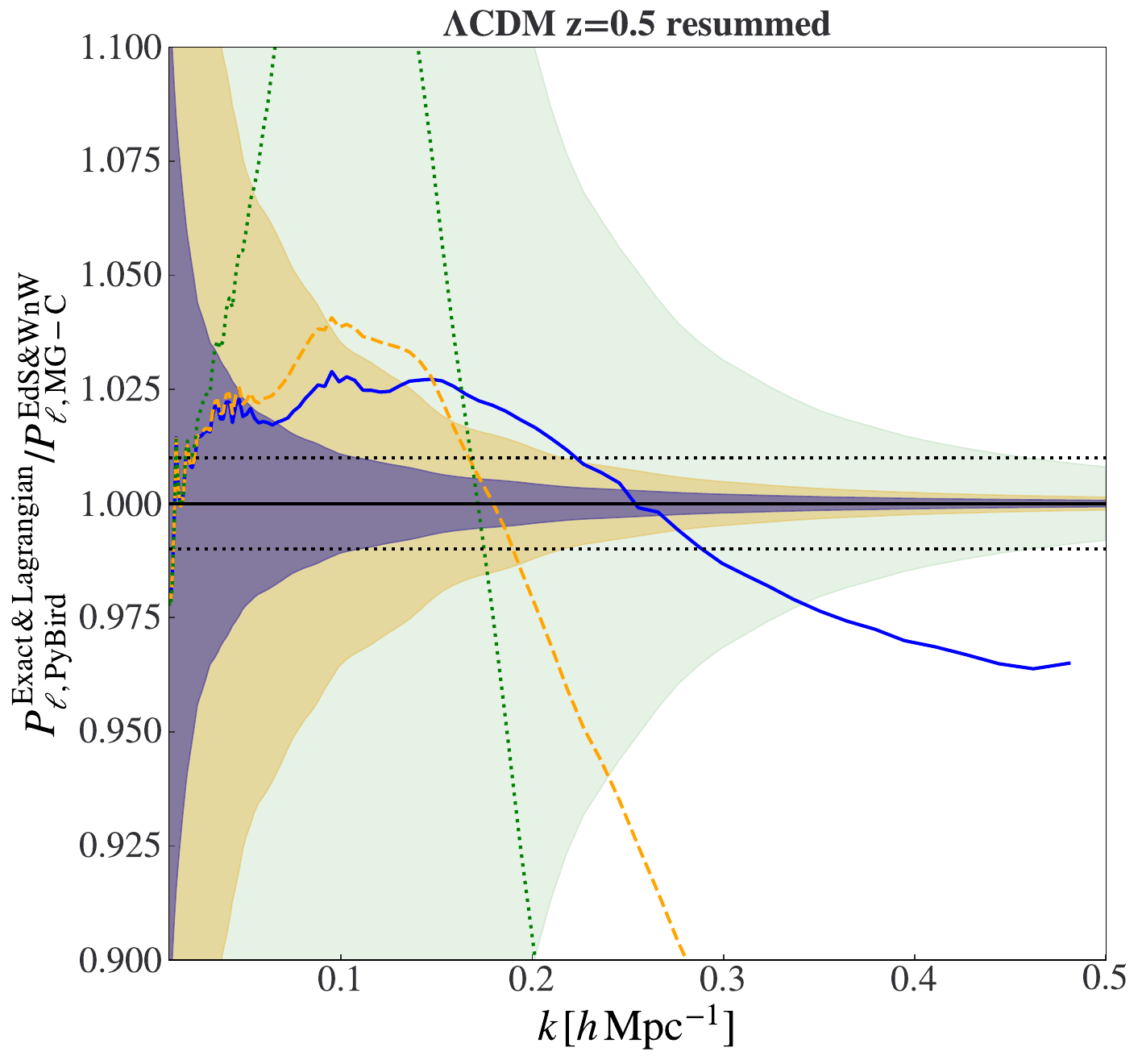}
    \caption{The ratio of the SPT 1-loop monopole (blue solid), quadrupole (orange dashed) and hexadecapole (green dotted) between different codes for $\Lambda$CDM at $z=0.5$. The \texttt{MG-Copter} prediction is taken as the reference in all plots. The left panels show the ratios using the WnW or $\textsc{optiresum}$ resummation. The top left panel shows the \texttt{PBJ} prediction and the bottom panel the \texttt{Pybird} prediction, both predictions being computed using the EdS approximation. The right panel shows the same as the bottom left  plot but now with \texttt{Pybird} applying the full Lagrangian resummation method. Blue and beige bands indicate errors on the monopole and quadrupole assuming a $\Lambda$CDM Gaussian covariance with $V = 8.8 \, h^{-3} \, {\rm Gpc}^3$ and no shot noise contribution. We note that the hexadecapole error fills the left plot and so we have omitted it but include it in the right hand plot as a green band.}
    \label{fig:validation2}
\end{figure*}

In \cref{fig:validation2} we move on to compare the different infrared-resummation implementations, considering only $\Lambda$CDM, and again for $k\leq0.5\, h \, {\rm Mpc}^{-1}$ at $z=0.5$. In this case, \texttt{PBJ} and \texttt{MG-Copter} only have the wiggle-no-wiggle (WnW) decomposition approach (see \cref{sec:resummation}), both within the EdS approximation, while \texttt{Pybird} has the option of both Lagrangian and the approximate $\textsc{optiresum}$ method. The former codes also have slightly different WnW implementations, but despite these differences, we still observe sub-1\% agreement. This suggests that the infrared-resummation prescription will have little impact on inferred parameter posteriors from upcoming data analyses. 

On the other hand, applying the Lagrangian resummation method leads to significant differences, far outside the estimated  measurement errors within the predictions validity range and for all multipoles. As discussed in \cref{scaledep} the presence of the baryon acoustic oscillations scale, $\ell_{\rm osc}$, provides a clear criterion to identify the leading contributions in $k^2$ at each order in perturbation theory. Both WnW and \texttt{PyBird}'s 
 $\textsc{optiresum}$  exploit this criterion and then resum the same class of contributions at each order. The full Lagrangian resummation, on the other hand, is applied to the full power spectrum, and therefore also includes subleading contributions like those in \cref{2der}. These are  genuine higher-order contributions but, on the other hand, there is no guiding principle on why to include them while neglecting other contributions of the same order. 
 
The difference between WnW and the Lagragian resummation may be degenerate with counterterms, which is left to be seen. A full comparison between \texttt{PBJ} and \texttt{Pybird} at the posterior level will be conducted in a forthcoming \Euclid paper (D'Amico et al., in prep.).

\end{appendix}

\end{document}